\documentclass[format=sigconf, nonacm, screen=True]{acmart}
\usepackage{balance}
\usepackage{amsmath} 
\usepackage{booktabs} 
\usepackage[ruled,linesnumbered]{algorithm2e} 

\SetAlFnt{\small}
\SetAlCapFnt{\small}
\SetAlCapNameFnt{\small}
\SetAlCapHSkip{0pt}

\usepackage{makecell}
\usepackage{lscape} 
\usepackage{float}
\usepackage{threeparttable}  
\usepackage{booktabs}
\usepackage{multirow}
\usepackage{makecell}
\usepackage{xcolor}

\hypersetup{colorlinks=true,
    linkcolor=blue,
    filecolor=blue,      
    urlcolor=blue,}
\newif\ifshowcomment
\showcommenttrue
\ifshowcomment
\newcommand{\luyao}[1]{\textcolor{blue}{[luyao] #1}}

\else
\newcommand{\luyao}[1]{}
\fi

\IncMargin{-\parindent}
{
    \theoremstyle{acmplain}
    \newtheorem*{assumption1}{Assumption 1}
    \newtheorem*{assumption2}{Assumption 2}
    \newtheorem*{assumption3}{Assumption 3}
    \newtheorem*{lemma1}{Lemma 1}
    \newtheorem*{lemma2}{Lemma 2}
    \newtheorem*{lemma3}{Lemma 3}
    \newtheorem*{proposition1}{Proposition 1}
    \newtheorem*{definition1}{Definition 1}
    \newtheorem*{definition2}{Definition 2}
    \newtheorem*{definition3}{Definition 3}
    \newtheorem*{definition4}{Definition 4}
    \newtheorem*{definition5}{Definition 5}
    \newtheorem*{definition6}{Definition 6}
    \newtheorem*{definition7}{Definition 7}
    \newtheorem*{definition8}{Definition 8}
    \newtheorem*{definition9}{Definition 9}
    
}

\setcitestyle{authoryear}
\begin{document}

\title[On Blockchain We Cooperate: An Evolutionary Game Perspective]{On Blockchain We Cooperate: An Evolutionary Game Perspective}

\author[Luyao Zhang*]{Luyao Zhang}
 \affiliation{%
  \department{Data Science Research Center and Social Science Division}
  \institution{Duke Kunshan University}
  \country{China}
 }

\authornote{Corresponding author: email: lz183@duke.edu, institutions: Data Science Research Center and Social Science Division, Duke Kunshan University. }
\authornote{Joint first authors. Duke Kunshan University, No. 8 Duke Ave, Kunshan, Suzhou, Jiangsu, China, 215316.}
\authornote{ORCID: \url{https://orcid.org/0000-0002-1183-2254}}

\author{Xinyu Tian}
 \affiliation{%
  \institution{Duke Kunshan University}
  \country{China}
 }

 \authornotemark[2]
 \authornotemark[3]


\begin{abstract}
Cooperation is fundamental for human prosperity. Blockchain, as a trust machine, is a cooperative institution in cyberspace that supports cooperation through distributed trust with consensus protocols. While studies in computer science focus on fault tolerance problems with consensus algorithms, economic research utilizes incentive designs to analyze agent behaviors. To achieve cooperation on blockchains, emerging interdisciplinary research introduces rationality and game-theoretical solution concepts to study the equilibrium outcomes of various consensus protocols. However, existing studies do not consider the possibility for agents to learn from historical observations. Therefore, we abstract a general consensus protocol as a dynamic game environment, apply a solution concept of bounded rationality to model agent behavior, and resolve the initial conditions for three different stable equilibria. In our game, agents imitatively learn the global history in an evolutionary process toward equilibria, for which we evaluate the outcomes from both computing and economic perspectives in terms of safety, liveness, validity, and social welfare. Our research contributes to the literature across disciplines, including distributed consensus in \textit{computer science}, game theory in \textit{economics} on blockchain consensus, evolutionary game theory at the intersection of \textit{biology} and \textit{economics}, bounded rationality at the interplay between \textit{psychology} and \textit{economics}, and cooperative AI with joint insights into \textit{computing} and \textit{social science}. Finally, we discuss that future protocol design can better achieve the most desired outcomes of our honest stable equilibria by increasing the reward-punishment ratio and lowering both the cost-punishment ratio and the pivotality rate.

\textit{We have no eternal allies, and we have no perpetual enemies. Our interests are eternal and perpetual, and those interests are our duty to follow.--- Lord Palmerston, the mid-19th century British Prime Minister}

\textit{The Master said, "When I walk along with two others, they may serve me as my teachers. I will select their good qualities and follow them, their bad qualities and avoid them."
---Then Confucius Analects}
\end{abstract}


\begin{teaserfigure}
    \centering
    \includegraphics[width=1.0\textwidth]{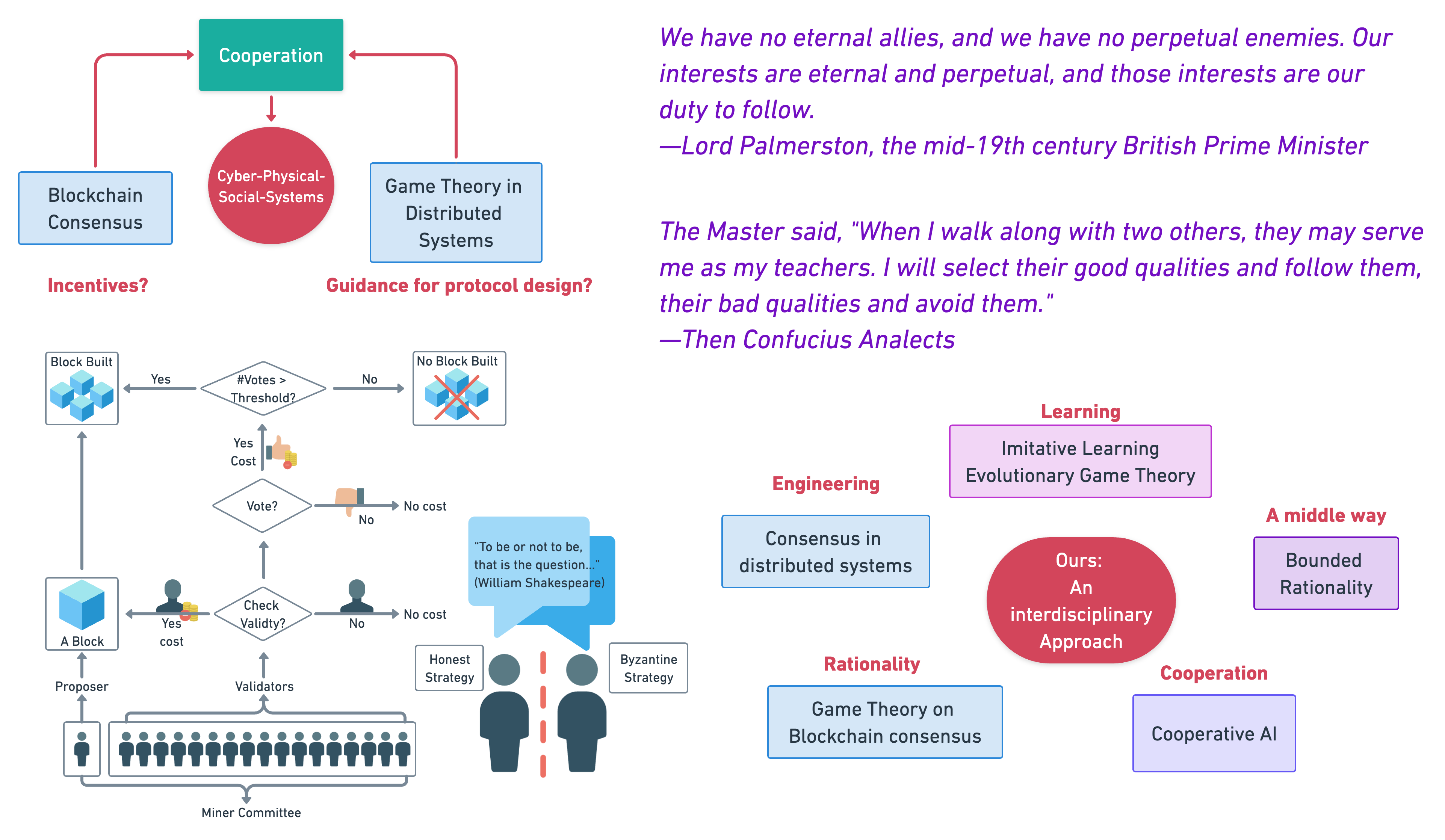}
    \caption{\textbf{On Blockchain We Cooperate: An Evolutionary Game Approach}}
    \label{fig:feature}
\end{teaserfigure}

\maketitle

\section*{acknowledgments}
We are forever indebted to Prof. Vincent Conitzer for introducing us to the cooperative AI literature, especially in the guest lecture entitled \href{https://ce.pubpub.org/pub/cs-econ}{"Computer Science Meets Economics"}. Luyao is grateful to Prof. Chun-Lei Yang and Prof. Gary Charness for introducing her to the assortative matching literature in evolutionary game theory. Luyao thanks the 33rd Stonybrook International Conference on Game Theory for hosting her presentation and helpful insights from the participants, including Prof. Hans Gersbach, Prof. Jiasun Li, and Manvir Schneider and inspirations from the IC3 Camp 2022 attendees including Ari Juels, Ittay Eyal, Andrew Miller, Dahlia Malkhi, and Fan Zhang. Xinyu is supported as a Teaching and Research Assistant by Social Science Support for Undergraduate Research Grant (Division Chairs: Prof. Bill Parson and Prof. Keping Wu) for three related interdisciplinary courses at Duke Kunshan University initiated by Prof. Luyao Zhang: "Intelligent Economics: An Explainable AI Approach," "Computational Microeconomics," and "Introduction to Machine Learning for Social Science." Xinyu Tian and Luyao Zhang are also with SciEcon CIC, an NPO registered in the U.K. aiming at cultivating integrated talents and interdisciplinary research. 

\section{Introduction}
Cooperation is an evolutionary process fundamental for human prosperity~\cite{naturehumanbehaviour_2018_the}. The mechanism of cooperation has been widely studied in a variety of disciplines, such as biology~\cite{koduri_2021_the}, psychology~\cite{henrich_origins_2020}, economics~\cite{fehr_2000_cooperation,fehr_2002_strong, fehr_2004_social}, and computer science~\cite{noamnisan_2007_algorithmic}. Human evolution in
the past one hundred years has witnessed marvelous advancements in artificial intelligence (AI)~\cite{horvitz2016one}. The advancement is integrating cyberspace (CS), physical space (PS), and social space (SS) into the cyber-physical-social system (CPSS), which expands the territories of human civilizations~\cite{wang2019data} extraordinarily. Blockchain further enables cooperative AI~\cite{dafoe_2020_open} by implementing a consensus process. Emerging blockchain technology and distributed systems have further empowered innovation in research and industry~\cite{cong_2018_blockchain} and therefore caught people's appreciation. Consensus protocols are utilized in blockchain  to achieve cooperation in cyberspace. Since~\citet{lamport_1982_the}, computer scientists have contributed considerably to designing consensus protocols that can tolerate a minority of malicious agents to achieve the goals of blockchain safety, liveness, and validity ~\cite{cachin2017blockchain, xiao2020survey}. While computer scientists focus on achieving fault tolerance with a consensus process on the distributed system, economists research how incentives influence  strategic agents' behaviors and affect the system. To further comprehend agent cooperation in the consensus process, emerging literature in game theory introduces rational agents and uses solution concepts to study equilibrium outcomes in the consensus process, including for the social welfare of various consensus protocols ~\cite{saleh2021blockchain}. However, the existing studies with game-theoretical implementations are limited in three facets: (1) they model the consensus process as a static game rather than considering the possibility of a dynamic consensus-building process, (2) ignore agents' abilities in imitative learning \cite{noamnisan_2007_algorithmic} from historical observations, and (3) lack a comprehensive evaluation of design choices considering both computing and economic properties. Moreover, the current blockchain consensus protocols are far from living up to their promise and still have great potential in promoting cooperation with rationality \cite{budish_2022_the}. To advance the literature and advise the designs, we strive to answer three research questions (RQs):

\begin{enumerate}
    \item \textbf{RQ1}: How can a consensus protocol be abstracted as a general game environment in extensive form~\cite{kroer2014extensive} that consists of the game tree, the set of agents, and the payoffs?
    \item \textbf{RQ2}: How does the behavior of boundedly rational agents evolve to stable equilibria that differ in desirable outcomes of validity, termination, and social welfare?
    \item \textbf{RQ3}: How can the game theoretical study of boundedly rational agents guide future designs of blockchain consensus protocols?
\end{enumerate}

We successfully abstract a general Byzantine consensus protocol as a game environment, apply bounded rationality and imitative learning strategies to model agent behaviors, and use the evolutionary stable strategy (ESS) \cite{smith_1973_the} to solve the three different stable equilibria. In our design, blockchain consensus is achieved within an evolutionary process when the strategies of bounded rational agents change under the imitative learning process along with the game rounds. In our model, agents learn imitatively by responding to incentives of strategies based on historical observations. The bounded-rational agents are able to choose their strategy before each round of the game. 

Our general Byzantine consensus protocol suggests three stable equilibria: the honest equilibrium when all agents cooperate, the Byzantine equilibrium when all agents defend, and the pooling equilibrium when a portion of agents cooperate while others defend. Among all three equilibria, our honest equilibria achieve the desired outcomes in both computing and economic aspects of safety, liveness, validity, and social welfare. 

Based on the three equilibria and the consensus process, our research also guides future designs of consensus protocols to achieve desired outcomes with four parameters related to the \textit{reward}-\textit{punishment} \textit{ratio}, \textit{cost-punishment ratio}, \textit{pivotality rate}, and \textit{assortative matching parameter}. We take Ethereum 2.0 upgrades as an example and explain how our work can suggest the possible optimization of some specific policy parameters. We show that future protocol design can better achieve the most desired outcomes of our honest and stable equilibria by increasing the reward-punishment ratio and lowering both the cost-punishment ratio and pivotality rate.

Fig. ~\ref{fig:feature} demonstrates the philosophy and perspective of our studies. We organize the paper as follows. In Section 2, we discuss the related literature in terms of five facets. In Section 3, we abstract the general Byzantine consensus protocol to an extensive form game. We define three equilibria and solve their initial conditions in Section 4. Section 5 discusses the implications of our results for future designs and research on blockchain consensus protocols. Sections 3, 4, and 5 answer research questions 1, 2, and 3, respectively. We provide Tables and Figures in Section 6. We also provide a glossary table and a notation table in the Appendix. 

\section{Related Literature}
Our research mainly contributes to five streams of literature across these disciplines: (1) the Byzantine consensus protocol in \textit{computer science}, (2) game theory in \textit{economics} on blockchain consensus,(3) evolutionary game theory at the intersection of \textit{biology} and \textit{economics},  (4) bounded rationality at the interplay between \textit{psychology} and \textit{economics}, and (5) cooperative AI  at the interface of \textit{computer science} and \textit{economics}.

\subsection{Consensus in Distributed Systems}
The Byzantine generals problem, proposed by \citet{lamport_1982_the}, describes the network failure in a distributed system caused by malicious agents or corrupt nodes. Many Byzantine fault tolerance (BFT) protocols have been developed thus far (i.e., Zyzzyva \cite{kotla_2007_zyzzyva}, HotStuff \cite{yin_2019_hotstuff}, RBFT \cite{aublin_2013_rbft}, and MinBFT \cite{veronese_2013_efficient}), and the practical Byzantine fault tolerance (PBFT) has become one of the most popular protocols. Much recent research has perfected the study of Byzantine generals in various settings \cite{khanchandani_2021_byzantine, lewispye_2022_byzantine}. By running Byzantine consensus protocols, the distributed systems are able to tolerate a proportion of existing Byzantine agents and still achieve consensus. Compared with other protocols (i.e., Proof-of-Work (PoW) \cite{nakamoto_2008_bitcoin}, Proof-of-Stake (PoS) \cite{saleh_2018_blockchain}), PBFT and its variants (\cite{tong_2019_trustpbft,wang_2021_an}) with $30\%$ fault tolerance  proved to have superiority confronting the Byzantine general problems. 
Systematized by \citet{bano_2019_sok}, consensus in a distributed system has been developed to satisfy multitudinous requirements. In contrast from others, \citet{bano_2019_sok} analyzed the evolution of consensus protocols toward a systematization of knowledge (SoK). The SoK highlights consensus protocols in blockchain to satisfy higher requirements and apply to multiple scenarios compared to those in classical distributed systems. 

Computer scientists endeavor to improve a consensus protocol by engineering approaches that, for example, reduce the cost and simplify the replicas in the state machine \cite{tong_2019_trustpbft} or optimize the efficiency in communications \cite{wang_2021_an}. In contrast, our research provides an evolutionary game perspective to evaluate the consensus process for learning agents driven by incentives. Our perspective can be applied to evaluate any consensus protocol with \textit{leader selections} and \textit{committee voting} by simple variations of the game environment, including proof-of-work (PoW) \cite{nakamoto_2008_bitcoin}, proof-of-stake (PoS) \cite{saleh_2018_blockchain}), hybrid single committees (\cite{gilad_2017_algorand,Pass2017HybridCE}), and hybrid multiple committees \cite{kokoriskogias_2018_omniledger,albassam_2017_chainspace}. \footnote{Committee-based blockchains have superiority in reducing fork risk \cite{biais_2019_the,amoussouguenou_2020_rational}, expanding scalability \cite{tron_2018_advanced}, and performing robustness and efficiency \cite{benhaim_2021_scaling,auer_2021_distributed}.} Among all the existing blockchain systems, Algorand is the closest application scenario~\cite{gilad_2017_algorand, chen_2017_algorand}. Moreover, our results guide optimizing the incentive scheme in consensus protocols to achieve desirable outcomes on top of computing designs. 

\subsection{Game Theory on Distributed Systems}
\begin{figure}
    \centering
    \includegraphics[width=8.5cm]{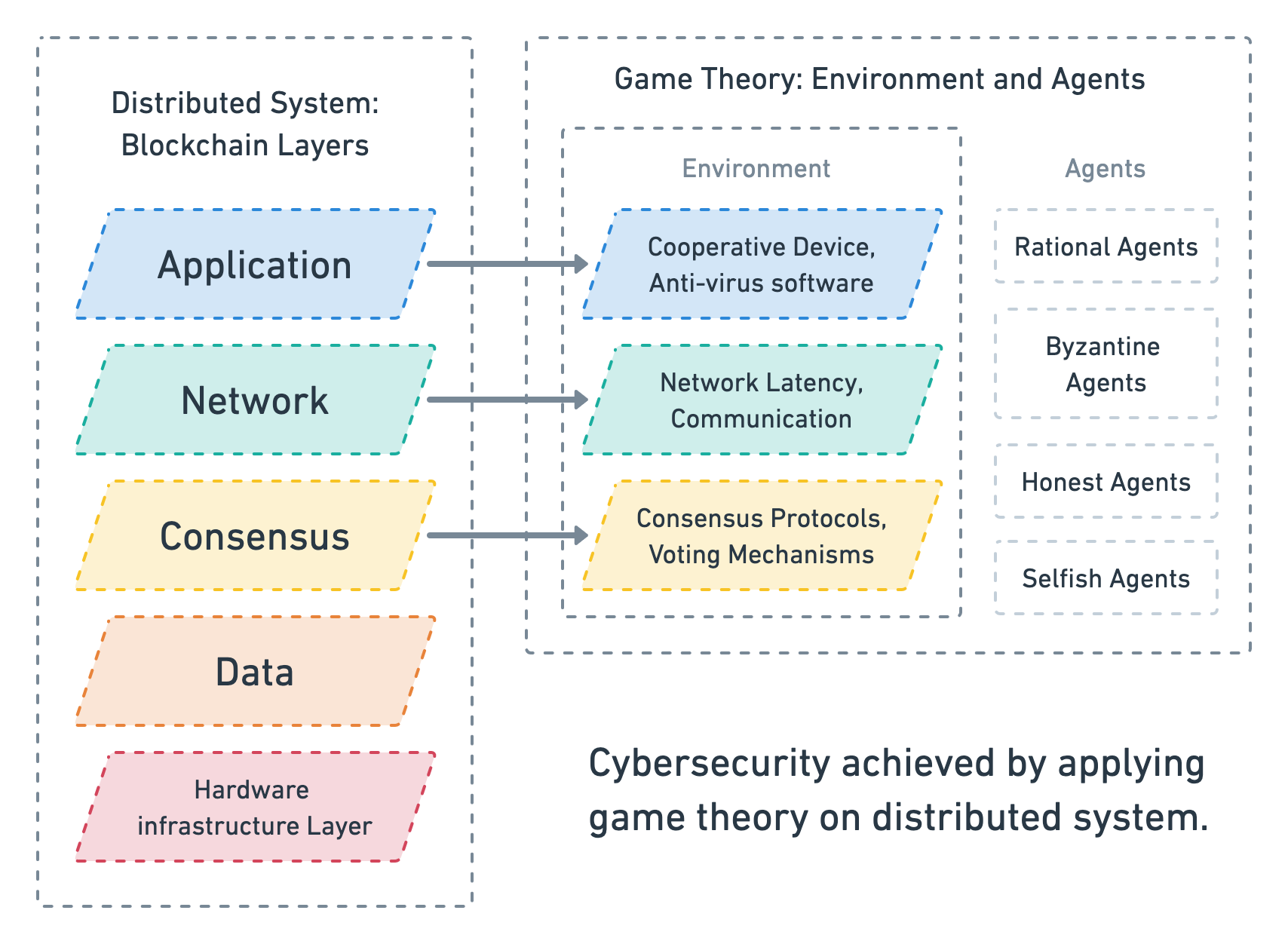}
    \caption{Game theory on distributed systems}
    \label{fig:8}
\end{figure}

Dating back to 2005, \citet{aiyer_2005_bar,abraham_2006_distributed}, and \citet{moscibroda_2006_when} are among the pioneers who applied game theory to study incentives in distributed systems for security concerns.  \citet{Azouvi2021SoK} analyzed the literature that further applies game-theoretical models to study the security of cryptocurrencies. Other studies \cite{xue_2020_incentive, chen_2019_an} may focus on miner utilities. According to the existing categorizations~\cite{zhang2022sok} of blockchain architectures, including the hardware infrastructure layer, data layer, consensus layer, network layer, and application layer, we review the studies of game theory on distributed systems in the taxonomy of consensus, network, and application layers. Figure ~\ref{fig:8} summarizes insights in the literature about how cybersecurity can be achieved by applying game theory to distributed systems in the consensus, network, and application layers, using the dichotomy of the game environment and agent.

\textbf{Application Layer:} The implementation of game theory on the application layer is mainly conducted using assistant devices or software. For instance, \citet{aiyer_2005_bar} utilized a cooperative service with rational users to achieve fault tolerance on BAR (Byzantine, Altruistic, Rational) model. 

\textbf{Network Layer:} Studies on the network layer mainly focus on tolerating the attacks that appear when long network latency occurs \cite{cryptoeprint:2022/404, neu_2022_longest, guo_2022_bitcoins} and the information forward process in network communication \cite{apostolaki_2018_sabre}. From this perspective, while studies on PBFT variants have contributed greatly to synchronous \cite{abraham_2017_efficient}, asynchronous \cite{wang_2020_asynchronous} and partial asynchronous systems, \citet{abraham_2006_distributed} used k-resilient Nash equilibria to analyze the information communication on distributed multiagent systems.

\textbf{Consensus Layer} Instead of the application and network layers, our research contributes to the specific literature focusing on the consensus layer. Existing studies on the consensus layer apply different game theory solution concepts or behavioral strategies (e.g., Byzantine, honest, selfish agents, etc.), each on different consensus protocols. Table~\ref{tab:1} represents several recent game theory studies on blockchain consensus considering the effect of incentives on miner behaviors in the facets of consensus protocols, game theory solution concepts, agent types, and aspects for evaluations. \citet{article} mentioned incentives in the design of consensus protocols but did not provide a formal game theory study to evaluate performance. \citet{cryptoeprint:2022/308} also pioneered the design of the first incentive-compatible blockchain consensus protocol named \textit{Colordag} that implements an epsilon-Name Equilibrium \cite{jackson2012epsilon}.  \citet{biais_2019_the,amoussouguenou_2020_rational, halaburda_2021_an} and \citet{saleh_2018_blockchain} abstracted the consensus protocol to a dynamic game of imperfect information~\cite{fudenberg1991game} and further introduced rational agents, game theory solution concepts, and welfare analysis to study a variety of consensus protocols, including the PoS~\cite{saleh_2018_blockchain}, the PoW ~\cite{biais_2019_the}, and the PBFT variants ~\cite{biais_2019_the,amoussouguenou_2020_rational,halaburda_2021_an}. The above studies focus on the solution concepts of perfect Bayesian equilibrium (PBE,~\cite{fudenberg1991perfect}) or its refinement, Markov perfect equilibrium (MPE,~\cite{maskin2001markov}.

Our research contributes to the literature by providing a generalized BFT consensus protocol as a game environment, introducing bounded-rational agents with learn ability to the system, and using an evolutionary stable strategy (ESS,~\cite{smith1973logic,smith1979game} to model agent behavior and evaluate equilibria.

\subsection{Evolutionary Game Theory}
Emerging from \citeauthor{darwin_1859_on}'s [\citeyear{darwin_1859_on}] evolutionary study, evolutionary models have been applied widely in many interdisciplinary studies. In \textit{macroeconomics}, evolutionary models were derived from the debate of Homo Economics and irrational humans to explain complex phenomena in financial markets \cite{levin_2021_introduction,henrich_2001_in}. In \textit{computer science},  evolutionary models are applied to solve graphical problems and estimate computational complexity \cite{noamnisan_2007_algorithmic,yoavshoham_2009_multiagent}. In game theory, the Prisoner's Dilemma, as the most classic example, has been studied with multiple evolutionary game designs \cite{axelrod_1981_the,yang_2019_cooperation,yoavshoham_2009_multiagent}, whose steady state proved to be an equilibrium convergence \cite{noamnisan_2007_algorithmic,yoavshoham_2009_multiagent} and is resilient to small mutual invasions \cite{yoavshoham_2009_multiagent}. Researchers have been applying evolutionary game theory to study cooperation mechanisms for a long time \cite{yoavshoham_2009_multiagent,axelrod_1981_the}. Our research advances the literature by extending the application scenario of evolutionary game theory to blockchain consensus protocols, especially in cooperation among agents. 

\subsection{Bounded Rationality}
Another core assumption of our approach is the bounded rationality of agents \cite{simon_1955_a,chase_1973_perception,rubinstein_2002_modeling}. Compared to rational agents, the boundedly rational is allowed to be  limited in either forming subjective beliefs consistent with the objective situations or best responding. In contrast to behavior agents, the boundedly rational deviates from the rational in a more systematic way and can learn to converge to the rational in extreme cases. Agents in our model are bounded-rational in three facets: 
\begin{enumerate}
    \item Agents are constrained to choose among a limited set of strategies, i.e., either honest strategy or byzantine strategy following classical literature in distributed systems, but the agents are able to learn to react to incentives based on historical observations;
    \item In each round of the game, agents choose strategies based on the current state that includes historical information, but without the foresight of the future state \cite{conlisk_1996_why};
    \item Agents are allowed to hold the subjective belief both consistent or inconsistent to the objective probability to meet agents of the same strategies, which is supported by research in bounded rationality and game theory ~\cite{levin_2020_bridging,rubinstein2016}, the false consensus effect in psychology \cite{ross_1977_the}, the interdisciplinary research about computer science and game theory \cite{halpern_2007_computer}, and the assortative matching literature in behavioral science~\cite{eshel_1982_assortment,angelucci_2017_assortative,durlauf_2003_is,shimer_2000_assortative,yang_2019_cooperation,eeckhout_2018_assortative}. 
\end{enumerate}
Our research extends bounded rationality to study the dynamics of miner behaviors in blockchain consensus. Classic game theory in \textit{economics} assumes full rationality. In contrast, consensus protocol studies in \textit{computer science} start with behavioral agents, the honest and the Byzantine. Following Aristotle's doctrine of the mean, ours is a middle way to bridge the two. 

\subsection{Cooperative AI}
Cooperative AI studies how artificial intelligence (AI) can contribute to solving cooperative problems in AI-AI, AI-human, and human-human interactions~\cite{conitzer_2019_designing,dafoe_2020_open,dafoe_2021_cooperative,conitzer_foundations}. Different from other AI technologies, blockchain is the coined decentralized AI that utilizes cryptography and on-chain governance to support research and innovations~\cite{wang_2007_formal,sen_2013_a,xing_2018_the,marwala_2018_blockchain,salah_2019_blockchain,harris_2019_decentralized,singh_2019_blockiotintelligence,cao_2022_decentralized}. However, applying the methods in the existing cooperative AI literature to study cooperation on the new application scenarios of blockchain is largely unexplored. 

On the blockchain, as a trust machine, miners originally had the commitment and had to work cooperatively. This feature comes from the nature of blockchain. As a distributed system, blockchain is built by miners. While the miners are incentivized by the block rewards and future transactions on the block, their welfare is maximized when the stability of the blockchain is also maximized. Any malicious behavior or system hack would cause a great loss in miner rewards. Therefore, it is the nature of blockchain that miners are to have a consensus and work cooperatively in the block-building process. Our research contributes to cooperative AI by providing an application scenario in the blockchain consensus process. 

Our research provides a game-theoretical environment and strategic reasoning for blockchain as a decentralized AI system to support cooperative dynamics to extend the application scenario of cooperative AI to the blockchain consenus design. 

\section{Model}
Our model is based on a generalized BFT blockchain where
\begin{enumerate}
    \item Agents are selected to form different parallel committees before the game to validate transactions.
    \item Each selected agent has exactly one unit of voting weight to participate in the mining process.
    \item Agents are able to observe the global history of aggregate payoffs and choose their strategy before each round of the game.
\end{enumerate}

In our model, we only focus on one miner committee and play an evolutionary game where a mining committee participates in an $n$ round mining game. The model can then be extended to multiple committees. Please note that in our paper, "miner" and "agent" are interchangeable. 

\subsection{Game Environment}
\begin{figure}[!htbp]
	\includegraphics[scale=0.135]{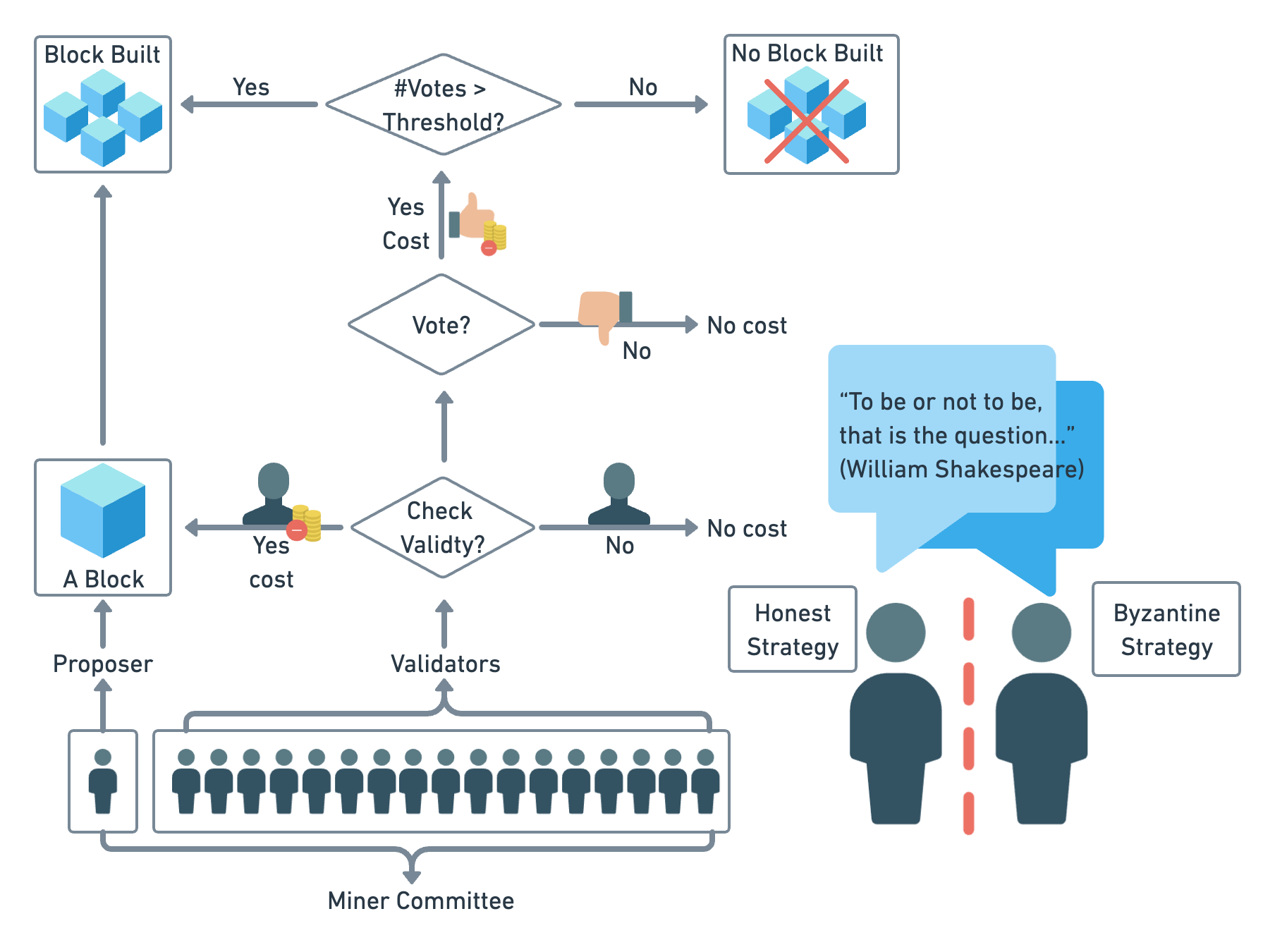}
	\caption{Byzantine consensus-based blockchain model.}
	\label{fig:1}
\end{figure}

We follow a general Byzantine consensus model with a PBFT consensus protocol. To facilitate interdisciplinary conversations, we further abstract the consensus protocol as a game environment. The environment consists of the game tree, the set of agents, and the payoffs, as shown in Fig. ~\ref{fig:1}.

\subsubsection{Set of agents}
\begin{definition1}[Miner Committee]
    We use an ordered set $\mathcal{A} = \{A_{i}\}_{i=1}^{N}$ to denote the committee of $N$ miners established at time $t = 0$. The committee set will not change until the end of the game.
\end{definition1}

\begin{definition2}[Proposer]
    In round $t, \text{ } t \in \{1,\cdots,n\}$, an agent $A_{i}$ is randomly selected to be proposer $P_{t}$. $P_{t}$ makes a proposal $h_{t}$ that contains a block and its validity. 
\end{definition2}

\begin{definition3}[Validator]
    In round $t, \text{ } t \in \{1,\cdots,n\}$, agents except for $P_{t}$ are validators. The validators  check the proposal's validity and vote for it when the block in the proposal is valid.
\end{definition3}

\subsubsection{Game tree}
\begin{figure}[!htbp]
	\includegraphics[scale=0.27]{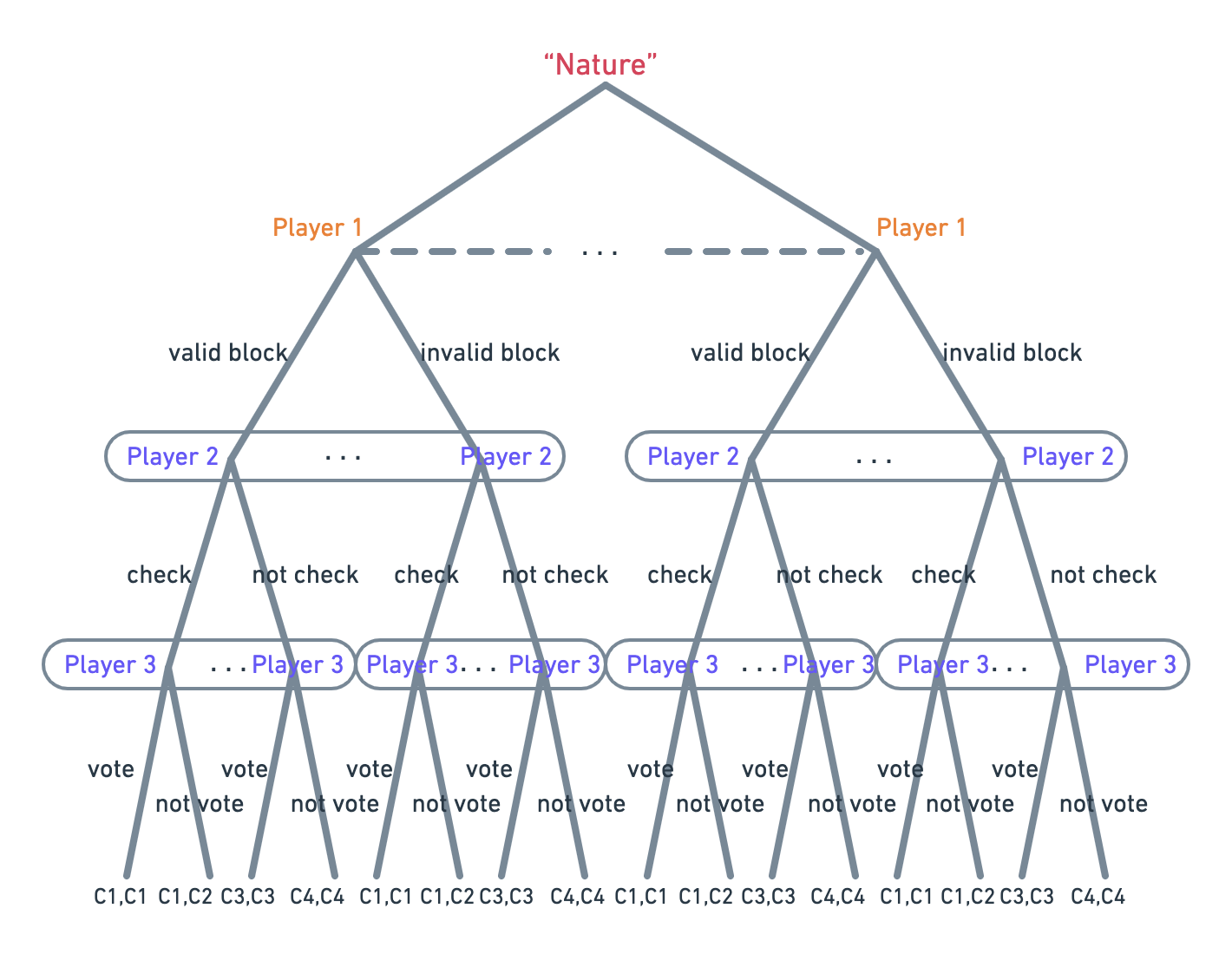}
	\caption{Game Tree. (notations: player 1: proposer; player 2, player 3: validator; $C1 = -(\text{cost of check}+\text{cost of vote})$; $C2 = -\text{cost of check}$; $C3 = -\text{cost of vote}$; $C4 = 0$.)}
	\label{fig:6}
\end{figure}

\
\newline
In  miner committee $\mathcal{A}$, in round $t$, proposer $P_{t}$ is first selected in a round-robin fashion, and she proposes  proposal $h_{t}$, which consists of a block and its validity value. The remaining miners in the committee, acting as validators, then choose to check the validity of $h_{t}$ and vote for it. The proposal will be accepted if the number of votes exceeds the majority threshold $\nu$ in a fixed period. The game tree is shown in Fig. ~\ref{fig:6} (since the final payoffs are relevant to the proportion of two strategies, Fig. ~\ref{fig:6} only shows the cost of actions).

\subsubsection{Payoffs}
\ 
\newline
The payoffs of the agents include $R$, the reward to the validators who send a message when the block is accepted; $c_{send}$, the cost to the validators who send a message; $c_{check}$, the cost to the validators who check the validity of the proposed block; and $\kappa$, the cost occurs to all validators with the honest strategy when an invalid block is accepted. In our model, we set $R>c_{check}>c_{send}>\kappa$. \footnote{ \citet{10.5555/3398761.3398772} assume $\kappa>R>c_{check}>c_{send}$. Ethereum~\cite{ethereumdevelopers_2022_ethereum, bitinfocharts_2019_ethereum} sets $R=\kappa > c_{send}$ without mentioning $c_{check}$ explicitly. In Section 5, we discuss the results of those variants of our assumption and show that our benchmark assumption is more desired considering both computing and economic properties.}

In practice, $R$ is the accumulation of multiple validator rewards (e.g., attestation rewards mentioned in \citet{ethereumdevelopers_2022_ethereum} and \citet{buterin_2020_combining}); $c_{check}$ is the cost of verifying each transaction by checking user signature and funds in an account \cite{ethereumdevelopers_2022_ethereum}; $c_{sent}$ is the cost of voting for transactions on a block; $k$ is the penalty for a validator's dishonest behaviors (e.g., attestation penalties mentioned in \citet{ethereumdevelopers_2022_ethereum} and \citet{buterin_2020_combining}).

\subsection{Bounded-Rational Agents}
We model the miners as bounded-rational agents in an evolutionary game who:
\begin{enumerate}
    \item in each round, only choose between the honest strategy and the byzantine strategy;
    \item hold subjective beliefs;
    \item follows an imitative updating rule.
\end{enumerate}

\subsubsection{Strategies}
\ 
\newline
In round $t$, each of the $N$ miners choose a strategy $s_{i} \in {\{S_{H},S_{B}\}}$, where $S_{H}$ is the honest strategy and $S_{B}$ is the byzantine strategy. To explicitly define the two strategies with the bounded-rational agents, we suppose that before round $t$ starts, each validator checks if the group of their congeners has pivotality \cite{hoffman_2013_one}, which means if the number of their congeners exceeds the majority threshold $\nu$. 

\begin{definition4}[Pivotal Strategy]
    In round $t$, $S_{H}$ is pivotal if and only if $Nx_{t-1} > \nu$; correspondingly, $S_{B}$ is pivotal if and only if $N(1-x_{t-1}) > \nu$.
\end{definition4}

\begin{figure}[!htbp]
	\includegraphics[scale=0.2]{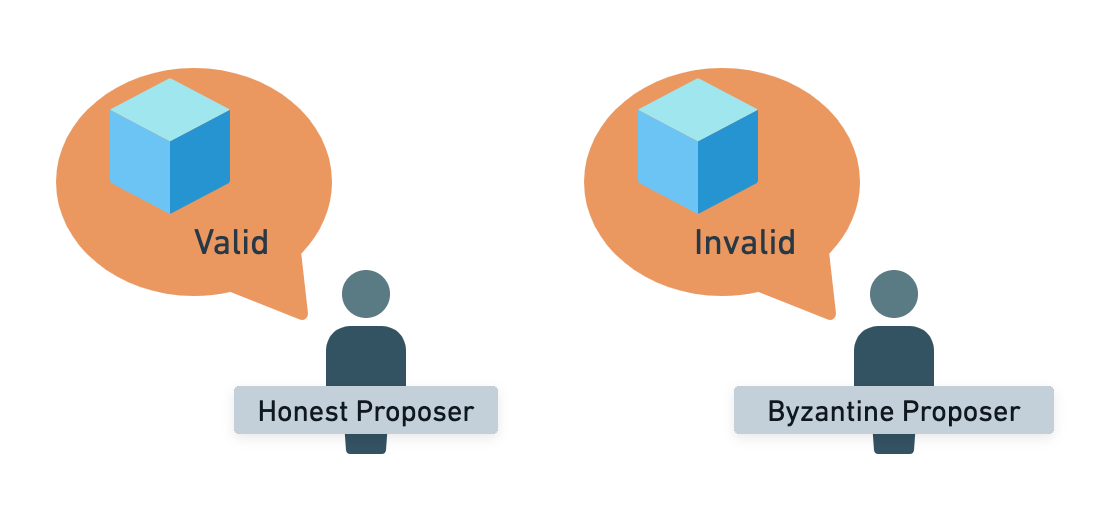}
	\caption{Proposer actions with different strategies.}
	\label{fig:2}
\end{figure}

\begin{figure}[!htbp]
\centering
\begin{minipage}[t]{0.5\textwidth}
\centering
\includegraphics[width=6.5cm]{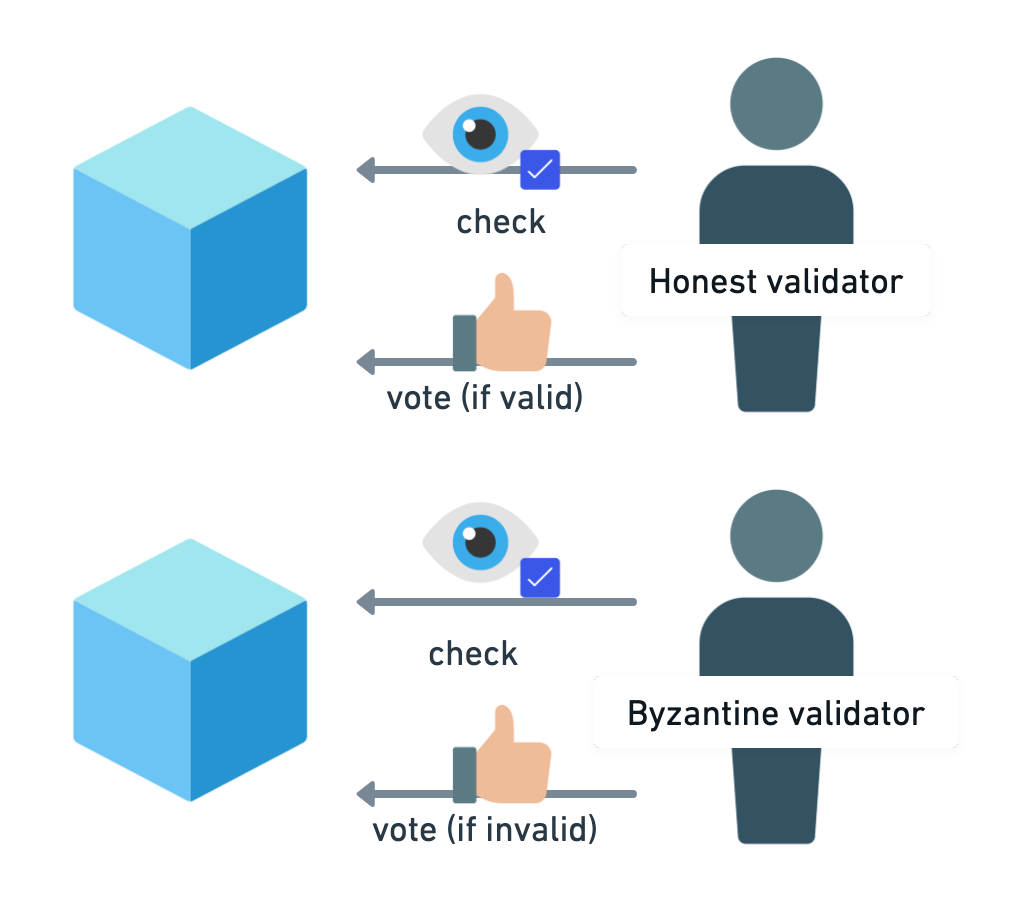}
\caption{Validator actions with pivotality.}
\label{fig:3}
\end{minipage}
\begin{minipage}[t]{0.5\textwidth}
\centering
\includegraphics[width=6cm]{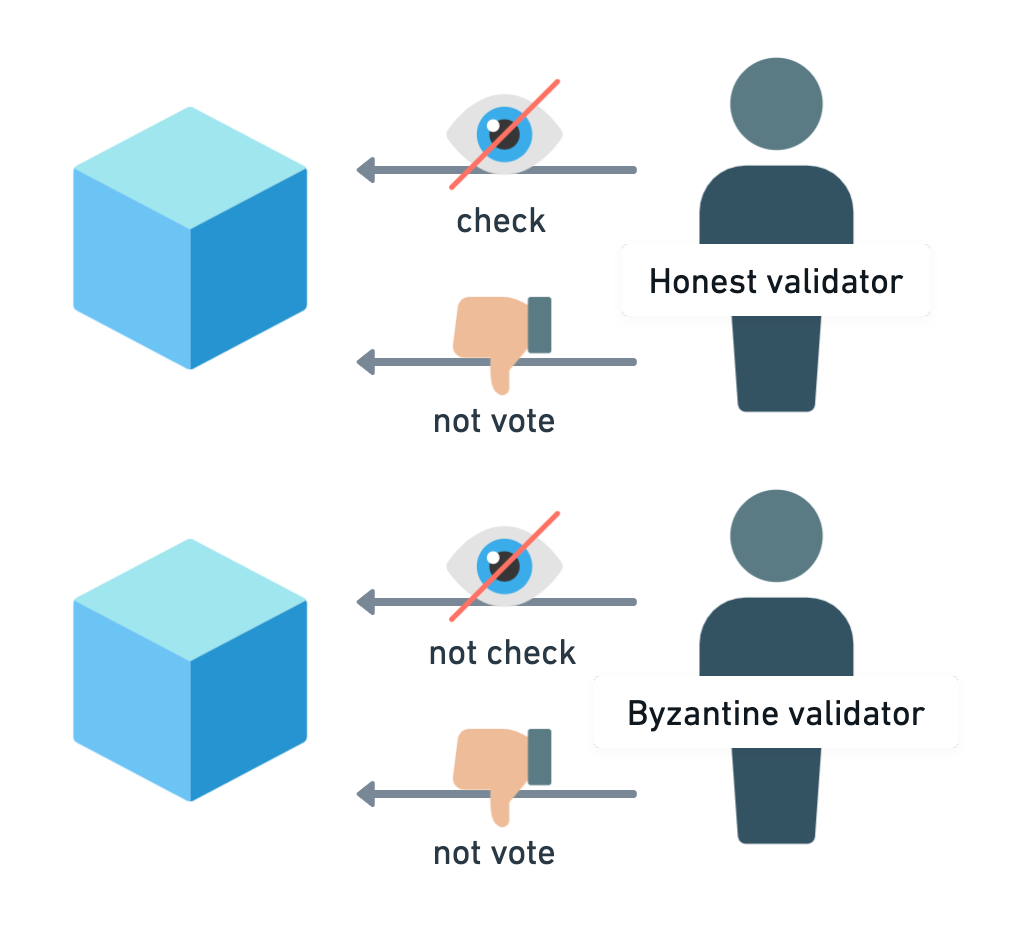}
\caption{Validator actions without pivotality.}
\label{fig:4}
\end{minipage}
\end{figure}

\begin{definition5}[Honest Strategy]
    We use $S_{H}$ to denote the honest strategy, which asks a miner to achieve the consensus protocol. When playing the honest strategy:
    \begin{enumerate}
        \item if pivotal, a proposer proposes a valid proposal (Fig.~\ref{fig:2}); if not pivotal, do nothing.
        \item if pivotal, a validator checks the proposal's validity and votes for it if the proposal is valid (Fig.~\ref{fig:3});
        \item If not pivotal, a validator neither checks the proposal's validity nor votes for it (Fig.~\ref{fig:4}). 
    \end{enumerate}
\end{definition5}

\begin{definition6}[Byzantine Strategy]
    We use $S_{B}$ to denote the Byzantine strategy, which asks a miner to damage the consensus protocol. An agent playing the Byzantine strategy in our paper has the representation of a malicious player. When playing the Byzantine strategy:
    \begin{enumerate}
        \item if, pivotal, a proposer proposes an invalid proposal (Fig.~\ref{fig:2}); if not pivotal, do nothing
        \item If pivotal, a validator checks the proposal's validity and votes for it if the proposal is invalid (Fig.~\ref{fig:3});
        \item If not pivotal, a validator neither checks the proposal's validity nor votes for it (Fig.~\ref{fig:4}). 
    \end{enumerate}
\end{definition6}

Proposers and validators of the same strategies and pivotal status have the same payoff. Other alternative strategies are dominated by the honest or the Byzantine strategy. 

\begin{assumption1}[Game Initials]
    We assume that all the participants are fixed before round $1$ so that no one would quit or be added to any round. We use $x_{t}$ to denote the proportion of miners with $S_{H}$ in round $t$. We assume that $x_{1}$ of the miners choose the honest strategy in the initial round.
\end{assumption1}

\subsubsection{Subjective Belief}
\ 
\newline
Inspired by the assortative matching theory \cite{eshel_1982_assortment,angelucci_2017_assortative,durlauf_2003_is,shimer_2000_assortative,yang_2019_cooperation,eeckhout_2018_assortative} from the evolutionary stable strategy (ESS) \cite{smith_1973_the}, we make assumption 2 about the validators' subjective belief to meet a proposer with the same strategy.
\begin{assumption2}[subjective belief with assortative matching theory]
    We  $\textit{use} m, \ m\in [0,1]$ to represent the portion of the rounds that a validator believes to meet a proposer with the same strategy in the game. We use Algorithm ~\ref{algorithm1} to describe the matching process in a mining game. For instance, there are two extreme cases: 
    \begin{itemize}
        \item When $m = 1$, a validator believes that in arbitrary round $t, \text{ } t \in \{1,\cdots,n\}$, proposer $P_{t}$ plays the same strategy as she does with probability 1;
        \item When $m = 0$, a validator believes that in arbitrary round $t, \text{ } t \in \{1,\cdots,n\}$, proposer $P_{t}$ plays the same strategy with the probability of a uniform random draw from the population.     
    \end{itemize} 
\end{assumption2}

\begin{algorithm}
\caption{Assortative Matching Algorithm}\label{algorithm1}
\KwData{assortative match probability $m$, round of game $t$, a committee $\mathcal{A}$ of $N$ agents that established at $t=0$}
\KwResult{In every round, each validator is believed to be matched either with a proposer of the same strategy with probability $m$ or with a proposer of a uniform random draw from the population with probability $(1-m)$}
    \For {$t$ $\leftarrow$ 1 $\KwTo$ $n$} {
        \For {$i$ $\leftarrow$ 1 $\KwTo$  $N$} {
	        num $\leftarrow$ rand(0,1)\;
		    \While {num $<$ m}{
                $A_{j}$ $\leftarrow$ rand($\mathcal{A}$)\;
                \If {$s_{i} == s_{j}$ and $i \neq j$}{
                    Match $A_{i}$ and $A_{j}$\;
                }
                break\;
            } 
            \While {num $>$ m}{
                $A_{j}$ $\leftarrow$ rand($\mathcal{A}$)\;
                \If {$i \neq j$}{
                    Match $A_{i}$ and $A_{j}$\;
                }
                break\;
            } 
	    }
    }
\end{algorithm}

Assumption 2 is also consistent with the false consensus effect \cite{ross_1977_the} as $m \geqslant 0$. Therefore, the validators' expected payoff will be updated according to assumption 2.

\begin{lemma1}[Subjective Meeting Probabilities]
We use $\pi_{ij}(x_{t}),$ $i,j \in \{H, B\}$ to denote the subjective meeting probability of one validator with strategy $S_{i}$ meeting one proposer with strategy $S_{j}$ in the function of $x_{t}$.
    \begin{align}
    \begin{aligned}\label{eq:1}
        \pi_{HH}(x_{t}) &= m + (1-m)x_{t}\\
        \pi_{HB}(x_{t}) &= (1-m)(1-x_{t})\\
        \pi_{BH}(x_{t}) &= (1-m)x_{t}\\
        \pi_{BB}(x_{t}) &= 1-(1-m)x_{t}
    \end{aligned}
    \end{align}
\end{lemma1}

\begin{lemma2}[Agents Expected Payoff]
We use $V_{ij}(x_{t})$ to denote the expected payoff that one validator with $S_{i}$ meets a proposer with $S_{B}$ in function of $x_{t}$, where $i,j \in \{H,B\}$. We also use $V_{i}(x_{t}), \text{ } i \in \{H, B\}$ to denote the expected validator payoff with the subjective belief of one validator with strategy $S_{i}$ in the function of $x_{t}$.
    \begin{align}\label{eq:2}
    \begin{aligned}
        V_{H}(x_{t}) &= \pi_{HB}(x_{t})V_{HB}(x_{t}) + \pi_{HH}(x_{t})V_{HH}(x_{i})\\
        V_{B}(x_{t}) &= \pi_{BB}(x_{t})V_{BB}(x_{t}) + \pi_{BH}(x_{t})V_{BH}(x_{i})
    \end{aligned}
    \end{align}
Moreover, as validators' behaviors are crucial to consensus building and evolutionary game updates, we treat the proposers' expected payoff the same as the validators' payoff as long as they play the same strategy.
\end{lemma2}

\subsubsection{Imitative Updating Rule}
\begin{assumption3}[Imitative Updating Rule \cite{noamnisan_2007_algorithmic}]
Before round $t$, for one individual bounded-rational miner, she has $P_{H_{t}}$ probability of choosing the honest strategy in round $t$ and $P_{B_{t}}$ probability of choosing the Byzantine strategy in round $t$.
    \begin{align*}
        P_{H_{t}} = \frac{x_{t-1}V_{H}(x_{t-1})}{x_{t-1}V_{H}(x_{t-1})+(1-x_{t-1})V_{B}(x_{t-1})}
    \end{align*}
    \begin{align*}
        P_{B_{t}}=\frac{(1-x_{t-1})V_{B}(x_{t-1})}{x_{t-1}V_{H}(x_{t-1})+(1-x_{t-1})V_{B}(x_{t-1})}
    \end{align*}
\end{assumption3}

Therefore, by summing up the individual validators' strategic choices, we can calculate the proportion of validators playing the honest strategy in round $t$ as follows:
\begin{align*}
    x_{t} = \frac{x_{t-1}V_{H}(x_{t-1})}{x_{t-1}V_{H}(x_{t-1})+(1-x_{t-1})V_{B}(x_{t-1})}
\end{align*}

\section{Stable Equilibrium}

\subsection{Definitions}

We use the evolutionary stable strategy (ESS) to define the stable equilibria in our model and apply steady-state equilibrium to explain the results. 

As proposed by \citet{smith_1973_the} and formalized by \citet{eshel_1982_assortment} and \citet{mckenzie_2009_evolutionary}, the ESS describes the dominance and stability of one strategy in the system even with a few invaders with other strategies. In our model, similar to the form in \citet{eshel_1982_assortment}, stable equilibrium is reached when: 
\begin{align}\label{eq:3}
    x_{t} = x_{t-1},\textit{where}~x_{t} = \frac{x_{t-1}V_{H}(x_{t-1})}{x_{t-1}V_{H}(x_{t-1})+(1-x_{t-1})V_{B}(x_{t-1})}.
\end{align}

Equation ~\ref{eq:3} results in three equilibria with different equilibrium strategies:
\begin{enumerate}
    \item Stable equilibrium 1: $x_{t} = x_{t-1} = 1$, $S_{H}$ becomes the equilibrium strategy;
    \item Stable equilibrium 2: $x_{t} = x_{t-1} = 0$, $S_{B}$ becomes the equilibrium strategy;
    \item Stable equilibrium 3: $V_{H} = V_{B}$, $x_{t} = x_{t-1} \in (0,1)$, the system reaches a steady-state equilibrium \cite{smith_1973_the}, representing that both strategies exist with a stable proportion because the numbers of agents changing between the two strategies are equal.
\end{enumerate}

To evaluate the validity of the steady state, we further define three types of stable equilibria:  Byzantine stable equilibrium, pooling stable equilibrium, and Byzantine equilibrium. 

\begin{definition7}[Honest Stable Equilibrium]
The honest stable equilibrium in our model is reached when the ESS is reached and none of the agents is playing the Byzantine strategy, which mathematically represents \footnote{This also satisfies the ESS in the form of \citet{mckenzie_2009_evolutionary}'s work by:$\pi(S_{H}|S_{H}) > \pi(S_{B}|S_{H})$, where $\pi(x|y)$ denotes the payoff an agent obtained when playing strategy $x$ against someone using the strategy $y$.}:
    \begin{align*}
        x_{t-1} = x_{t} = 1
    \end{align*}
\end{definition7}

\begin{definition8}[Byzantine Stable Equilibrium]

The Byzantine stable equilibrium in our model is reached when the ESS is reached and all of the agents are playing the Byzantine strategy, which mathematically represents \footnote{This also satisfies the ESS in the form of \citet{mckenzie_2009_evolutionary}'s work by: $\pi(S_{B}|S_{B}) > \pi(S_{H}|S_{B})$, where$\pi(x|y)$ denotes the payoff an agent obtained when playing strategy $x$ against someone using the strategy $y$.}:
    \begin{align*}
        x_{t-1} = x_{t} = 0
    \end{align*}
\end{definition8}

\begin{definition9}[Pooling Stable Equilibrium]
The pooling stable equilibrium in our model is reached when the stable equilibrium is reached and both strategies exist, which mathematically represents the following:
    \begin{align*}
        x_{t-1} = x_{t} \in (0,1)
    \end{align*}
This also satisfies the steady-state equilibrium in the sense that the number of agents changing from $S_{H}$ to $S_{B}$ is equal to the number of agents changing from $S_{B}$ to $S_{H}$. As defined by \citet{gagniuc_2017_markov}, a steady-state equilibrium is reached when the variables defining the behavior of the system or the process are unchanged in time. In our pooling stable equilibrium, the numbers of the agents playing the two strategies are unchanged in the game.
\end{definition9}

\subsection{Propositions}
Proposition 1 demonstrates the initial conditions for three stable equilibria. 

\begin{proposition1}[Three Stable Equilibria]
    \
    \newline
    \begin{enumerate}
        \item The honest stable equilibrium can be reached if and only if: 
        \begin{align*}
            1 - \frac{\nu}{N} \geqslant x_{1} & > \max(\frac{R-c_{send}+\kappa}{2R-2c_{send}+\kappa}, \frac{\nu}{N})
            \quad\text{and}\quad m \neq 1,\\
            \text{or}\quad
           \Bigl( x_{1} & \geqslant \frac{\nu}{N} \quad\text{and}\quad x_{1} > 1- \frac{\nu}{N} \Bigl)
        \end{align*}
        is satisfied.
        \
        \newline
        \item The Byzantine stable equilibrium can be reached if and only if:
        \begin{align*}
            \frac{\nu}{N} \leqslant x_{1} &< min(\frac{R-c_{send}+\kappa}{2R-2c_{send}+\kappa},1 - \frac{\nu}{n})
            \quad\text{and}\quad m \neq 1, \\
            \text{or}\quad
            \Bigl( x_{1} & < \frac{\nu}{N} \quad\text{and}\quad x_{1} \leqslant 1- \frac{\nu}{N} \Bigl)
        \end{align*}
        is satisfied.
        \
        \newline
        \item The pooling stable equilibrium can be reached if and only if:
        \begin{align*}
            1-\frac{\nu}{N} \geqslant x_{1} > \frac{\nu}{N} \text{ ,}
            \quad\text{and}\quad
            \Bigl(x_{1} = \frac{R-c_{send}+\kappa}{2R-2c_{send}+\kappa}
            \quad\text{or}\quad
            m = 1 \Bigl)
        \end{align*}
        is satisfied.
    \end{enumerate}
\end{proposition1}

\textbf{Lemma 3} further elaborates on the initial conditions of Proposition 1 in facets of the initial pivotality of two strategies and the agents’ subjective beliefs. 

\begin{lemma3}[Initial Conditions for Three Stable Equilibria]
    \
    \newline
    \begin{enumerate}
        \item If initially $N(1-x_{1}) \geqslant \nu$, then the initial number of agents with the Byzantine strategy is equal to or larger than the majority threshold, if and only if $x_{1} > \max(\frac{R-c_{send}+\kappa}{2R-2c_{send}+\kappa}, \frac{\nu}{N})$ and $m \neq 1$, the honest stable equilibrium can be reached.
        \
        \newline
        \item If initially $N(1-x_{1}) < \nu$, which represents that the initial number of agents with the Byzantine strategy is smaller than the threshold, if and only if $x_{1} \geqslant \frac{\nu}{N}$, meaning that the initial number of agents with the honest strategy is larger than the threshold, and the honest stable equilibrium can be reached.
        \
        \newline
        \item If initially $N(1-x_{1}) \geqslant \nu$, then the initial number of agents with the Byzantine strategy is equal or larger than the majority threshold, if and only if $x_{1} < \min(\frac{R-c_{send}+\kappa}{2R-2c_{send}+\kappa}, \frac{\nu}{N})$ and $m \neq 1$, or $x < min(1-\frac{\nu}{N}, \frac{\nu}{N})$, the Byzantine stable equilibrium can be reached.
        \
        \newline
        \item If initially $N(1-x_{1}) \geqslant \nu$, then initial number of agents with the Byzantine strategy is equal to or larger than the majority threshold, if and only if $x_{1} = \frac{R-c_{send}+\kappa}{2R-2c_{send}+\kappa}$ or $m \neq 1$, the pooling stable equilibrium can be reached.
        \
        \newline
        \item If initially $N(1-x_{1})<\nu$, which represents that the initial number of agents with the Byzantine strategy is smaller than the threshold, if and only if $Nx_{1}<\nu$, meaning that the initial number of agents with the honest strategy is also smaller than the threshold, the protocol will never converge to one single strategy, and the protocol liveness cannot be guaranteed.
        \
        \newline
        \item If $m=1$, which represents all the agents believe that all other agents are choosing the same strategy as theirs, the equilibrium outcomes depend on the real pivotality of strategies. An honest stable equilibrium can be reached if and only if $x_{1} \geqslant \frac{\nu}{N}$ and $x_{1} > 1-\frac{\nu}{N}$, meaning that only the honest strategy has initial pivotality. The Byzantine stable equilibrium can be reached if and only if $x_{1} < \frac{\nu}{N}$ and $x_{1} \leqslant 1-\frac{\nu}{N}$, meaning that the Byzantine strategy has initial pivotality. The pooling stable equilibrium can be reached if and only if $\frac{\nu}{N} \leqslant x_{1} \leqslant 1-\frac{\nu}{N}$, meaning that both strategies have initial pivotality. 
    \end{enumerate}
\end{lemma3}

\begin{proof} [Proof for Lemma 3: see Appendix A]
    \
    \newline
    
\end{proof}

\subsection{Evaluations}
We evaluate our model from the two perspectives of blockchain security \cite{zhang_2019_security} and economic welfare. Our evaluation criteria for blockchain security are \textit{safety}, \textit{liveness}, and \textit{validity}.  

\textbf{Safety}: We describe the achievement of safety as "bad things will not happen" \cite{alpern_1987_recognizing}. As defined by \citet{castro_2002_practical} and further interpreted by \citet{cachin_2017_blockchain, lone_2019_consensus}, the safety of a protocol should guarantee that all the agents have the same output. In our model, when either the honest stable equilibrium or the Byzantine stable equilibrium is reached, all the agents play the same strategy in the validating process; then, the protocol's safety is satisfied.

\textbf{Liveness}: As defined by \citet{castro_2002_practical} and further interpreted by \citet{cachin_2017_blockchain} and \citet{lone_2019_consensus}, the liveness of a protocol should guarantee that all nonfaulty agents have output. In our model, as long as the agents playing honest strategy vote for the proposals, immediate liveness is attained. When the honest stable equilibrium is reached, all the agents would play an honest strategy and agree on the same valid proposals; then, the protocol's eventual liveness is satisfied. The condition under which eventual liveness would fail is specified in LEMMA 3(3), (5) and the Byzantine stable equilibrium in (6)\footnote{In our Byzantine stable equilibrium, the nonfaulty agents are an empty set. Strictly, we can neither falsify nor verify whether nonfaulty agents have output. Let us call the tricky case not satisfying liveness based on the reasoning that the verification of liveness is impossible.}.

\textbf{Validity}: We describe the achievement of safety and validity together as "good things will eventually happen" \cite{alpern_1987_recognizing}. As discussed in \citet{cryptoeprint:2022/796}, the validity of a protocol should guarantee that all the agents have the same and valid output. In our model, when the honest stable equilibrium is finally achieved, all the agents play the honest strategy that they vote for a valid proposal, and the validity of the protocol is satisfied. 

Table ~\ref{tab:6} summarizes the results attained by our protocol in every stable equilibrium with different initial conditions corresponding to protocol safety, liveness, and validity. It shows that only the honest stable equilibrium could support eventual safety and liveness simultaneously (also support validity), and the social welfare obtained by the agents playing the honest strategy is optimal compared with other strategies or stable equilibria. Thus, the honest stable equilibrium is ideal considering both computing and economic properties. 

 To further discuss the policy implications of our model, we define the policy parameters $\alpha$, $\beta$, and $\gamma$. 
 \begin{enumerate}
    \item \textbf{$\alpha = \frac{R}{\kappa}$} denotes the reward-punishment ratio. 
    \item \textbf{$\beta = \frac{c_{send}}{\kappa}$} denotes the cost-punishment ratio.
    \item \textbf{$\gamma = \frac{\nu}{N}$} denotes the ratio of the majority threshold, the signal of pivotality, or the majority rate. 
\end{enumerate}
Therefore, the initial condition to achieve the honest stable equilibrium is written as Formula \ref{eq:4}.
\begin{align}
    \begin{aligned}~\label{eq:4}
    1 - \gamma \geqslant x_{1} & > \max(\frac{1}{2}+\frac{1}{4\alpha-4\beta+2}, \gamma) \quad\text{and}\quad m \neq 1,\\
    \text{or}\quad\Bigl( x_{1} & \geqslant \gamma \quad\text{and}\quad x_{1} > 1- \gamma \Bigl).
     \end{aligned}
\end{align}
Therefore, the stable equilibria conditions are rewritten as Table ~\ref{tab:5}, which affirms that the range of $x_{1}$ that achieves each equilibrium only depends on $\alpha$, $\beta$, and $\gamma$. We further interpret the policy implications in Table~\ref{tab:5}. By taking the first derivatives, the effects of $\alpha$, $\beta$, and $\gamma$ on agent incentives and influences on blockchain safety and liveness are concluded as follows:
\begin{enumerate}
    \item \textbf{Per Ceteris Paribus, if we increase the reward-punishment ratio $\alpha$, the security of the blockchain is enhanced, and the social welfare for agents playing an honest strategy increases.} If $\alpha$ increases, the conditions in Table ~\ref{tab:5} of honest stable equilibrium become easier for agents to achieve. In honest stable equilibrium, the negative effect of $\kappa$ on agents with honest strategy decreases as $\alpha$ increases, making the equilibrium easier to achieve.
    \item \textbf{Per Ceteris Paribus, if we increase the send cost-punishment ratio $\beta$, the eventual safety and liveness (both immediate and eventual) of blockchain are threatened.} If $\beta$ increases, the conditions in Table ~\ref{tab:5} of honest stable equilibrium become harder to achieve, while those of Byzantine stable equilibrium become easier to access. In one round of games, the cost of voting for a proposal becomes larger, so it is harder for agents with honest strategies to output a value. Therefore, blockchain security is threatened.
    \item \textbf{Per Ceteris Paribus, if we increase the pivotality rate, the requirements for achieving all three stable equilibria become tighter, and the risk of that blockchain liveness increases.} If $\gamma$ increases, the conditions in Table ~\ref{tab:5} of all three stable equilibria become harder to achieve. Therefore, the increase in $\gamma$ offers a higher risk for the situation in Lemma 3(6) that neither immediate nor eventual liveness can be achieved.
\end{enumerate}

Moreover, different from \citeauthor{castro_2002_practical}'s [\citeyear{castro_2002_practical}] study about peer-to-peer communication in the network layer, when blockchain safety, liveness, and validity are achieved by PBFT under the condition that no more than 1/3 nodes are faulty, our research focuses on the consensus layer, and the equilibrium results depend on the values of $\alpha, \beta, \gamma$. For instance, if $\gamma = \frac{2}{3}$, which is consistent with the PoS Ethereum \cite{buterin_2016_on,ethereumdevelopers_2022_proofofstake}, our honest stable equilibrium shows that as long as the initial proportion of honest agents $x_{1}$ is larger than $\frac{2}{3}$, an obviously inference from Formula (4), our model could incentivize all the agents to ultimately choose the honest strategy. Compared to other models of consensus mechanisms, ours provides unique policy advice on incentive design that converts the Byzantine strategy to an honest strategy in an evolutionary process. Our honest stable equilibrium is robust to the perturbations in strategies, such as an unexpectedly small increase in Byzantine strategies. 

\section{Discussion and Future Research}
In this paper, we demonstrate a general Byzantine consensus protocol as a game environment, apply bounded rationality to model the imitative learning dynamics of agent behavior, and solve the initial conditions for three stable equilibria. In our study, the miners’ cooperation has embodied consensus achievement on the blockchain and the social welfare maximization of the miners' rewards. 
\subsection{Discussion}
\subsubsection{Cooperative AI}
Our work reflects three key features of cooperative AI in blockchain consensus:-
\begin{enumerate}
    \item \textbf{Cooperation, reflected as committing to honest strategies, leads to desired outcomes in both computing and economic perspectives in the blockchain consensus.} By constructing an evolutionary game, our paper shows the merits of cooperation in block building. Similar to the analysis in \citeauthor{schneider_2013_longterm}'s [\citeyear{schneider_2013_longterm}] work, our work uses the bounded-rational agents' choices of strategy to represent the miner's willingness to interact with other participants and have a commitment with the committee in long-term cooperation. Consistent with prior research, our results also show that the miner's willingness to a longer commitment of cooperation, choosing the honest strategy, will provide higher quality welfare when the global expected outcome is to achieve consensus and support block building.
    \item \textbf{The blockchain committees can be modeled as coalitions, and competitions between coalitions can support the cooperation within each committee.} Since our research only focuses on consensus achievement within one committee, considering the competition effects between coalitions \cite{staatz_1983_the}, our model can be extended to verify if the competition between coalitions (e.g., for computing power, etc.) will eventually strengthen cooperation within the committee such that the whole coalition becomes more competitive. We suggest that instead of practicing the same model on all parallel committees, future research should consider analyzing how competitions between miner committees support global consensus achievement.
    \item \textbf{Problems and hypothesis in cooperative AI:} Our result shows the potential scenario when one of the downsides of cooperative AI revealed in cooperation on the blockchain. As mentioned and explained by \citet{dafoe_2020_open}, one of the potential downsides of cooperative AI is the appearance of exclusion and collusion, which happens when the agents’ cooperation  harms the external environment or other agents not involved in the cooperation, or when the cooperation undermines prosocial competition. In our stable equilibria analysis, we show that when the system reaches the Byzantine equilibrium, all the agents play the Byzantine strategy that harms the blockchain consensus and security. Therefore, the convergence to Byzantine equilibrium shows the potential scenario of harmful cooperation in the blockchain system. To mitigate such cooperation downsides, \citet{dafoe_2020_open} offered the \textit{hypothesis that broad cooperative competence is beneficial}, which we partially verify with the model simulation. When any of the values of $\alpha$ and $\beta$ increase, which represents a larger incentive in the global environment and  raises  broad cooperative competence, the system is able to transfer a larger proportion of faulty agents with incentives.
\end{enumerate}
\subsubsection{Assortative Matching}
Introducing an evolutionary game perspective to the application scenario of blockchain consensus, our results provide insights that are different from the application scenario of assortative matching in the existing evolutionary game literature. In assortative matching, a positive value of $m$ of inconsistent belief is crucial to sustaining cooperation~\cite{eshel_1982_assortment,angelucci_2017_assortative,durlauf_2003_is,shimer_2000_assortative,yang_2019_cooperation,eeckhout_2018_assortative}. In contrast, the value of $m$ does not influence the achievement of cooperation and the honest stable equilibrium, as long as it does not equal $1$. When $m=1$, the strongest false consensus effect of inconsistent beliefs is harmful to cooperation: protocol liveness cannot be guaranteed.

\subsubsection{Game Theory in the BFT consensus protocol}
Compared with \citet{halaburda_2021_an}, which also applies game theory to a BFT consensus protocol and highlights the potential application of incentives in protocol designs, our model differs in two crucial aspects:
\begin{enumerate}
    \item \citet{halaburda_2021_an} supposed that all non-Byzantine agents are rational, ambiguity averse, and Knightian uncertain about Byzantine actions. In contrast, we allow our agents to be bounded rationally, choose between the honest and Byzantine strategies as specified in the classic distributed system literature, and learn imitatively by responding to expected payoffs based on historical observations.
    \item \citet{halaburda_2021_an} apply PBE to analyze the strategic interactions in blockchain consensus. In comparison, we apply ESS to model the stable equilibrium in an evolutionary consensus process. \citet{halaburda_2021_an} concluded a prediction of multiple equilibria. In contrast, we specify the initial condition for an ideal honest stable equilibrium that achieves the desired properties of liveness, safety, and optimal social welfare.
\end{enumerate}

\subsection{Future research}
Fundamentally, our research suggests a new perspective to evaluate the performance of consensus protocols and suggest future design choices to achieve desired outcomes of security from the computer science perspective and welfare from economics concerns. Our study can be easily extended in three directions: 1) the mechanism design of blockchain consensus; 2) the game theoretical approach and bounded rationality, and 3) learning the agents’ algorithms.

\subsubsection{The Mechanism Design of Blockchain Consensus}
Our research contributes to informing the mechanism design of blockchain~\cite{zhang2023design} in the facet of consensus. Future research can adapt our approach to the design of consensus protocols for various blockchains.  We use the Ethereum 2.0 proof-of-stake as an example. Table~\ref{tab:7} shows the parameters in our model that affect equilibrium outcomes and their related policy parameters in Ethereum 2.0. 
\begin{enumerate}
    \item The $\alpha$ and $\beta$ (or parameters related to rewards and cost: $R$, $c_{send}$, $c_{check}$) suggest that increasing the rewards and reducing the cost of voting for proposals or reducing the punishment for honest players can encourage the achievement of the honest stable equilibrium. For example, per Ceteris Paribus, increasing blockchain rewards \cite{bitinfocharts_2019_ethereum} would increase $\alpha$. Thus our results inspire future research of empirical evaluations for blockchain reward designs. 
    \item The $\kappa$ parameter is related to but also different from slashing implemented on Ethereum~\cite{buterin_2020_combining,he_2022_contract}. On the Ethereum beacon chain \cite{wackerow_2022_proofofstake}, slashing is defined as a correlated penalty for misbehaving validators. It can range from a tiny amount to as large as the deposited fee.  \citet{he_2022_contract} defined the slashing mechanism as the punishment for the violators on blockchain that they will be prohibited from acting as validators anymore and get a deduction on the stakes they hold. $\kappa$, instead, represents the punishment an honest player receives when an invalid proposal is accepted, and its increase would cause negative effects on honest stable equilibrium. 
    When the value of $\kappa$ increases to as large as $R$, our proposition in Section 4.2 shows that the initial condition for the honest stable equilibrium is harder to achieve. In addition, when $\kappa$ increases to infinity, relating to but also different from an agent being slashed over on the Ethereum Beacon chain, Equation ~\ref{eq:10} shows the payoffs of the honest strategy are forever smaller than the Byzantine strategy if those with the honest strategies are punished. Thus, our analysis of $\kappa$ suggests reducing the punishment of honest agents while punishing  Byzantine agents. Adapting our model to slashing, a mechanism that identifies and removes Byzantine agents, would increase the chance and accelerate the convergence to the honest stable equilibrium.
   
    \item The $\gamma$ parameter sets a criterion for pivotality. Our analysis suggests that setting $\gamma$ too high might harm blockchain liveness. We relate this to the block finality on Ethereum \cite{wackerow_2022_proofofstake}. On the PoS Ethereum blockchain, the voting threshold is $\frac{2}{3}$ \cite{buterin_2016_on,ethereumdevelopers_2022_proofofstake}. We suggest that future research elaborate on how the change in $\gamma$ would affect blockchain security in the face of attacks in reality.
    
    \item The $x_{1}$ parameter represents the initial proportion of agents who choose the honest strategy. While most of the existing research and blockchain applications consider the majority of their on-chain agents to be honest ones, we created a generalized discussion for the initial condition of $x_{1}$. For example, $x_{1}$ could be affected by the depositing fee on Ethereum \cite{ethereumfoundation_2022_solo}, which affects agents' incentives to participate as a validator and thus the initial validator set and distributions of types. We suggest that future research investigate the design features of a consensus protocol that affect the initial percentage of honest strategy, taken as exogenously and arbitrarily in the literature. 
\end{enumerate}
\subsubsection{Game theoretical approach and bounded rationality}
\begin{figure}[!htbp]
	\includegraphics[scale=0.2]{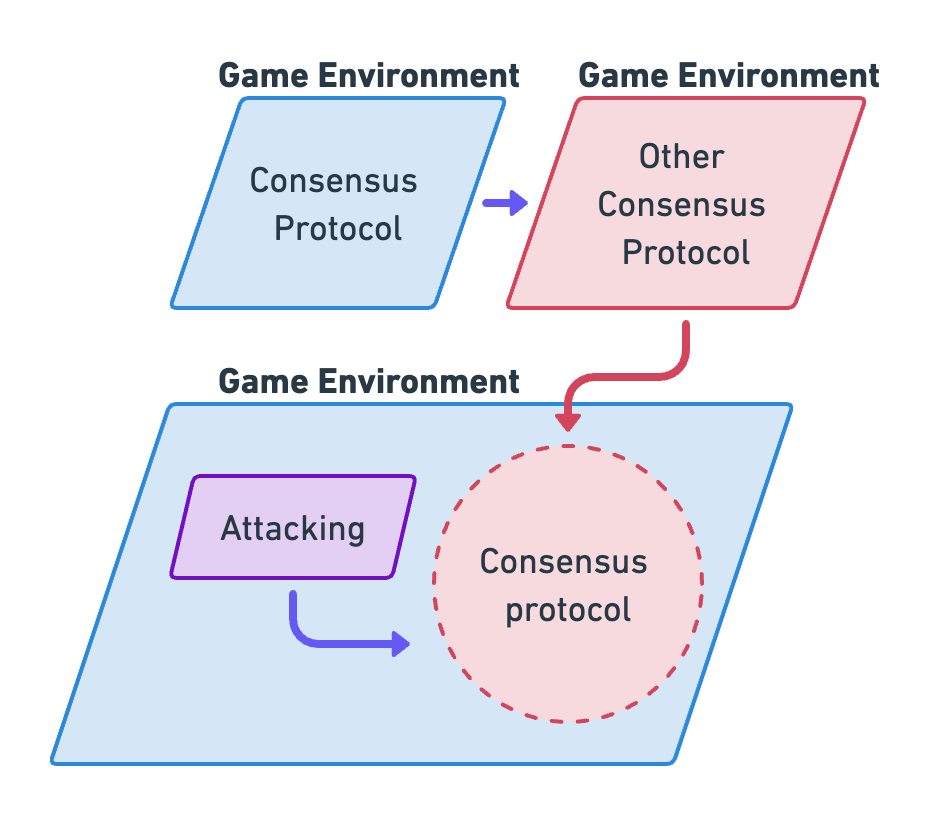}
	\caption{Game Environment Considering Attacking Agents and Strategies Not Specified in the Consensus Protocol.}
	\label{fig:9}
\end{figure}

Our research is seminal in the game theoretical approach and bounded rationality for the blockchain consensus problem that can be extended in the three facets.
\begin{enumerate}
    \item \textbf{agent}: One of the distinct highlights of our research is the application of bounded-rational agents with imitative learning abilities. Future research can consider including other bounded rational agents such as the  prominent Level-K~\cite{crawford2007level,arad201211,levin_2020_bridging} or the limited foresight~\cite{jehiel_2001_limited} agent in behavioral economics. Future research can also compare the results of a variety of game theory solution concepts~\cite{tardos2007basic} and test the performance in empirical or experimental studies. 
    
    \item \textbf{game environment: variants of consensus protocols}: On blockchains, there is a variety of consensus protocols~\cite{xiao2020survey,cachin2017blockchain,sankar2017survey}, each provides a game environment. Future research can follow our approach to abstract other consensus protocols into a new game environment for performance evaluation and comparative studies. 

    \item \textbf{Game environment: attacks}: Figure~\ref{fig:9} shows that, other than considering the game environment of other consensus protocols, future research can also consider attacking strategies outside the protocol. For example, \citet{cryptoeprint:2022/1020} identifies a novel and riskless attack on the Ethereum blockchain. In the study of Hashed Time-Locked Contracts (HTLC), ~\citet{tsabary_2021_madhtlc} and \citet{cryptoeprint:2022/546} discussed the vulnerability of HTLC subject to different bribery attacks outside the protocol. Future research can consider more complex attacks in the game environment abstraction of consensus protocols.
    \end{enumerate}

\subsubsection{Learning Algorithms for Agents}
One of the distinct highlights of our research is the Application of bounded-rational agents with imitative learning abilities. As described in Assumption 3, the agents are able to form expectations based on historical observations and adjust choices accordingly. We could extend our methods to other learning methods~\cite{mallouh_2021_a, cryptoeprint:2022/175}. For example, we could consider the trending reinforcement learning agent~\cite{sutton2018reinforcement,ivanov2022reinforcement} in computational science. Among the few pioneering studies that implement reinforcement learning to study blockchain security, \citet{tsabary_2021_madhtlc} and \citet{cryptoeprint:2022/175} provided the effectivity of a reinforcement learning algorithm called \textit{WeRLman} in analyzing blockchain security.~\citet{hou_2020_squirrl} also applied learning algorithms to defend against on blockchain systems.

\balance
\bibliographystyle{ACM-Reference-Format}
\bibliography{citation}


\begin{thebibliography}{133}


\ifx \showCODEN    \undefined \def \showCODEN     #1{\unskip}     \fi
\ifx \showDOI      \undefined \def \showDOI       #1{#1}\fi
\ifx \showISBNx    \undefined \def \showISBNx     #1{\unskip}     \fi
\ifx \showISBNxiii \undefined \def \showISBNxiii  #1{\unskip}     \fi
\ifx \showISSN     \undefined \def \showISSN      #1{\unskip}     \fi
\ifx \showLCCN     \undefined \def \showLCCN      #1{\unskip}     \fi
\ifx \shownote     \undefined \def \shownote      #1{#1}          \fi
\ifx \showarticletitle \undefined \def \showarticletitle #1{#1}   \fi
\ifx \showURL      \undefined \def \showURL       {\relax}        \fi
\providecommand\bibfield[2]{#2}
\providecommand\bibinfo[2]{#2}
\providecommand\natexlab[1]{#1}
\providecommand\showeprint[2][]{arXiv:#2}

\bibitem[\protect\citeauthoryear{Abraham, Devadas, Dolev, Nayak, and
  Ren}{Abraham et~al\mbox{.}}{2017}]%
        {abraham_2017_efficient}
\bibfield{author}{\bibinfo{person}{Ittai Abraham}, \bibinfo{person}{Srinivas
  Devadas}, \bibinfo{person}{Danny Dolev}, \bibinfo{person}{Kartik Nayak},
  {and} \bibinfo{person}{Ling Ren}.} \bibinfo{year}{2017}\natexlab{}.
\newblock \showarticletitle{Efficient Synchronous Byzantine Consensus}.
\newblock \bibinfo{journal}{\emph{arXiv:1704.02397 [cs]}} (\bibinfo{date}{09}
  \bibinfo{year}{2017}).
\newblock
\urldef\tempurl%
\url{https://arxiv.org/abs/1704.02397}
\showURL{%
\tempurl}


\bibitem[\protect\citeauthoryear{Abraham, Dolev, Eyal, and Halpern}{Abraham
  et~al\mbox{.}}{2022}]%
        {cryptoeprint:2022/308}
\bibfield{author}{\bibinfo{person}{Ittai Abraham}, \bibinfo{person}{Danny
  Dolev}, \bibinfo{person}{Ittay Eyal}, {and} \bibinfo{person}{Joseph~Y.
  Halpern}.} \bibinfo{year}{2022}\natexlab{}.
\newblock \bibinfo{title}{Colordag: An Incentive-Compatible Blockchain}.
\newblock \bibinfo{howpublished}{Cryptology ePrint Archive, Paper 2022/308}.
\newblock
\urldef\tempurl%
\url{https://eprint.iacr.org/2022/308}
\showURL{%
\tempurl}
\newblock
\shownote{\url{https://eprint.iacr.org/2022/308}}.


\bibitem[\protect\citeauthoryear{Abraham, Dolev, Gonen, and Halpern}{Abraham
  et~al\mbox{.}}{2006}]%
        {abraham_2006_distributed}
\bibfield{author}{\bibinfo{person}{Ittai Abraham}, \bibinfo{person}{Danny
  Dolev}, \bibinfo{person}{Rica Gonen}, {and} \bibinfo{person}{Joe Halpern}.}
  \bibinfo{year}{2006}\natexlab{}.
\newblock \showarticletitle{Distributed computing meets game theory}.
\newblock \bibinfo{journal}{\emph{Proceedings of the twenty-fifth annual ACM
  symposium on Principles of distributed computing - PODC '06}}
  (\bibinfo{year}{2006}).
\newblock
\urldef\tempurl%
\url{https://doi.org/10.1145/1146381.1146393}
\showDOI{\tempurl}


\bibitem[\protect\citeauthoryear{Abraham, Malkhi, Nayak, Ren, and
  Spiegelman}{Abraham et~al\mbox{.}}{2016}]%
        {article}
\bibfield{author}{\bibinfo{person}{Ittai Abraham}, \bibinfo{person}{Dahlia
  Malkhi}, \bibinfo{person}{Kartik Nayak}, \bibinfo{person}{Ling Ren}, {and}
  \bibinfo{person}{Alexander Spiegelman}.} \bibinfo{year}{2016}\natexlab{}.
\newblock \showarticletitle{Solidus: An Incentive-compatible Cryptocurrency
  Based on Permissionless Byzantine Consensus}.
\newblock  (\bibinfo{date}{12} \bibinfo{year}{2016}).
\newblock


\bibitem[\protect\citeauthoryear{Aiyer, Alvisi, Clement, Dahlin, Martin, and
  Porth}{Aiyer et~al\mbox{.}}{2005}]%
        {aiyer_2005_bar}
\bibfield{author}{\bibinfo{person}{Amitanand~S. Aiyer},
  \bibinfo{person}{Lorenzo Alvisi}, \bibinfo{person}{Allen Clement},
  \bibinfo{person}{Mike Dahlin}, \bibinfo{person}{Jean-Philippe Martin}, {and}
  \bibinfo{person}{Carl Porth}.} \bibinfo{year}{2005}\natexlab{}.
\newblock \showarticletitle{BAR fault tolerance for cooperative services}.
\newblock \bibinfo{journal}{\emph{Proceedings of the twentieth ACM symposium on
  Operating systems principles - SOSP '05}} (\bibinfo{year}{2005}).
\newblock
\urldef\tempurl%
\url{https://doi.org/10.1145/1095810.1095816}
\showDOI{\tempurl}


\bibitem[\protect\citeauthoryear{Al-Bassam, Sonnino, Bano, Hrycyszyn, and
  Danezis}{Al-Bassam et~al\mbox{.}}{2017}]%
        {albassam_2017_chainspace}
\bibfield{author}{\bibinfo{person}{Mustafa Al-Bassam}, \bibinfo{person}{Alberto
  Sonnino}, \bibinfo{person}{Shehar Bano}, \bibinfo{person}{Dave Hrycyszyn},
  {and} \bibinfo{person}{George Danezis}.} \bibinfo{year}{2017}\natexlab{}.
\newblock \showarticletitle{Chainspace: A Sharded Smart Contracts Platform}.
\newblock \bibinfo{journal}{\emph{arXiv:1708.03778 [cs]}} (\bibinfo{date}{08}
  \bibinfo{year}{2017}).
\newblock
\urldef\tempurl%
\url{https://arxiv.org/abs/1708.03778}
\showURL{%
\tempurl}


\bibitem[\protect\citeauthoryear{Alpern and Schneider}{Alpern and
  Schneider}{1987}]%
        {alpern_1987_recognizing}
\bibfield{author}{\bibinfo{person}{Bowen Alpern} {and} \bibinfo{person}{Fred~B.
  Schneider}.} \bibinfo{year}{1987}\natexlab{}.
\newblock \showarticletitle{Recognizing safety and liveness}.
\newblock \bibinfo{journal}{\emph{Distributed Computing}}  \bibinfo{volume}{2}
  (\bibinfo{date}{09} \bibinfo{year}{1987}), \bibinfo{pages}{117--126}.
\newblock
\urldef\tempurl%
\url{https://doi.org/10.1007/bf01782772}
\showDOI{\tempurl}


\bibitem[\protect\citeauthoryear{Amoussou-Guenou, Biais, Potop-Butucaru, and
  Tucci-Piergiovanni}{Amoussou-Guenou et~al\mbox{.}}{2020a}]%
        {amoussouguenou_2020_rational}
\bibfield{author}{\bibinfo{person}{Yackolley Amoussou-Guenou},
  \bibinfo{person}{Bruno Biais}, \bibinfo{person}{Maria Potop-Butucaru}, {and}
  \bibinfo{person}{Sara Tucci-Piergiovanni}.} \bibinfo{year}{2020}\natexlab{a}.
\newblock \bibinfo{title}{Rational Behavior in Committee-Based Blockchains}.
\newblock
\newblock
\urldef\tempurl%
\url{https://hal.archives-ouvertes.fr/hal-02867095}
\showURL{%
\tempurl}


\bibitem[\protect\citeauthoryear{Amoussou-Guenou, Biais, Potop-Butucaru, and
  Tucci-Piergiovanni}{Amoussou-Guenou et~al\mbox{.}}{2020b}]%
        {10.5555/3398761.3398772}
\bibfield{author}{\bibinfo{person}{Yackolley Amoussou-Guenou},
  \bibinfo{person}{Bruno Biais}, \bibinfo{person}{Maria Potop-Butucaru}, {and}
  \bibinfo{person}{Sara Tucci-Piergiovanni}.} \bibinfo{year}{2020}\natexlab{b}.
\newblock \showarticletitle{Rational vs Byzantine Players in Consensus-Based
  Blockchains}. In \bibinfo{booktitle}{\emph{Proceedings of the 19th
  International Conference on Autonomous Agents and MultiAgent Systems}}
  (Auckland, New Zealand) \emph{(\bibinfo{series}{AAMAS '20})}.
  \bibinfo{publisher}{International Foundation for Autonomous Agents and
  Multiagent Systems}, \bibinfo{address}{Richland, SC},
  \bibinfo{pages}{43–51}.
\newblock
\showISBNx{9781450375184}


\bibitem[\protect\citeauthoryear{Angelucci and Bennett}{Angelucci and
  Bennett}{2017}]%
        {angelucci_2017_assortative}
\bibfield{author}{\bibinfo{person}{Manuela Angelucci} {and}
  \bibinfo{person}{Daniel Bennett}.} \bibinfo{year}{2017}\natexlab{}.
\newblock \showarticletitle{Assortative Matching under Asymmetric Information:
  Evidence from Malawi}.
\newblock \bibinfo{journal}{\emph{American Economic Review}}
  \bibinfo{volume}{107} (\bibinfo{date}{05} \bibinfo{year}{2017}),
  \bibinfo{pages}{154--157}.
\newblock
\urldef\tempurl%
\url{https://doi.org/10.1257/aer.p20171055}
\showDOI{\tempurl}


\bibitem[\protect\citeauthoryear{Apostolaki, Marti, Müller, and
  Vanbever}{Apostolaki et~al\mbox{.}}{2018}]%
        {apostolaki_2018_sabre}
\bibfield{author}{\bibinfo{person}{Maria Apostolaki}, \bibinfo{person}{Gian
  Marti}, \bibinfo{person}{Jan Müller}, {and} \bibinfo{person}{Laurent
  Vanbever}.} \bibinfo{year}{2018}\natexlab{}.
\newblock \showarticletitle{SABRE: Protecting Bitcoin against Routing Attacks}.
\newblock \bibinfo{journal}{\emph{arXiv:1808.06254 [cs]}} (\bibinfo{date}{08}
  \bibinfo{year}{2018}).
\newblock
\urldef\tempurl%
\url{https://arxiv.org/abs/1808.06254}
\showURL{%
\tempurl}


\bibitem[\protect\citeauthoryear{Arad and Rubinstein}{Arad and
  Rubinstein}{2012}]%
        {arad201211}
\bibfield{author}{\bibinfo{person}{Ayala Arad} {and} \bibinfo{person}{Ariel
  Rubinstein}.} \bibinfo{year}{2012}\natexlab{}.
\newblock \showarticletitle{The 11-20 money request game: A level-k reasoning
  study}.
\newblock \bibinfo{journal}{\emph{American Economic Review}}
  \bibinfo{volume}{102}, \bibinfo{number}{7} (\bibinfo{year}{2012}),
  \bibinfo{pages}{3561--73}.
\newblock


\bibitem[\protect\citeauthoryear{Aublin, Mokhtar, and Quema}{Aublin
  et~al\mbox{.}}{2013}]%
        {aublin_2013_rbft}
\bibfield{author}{\bibinfo{person}{Pierre-Louis Aublin},
  \bibinfo{person}{Sonia~Ben Mokhtar}, {and} \bibinfo{person}{Vivien Quema}.}
  \bibinfo{year}{2013}\natexlab{}.
\newblock \showarticletitle{RBFT: Redundant Byzantine Fault Tolerance}.
\newblock \bibinfo{journal}{\emph{2013 IEEE 33rd International Conference on
  Distributed Computing Systems}} (\bibinfo{date}{07} \bibinfo{year}{2013}).
\newblock
\urldef\tempurl%
\url{https://doi.org/10.1109/icdcs.2013.53}
\showDOI{\tempurl}


\bibitem[\protect\citeauthoryear{Auer, Monnet, and Shin}{Auer
  et~al\mbox{.}}{2021}]%
        {auer_2021_distributed}
\bibfield{author}{\bibinfo{person}{Raphael Auer}, \bibinfo{person}{Cyril
  Monnet}, {and} \bibinfo{person}{Hyun~Song Shin}.}
  \bibinfo{year}{2021}\natexlab{}.
\newblock \showarticletitle{Distributed ledgers and the governance of money}.
\newblock \bibinfo{journal}{\emph{www.bis.org}} (\bibinfo{date}{01}
  \bibinfo{year}{2021}).
\newblock
\urldef\tempurl%
\url{https://www.bis.org/publ/work924.htm}
\showURL{%
\tempurl}


\bibitem[\protect\citeauthoryear{Axelrod and Hamilton}{Axelrod and
  Hamilton}{1981}]%
        {axelrod_1981_the}
\bibfield{author}{\bibinfo{person}{R Axelrod} {and} \bibinfo{person}{W.
  Hamilton}.} \bibinfo{year}{1981}\natexlab{}.
\newblock \showarticletitle{The evolution of cooperation}.
\newblock \bibinfo{journal}{\emph{Science}}  \bibinfo{volume}{211}
  (\bibinfo{date}{03} \bibinfo{year}{1981}), \bibinfo{pages}{1390--1396}.
\newblock
\urldef\tempurl%
\url{https://doi.org/10.1126/science.7466396}
\showDOI{\tempurl}


\bibitem[\protect\citeauthoryear{Azouvi and Hicks}{Azouvi and Hicks}{2021}]%
        {Azouvi2021SoK}
\bibfield{author}{\bibinfo{person}{Sarah Azouvi} {and}
  \bibinfo{person}{Alexander Hicks}.} \bibinfo{year}{2021}\natexlab{}.
\newblock \showarticletitle{SoK: Tools for {Game} {Theoretic} {Models} of
  {Security} for {Cryptocurrencies}}.
\newblock \bibinfo{journal}{\emph{Cryptoeconomic Systems}} \bibinfo{volume}{0},
  \bibinfo{number}{1} (\bibinfo{date}{apr 5} \bibinfo{year}{2021}).
\newblock
\newblock
\shownote{https://cryptoeconomicsystems.pubpub.org/pub/azouvi-sok-security}.


\bibitem[\protect\citeauthoryear{Bano, Sonnino, Al-Bassam, Azouvi, McCorry,
  Meiklejohn, and Danezis}{Bano et~al\mbox{.}}{2019}]%
        {bano_2019_sok}
\bibfield{author}{\bibinfo{person}{Shehar Bano}, \bibinfo{person}{Alberto
  Sonnino}, \bibinfo{person}{Mustafa Al-Bassam}, \bibinfo{person}{Sarah
  Azouvi}, \bibinfo{person}{Patrick McCorry}, \bibinfo{person}{Sarah
  Meiklejohn}, {and} \bibinfo{person}{George Danezis}.}
  \bibinfo{year}{2019}\natexlab{}.
\newblock \showarticletitle{SoK: Consensus in the Age of Blockchains}.
\newblock \bibinfo{journal}{\emph{Proceedings of the 1st ACM Conference on
  Advances in Financial Technologies}} (\bibinfo{date}{10}
  \bibinfo{year}{2019}).
\newblock
\urldef\tempurl%
\url{https://doi.org/10.1145/3318041.3355458}
\showDOI{\tempurl}


\bibitem[\protect\citeauthoryear{Bar-Zur, Abu-Hanna, Eyal, and Tamar}{Bar-Zur
  et~al\mbox{.}}{2022}]%
        {cryptoeprint:2022/175}
\bibfield{author}{\bibinfo{person}{Roi Bar-Zur}, \bibinfo{person}{Ameer
  Abu-Hanna}, \bibinfo{person}{Ittay Eyal}, {and} \bibinfo{person}{Aviv
  Tamar}.} \bibinfo{year}{2022}\natexlab{}.
\newblock \bibinfo{title}{WeRLman: To Tackle Whale (Transactions), Go Deep
  (RL)}.
\newblock \bibinfo{howpublished}{Cryptology ePrint Archive, Paper 2022/175}.
\newblock
\urldef\tempurl%
\url{https://eprint.iacr.org/2022/175}
\showURL{%
\tempurl}
\newblock
\shownote{\url{https://eprint.iacr.org/2022/175}}.


\bibitem[\protect\citeauthoryear{Behaviour}{Behaviour}{2018}]%
        {naturehumanbehaviour_2018_the}
\bibfield{author}{\bibinfo{person}{Nature~Human Behaviour}.}
  \bibinfo{year}{2018}\natexlab{}.
\newblock \showarticletitle{The cooperative human}.
\newblock \bibinfo{journal}{\emph{Nature Human Behaviour}}  \bibinfo{volume}{2}
  (\bibinfo{date}{07} \bibinfo{year}{2018}), \bibinfo{pages}{427--428}.
\newblock
\urldef\tempurl%
\url{https://doi.org/10.1038/s41562-018-0389-1}
\showDOI{\tempurl}


\bibitem[\protect\citeauthoryear{Benhaim, Hemenway~Falk, and Tsoukalas}{Benhaim
  et~al\mbox{.}}{2021}]%
        {benhaim_2021_scaling}
\bibfield{author}{\bibinfo{person}{Alon Benhaim}, \bibinfo{person}{Brett
  Hemenway~Falk}, {and} \bibinfo{person}{Gerry Tsoukalas}.}
  \bibinfo{year}{2021}\natexlab{}.
\newblock \showarticletitle{Scaling Blockchains: Can Elected Committees Help?}
\newblock \bibinfo{journal}{\emph{SSRN Electronic Journal}}
  (\bibinfo{year}{2021}).
\newblock
\urldef\tempurl%
\url{https://doi.org/10.2139/ssrn.3914471}
\showDOI{\tempurl}


\bibitem[\protect\citeauthoryear{Biais, Bisière, Bouvard, and Casamatta}{Biais
  et~al\mbox{.}}{2019}]%
        {biais_2019_the}
\bibfield{author}{\bibinfo{person}{Bruno Biais}, \bibinfo{person}{Christophe
  Bisière}, \bibinfo{person}{Matthieu Bouvard}, {and}
  \bibinfo{person}{Catherine Casamatta}.} \bibinfo{year}{2019}\natexlab{}.
\newblock \showarticletitle{The Blockchain Folk Theorem}.
\newblock \bibinfo{journal}{\emph{The Review of Financial Studies}}
  \bibinfo{volume}{32} (\bibinfo{date}{04} \bibinfo{year}{2019}),
  \bibinfo{pages}{1662--1715}.
\newblock
\urldef\tempurl%
\url{https://doi.org/10.1093/rfs/hhy095}
\showDOI{\tempurl}


\bibitem[\protect\citeauthoryear{BitInfoCharts}{BitInfoCharts}{2019}]%
        {bitinfocharts_2019_ethereum}
\bibfield{author}{\bibinfo{person}{BitInfoCharts}.}
  \bibinfo{year}{2019}\natexlab{}.
\newblock \bibinfo{title}{Ethereum / Ether (ETH) statistics - Price, Blocks
  Count, Difficulty, Hashrate, Value}.
\newblock
\newblock
\urldef\tempurl%
\url{https://bitinfocharts.com/ethereum/}
\showURL{%
\tempurl}


\bibitem[\protect\citeauthoryear{Budish}{Budish}{2022}]%
        {budish_2022_the}
\bibfield{author}{\bibinfo{person}{Eric~B. Budish}.}
  \bibinfo{year}{2022}\natexlab{}.
\newblock \bibinfo{title}{The Economic Limits of Bitcoin and Anonymous,
  Decentralized Trust on the Blockchain}.
\newblock
\newblock
\urldef\tempurl%
\url{https://papers.ssrn.com/sol3/papers.cfm?abstract_id=4148014}
\showURL{%
\tempurl}


\bibitem[\protect\citeauthoryear{Buterin}{Buterin}{2016}]%
        {buterin_2016_on}
\bibfield{author}{\bibinfo{person}{Vitalik Buterin}.}
  \bibinfo{year}{2016}\natexlab{}.
\newblock \bibinfo{title}{On Settlement Finality}.
\newblock
\newblock
\urldef\tempurl%
\url{https://blog.ethereum.org/2016/05/09/on-settlement-finality}
\showURL{%
\tempurl}


\bibitem[\protect\citeauthoryear{Buterin, Hernandez, Kamphefner, Pham, Qiao,
  Ryan, Sin, Wang, and Zhang}{Buterin et~al\mbox{.}}{2020}]%
        {buterin_2020_combining}
\bibfield{author}{\bibinfo{person}{Vitalik Buterin}, \bibinfo{person}{Diego
  Hernandez}, \bibinfo{person}{Thor Kamphefner}, \bibinfo{person}{Khiem Pham},
  \bibinfo{person}{Zhi Qiao}, \bibinfo{person}{Danny Ryan},
  \bibinfo{person}{Juhyeok Sin}, \bibinfo{person}{Ying Wang}, {and}
  \bibinfo{person}{Yan~X. Zhang}.} \bibinfo{year}{2020}\natexlab{}.
\newblock \showarticletitle{Combining GHOST and Casper}.
\newblock \bibinfo{journal}{\emph{arXiv:2003.03052 [cs]}} (\bibinfo{date}{05}
  \bibinfo{year}{2020}).
\newblock
\urldef\tempurl%
\url{https://arxiv.org/abs/2003.03052}
\showURL{%
\tempurl}


\bibitem[\protect\citeauthoryear{Cachin and Vukoli{\'c}}{Cachin and
  Vukoli{\'c}}{2017}]%
        {cachin2017blockchain}
\bibfield{author}{\bibinfo{person}{Christian Cachin} {and}
  \bibinfo{person}{Marko Vukoli{\'c}}.} \bibinfo{year}{2017}\natexlab{}.
\newblock \showarticletitle{Blockchain consensus protocols in the wild}.
\newblock \bibinfo{journal}{\emph{arXiv preprint arXiv:1707.01873}}
  (\bibinfo{year}{2017}).
\newblock


\bibitem[\protect\citeauthoryear{Cachin and Vukolić}{Cachin and
  Vukolić}{2017}]%
        {cachin_2017_blockchain}
\bibfield{author}{\bibinfo{person}{Christian Cachin} {and}
  \bibinfo{person}{Marko Vukolić}.} \bibinfo{year}{2017}\natexlab{}.
\newblock \bibinfo{title}{Blockchain Consensus Protocols in the Wild *}.
\newblock
\newblock
\urldef\tempurl%
\url{https://doi.org/10.4230/LIPIcs.DISC.2017.1}
\showDOI{\tempurl}


\bibitem[\protect\citeauthoryear{Cao}{Cao}{2022}]%
        {cao_2022_decentralized}
\bibfield{author}{\bibinfo{person}{Longbing Cao}.}
  \bibinfo{year}{2022}\natexlab{}.
\newblock \showarticletitle{Decentralized AI: Edge Intelligence and Smart
  Blockchain, Metaverse, Web3, and DeSci}.
\newblock \bibinfo{journal}{\emph{IEEE Intelligent Systems}}
  \bibinfo{volume}{37} (\bibinfo{date}{05} \bibinfo{year}{2022}),
  \bibinfo{pages}{6--19}.
\newblock
\urldef\tempurl%
\url{https://doi.org/10.1109/mis.2022.3181504}
\showDOI{\tempurl}


\bibitem[\protect\citeauthoryear{Castro and Liskov}{Castro and Liskov}{2002}]%
        {castro_2002_practical}
\bibfield{author}{\bibinfo{person}{Miguel Castro} {and}
  \bibinfo{person}{Barbara Liskov}.} \bibinfo{year}{2002}\natexlab{}.
\newblock \showarticletitle{Practical byzantine fault tolerance and proactive
  recovery}.
\newblock \bibinfo{journal}{\emph{ACM Transactions on Computer Systems}}
  \bibinfo{volume}{20} (\bibinfo{date}{11} \bibinfo{year}{2002}),
  \bibinfo{pages}{398--461}.
\newblock
\urldef\tempurl%
\url{https://doi.org/10.1145/571637.571640}
\showDOI{\tempurl}


\bibitem[\protect\citeauthoryear{Chase and Simon}{Chase and Simon}{1973}]%
        {chase_1973_perception}
\bibfield{author}{\bibinfo{person}{William~G. Chase} {and}
  \bibinfo{person}{Herbert~A. Simon}.} \bibinfo{year}{1973}\natexlab{}.
\newblock \showarticletitle{Perception in chess}.
\newblock \bibinfo{journal}{\emph{Cognitive Psychology}}  \bibinfo{volume}{4}
  (\bibinfo{date}{01} \bibinfo{year}{1973}), \bibinfo{pages}{55--81}.
\newblock
\urldef\tempurl%
\url{https://doi.org/10.1016/0010-0285(73)90004-2}
\showDOI{\tempurl}


\bibitem[\protect\citeauthoryear{Chen and Micali}{Chen and Micali}{2017}]%
        {chen_2017_algorand}
\bibfield{author}{\bibinfo{person}{Jing Chen} {and} \bibinfo{person}{Silvio
  Micali}.} \bibinfo{year}{2017}\natexlab{}.
\newblock \showarticletitle{Algorand}.
\newblock \bibinfo{journal}{\emph{arXiv:1607.01341 [cs]}} (\bibinfo{date}{05}
  \bibinfo{year}{2017}).
\newblock
\urldef\tempurl%
\url{https://arxiv.org/abs/1607.01341}
\showURL{%
\tempurl}


\bibitem[\protect\citeauthoryear{Chen, Papadimitriou, and Roughgarden}{Chen
  et~al\mbox{.}}{2019}]%
        {chen_2019_an}
\bibfield{author}{\bibinfo{person}{Xi Chen}, \bibinfo{person}{Christos
  Papadimitriou}, {and} \bibinfo{person}{Tim Roughgarden}.}
  \bibinfo{year}{2019}\natexlab{}.
\newblock \showarticletitle{An Axiomatic Approach to Block Rewards}.
\newblock \bibinfo{journal}{\emph{Proceedings of the 1st ACM Conference on
  Advances in Financial Technologies}} (\bibinfo{date}{10}
  \bibinfo{year}{2019}).
\newblock
\urldef\tempurl%
\url{https://doi.org/10.1145/3318041.3355470}
\showDOI{\tempurl}


\bibitem[\protect\citeauthoryear{Cong and He}{Cong and He}{2018}]%
        {cong_2018_blockchain}
\bibfield{author}{\bibinfo{person}{Lin~William Cong} {and}
  \bibinfo{person}{Zhiguo He}.} \bibinfo{year}{2018}\natexlab{}.
\newblock \bibinfo{title}{Blockchain Disruption and Smart Contracts}.
\newblock
\newblock
\urldef\tempurl%
\url{https://www.nber.org/papers/w24399}
\showURL{%
\tempurl}


\bibitem[\protect\citeauthoryear{Cong, He, and Li}{Cong et~al\mbox{.}}{2019}]%
        {cong_2019_decentralized}
\bibfield{author}{\bibinfo{person}{Lin~William Cong}, \bibinfo{person}{Zhiguo
  He}, {and} \bibinfo{person}{Jiasun Li}.} \bibinfo{year}{2019}\natexlab{}.
\newblock \bibinfo{title}{Decentralized Mining in Centralized Pools}.
\newblock
\newblock
\urldef\tempurl%
\url{https://www.nber.org/papers/w25592}
\showURL{%
\tempurl}


\bibitem[\protect\citeauthoryear{Conitzer}{Conitzer}{2019}]%
        {conitzer_2019_designing}
\bibfield{author}{\bibinfo{person}{Vincent Conitzer}.}
  \bibinfo{year}{2019}\natexlab{}.
\newblock \showarticletitle{Designing Preferences, Beliefs, and Identities for
  Artificial Intelligence}.
\newblock \bibinfo{journal}{\emph{Proceedings of the AAAI Conference on
  Artificial Intelligence}}  \bibinfo{volume}{33} (\bibinfo{date}{07}
  \bibinfo{year}{2019}), \bibinfo{pages}{9755--9759}.
\newblock
\urldef\tempurl%
\url{https://doi.org/10.1609/aaai.v33i01.33019755}
\showDOI{\tempurl}


\bibitem[\protect\citeauthoryear{Conitzer and Oesterheld}{Conitzer and
  Oesterheld}{[n.\,d.]}]%
        {conitzer_foundations}
\bibfield{author}{\bibinfo{person}{Vincent Conitzer} {and}
  \bibinfo{person}{Caspar Oesterheld}.} \bibinfo{year}{[n.\,d.]}\natexlab{}.
\newblock \bibinfo{title}{Foundations of Cooperative AI}.
\newblock
\newblock
\urldef\tempurl%
\url{https://www.cs.cmu.edu/~15784/focal_paper.pdf}
\showURL{%
\tempurl}


\bibitem[\protect\citeauthoryear{Conlisk}{Conlisk}{1996}]%
        {conlisk_1996_why}
\bibfield{author}{\bibinfo{person}{John Conlisk}.}
  \bibinfo{year}{1996}\natexlab{}.
\newblock \showarticletitle{Why Bounded Rationality?}
\newblock \bibinfo{journal}{\emph{Journal of Economic Literature}}
  \bibinfo{volume}{34} (\bibinfo{year}{1996}), \bibinfo{pages}{669–700}.
\newblock
\urldef\tempurl%
\url{https://www.jstor.org/stable/2729218}
\showURL{%
\tempurl}


\bibitem[\protect\citeauthoryear{Crawford and Iriberri}{Crawford and
  Iriberri}{2007}]%
        {crawford2007level}
\bibfield{author}{\bibinfo{person}{Vincent~P Crawford} {and}
  \bibinfo{person}{Nagore Iriberri}.} \bibinfo{year}{2007}\natexlab{}.
\newblock \showarticletitle{Level-k auctions: Can a nonequilibrium model of
  strategic thinking explain the winner's curse and overbidding in
  private-value auctions?}
\newblock \bibinfo{journal}{\emph{Econometrica}} \bibinfo{volume}{75},
  \bibinfo{number}{6} (\bibinfo{year}{2007}), \bibinfo{pages}{1721--1770}.
\newblock


\bibitem[\protect\citeauthoryear{Dafoe, Bachrach, Hadfield, Horvitz, Larson,
  and Graepel}{Dafoe et~al\mbox{.}}{2021}]%
        {dafoe_2021_cooperative}
\bibfield{author}{\bibinfo{person}{Allan Dafoe}, \bibinfo{person}{Yoram
  Bachrach}, \bibinfo{person}{Gillian Hadfield}, \bibinfo{person}{Eric
  Horvitz}, \bibinfo{person}{Kate Larson}, {and} \bibinfo{person}{Thore
  Graepel}.} \bibinfo{year}{2021}\natexlab{}.
\newblock \showarticletitle{Cooperative AI: machines must learn to find common
  ground}.
\newblock \bibinfo{journal}{\emph{Nature}}  \bibinfo{volume}{593}
  (\bibinfo{date}{05} \bibinfo{year}{2021}), \bibinfo{pages}{33–36}.
\newblock
\urldef\tempurl%
\url{https://doi.org/10.1038/d41586-021-01170-0}
\showDOI{\tempurl}


\bibitem[\protect\citeauthoryear{Dafoe, Hughes, Bachrach, Collins, McKee,
  Leibo, Larson, and Graepel}{Dafoe et~al\mbox{.}}{2020}]%
        {dafoe_2020_open}
\bibfield{author}{\bibinfo{person}{Allan Dafoe}, \bibinfo{person}{Edward
  Hughes}, \bibinfo{person}{Yoram Bachrach}, \bibinfo{person}{Tantum Collins},
  \bibinfo{person}{Kevin~R. McKee}, \bibinfo{person}{Joel~Z. Leibo},
  \bibinfo{person}{Kate Larson}, {and} \bibinfo{person}{Thore Graepel}.}
  \bibinfo{year}{2020}\natexlab{}.
\newblock \showarticletitle{Open Problems in Cooperative AI}.
\newblock \bibinfo{journal}{\emph{arXiv:2012.08630 [cs]}} (\bibinfo{date}{12}
  \bibinfo{year}{2020}).
\newblock
\urldef\tempurl%
\url{https://arxiv.org/abs/2012.08630}
\showURL{%
\tempurl}


\bibitem[\protect\citeauthoryear{Darwin}{Darwin}{1859}]%
        {darwin_1859_on}
\bibfield{author}{\bibinfo{person}{Charles Darwin}.}
  \bibinfo{year}{1859}\natexlab{}.
\newblock \bibinfo{booktitle}{\emph{On the Origin of Species.}}
\newblock \bibinfo{publisher}{Natural History Museum}.
\newblock


\bibitem[\protect\citeauthoryear{Developers}{Developers}{2022a}]%
        {ethereumdevelopers_2022_ethereum}
\bibfield{author}{\bibinfo{person}{Ethereum Developers}.}
  \bibinfo{year}{2022}\natexlab{a}.
\newblock \bibinfo{title}{Ethereum Proof-of-Stake Consensus Specifications}.
\newblock
\newblock
\urldef\tempurl%
\url{https://github.com/ethereum/consensus-specs/blob/dev/specs/phase0/beacon-chain.md}
\showURL{%
\tempurl}


\bibitem[\protect\citeauthoryear{Developers}{Developers}{2022b}]%
        {ethereumdevelopers_2022_proofofstake}
\bibfield{author}{\bibinfo{person}{Ethereum Developers}.}
  \bibinfo{year}{2022}\natexlab{b}.
\newblock \bibinfo{title}{Proof-of-stake (PoS)}.
\newblock
\newblock
\urldef\tempurl%
\url{https://ethereum.org/en/developers/docs/consensus-mechanisms/pos/#finality}
\showURL{%
\tempurl}


\bibitem[\protect\citeauthoryear{Durlauf and Seshadri}{Durlauf and
  Seshadri}{2003}]%
        {durlauf_2003_is}
\bibfield{author}{\bibinfo{person}{Steven~N. Durlauf} {and}
  \bibinfo{person}{Ananth Seshadri}.} \bibinfo{year}{2003}\natexlab{}.
\newblock \showarticletitle{Is assortative matching efficient?}
\newblock \bibinfo{journal}{\emph{Economic Theory}}  \bibinfo{volume}{21}
  (\bibinfo{date}{03} \bibinfo{year}{2003}), \bibinfo{pages}{475--493}.
\newblock
\urldef\tempurl%
\url{https://doi.org/10.1007/s00199-002-0269-8}
\showDOI{\tempurl}


\bibitem[\protect\citeauthoryear{Eeckhout and Kircher}{Eeckhout and
  Kircher}{2018}]%
        {eeckhout_2018_assortative}
\bibfield{author}{\bibinfo{person}{Jan Eeckhout} {and} \bibinfo{person}{Philipp
  Kircher}.} \bibinfo{year}{2018}\natexlab{}.
\newblock \showarticletitle{Assortative Matching With Large Firms}.
\newblock \bibinfo{journal}{\emph{Econometrica}}  \bibinfo{volume}{86}
  (\bibinfo{year}{2018}), \bibinfo{pages}{85--132}.
\newblock
\urldef\tempurl%
\url{https://doi.org/10.3982/ecta14450}
\showDOI{\tempurl}


\bibitem[\protect\citeauthoryear{Eshel and Cavalli-Sforza}{Eshel and
  Cavalli-Sforza}{1982}]%
        {eshel_1982_assortment}
\bibfield{author}{\bibinfo{person}{Ilan Eshel} {and} \bibinfo{person}{L.~L.
  Cavalli-Sforza}.} \bibinfo{year}{1982}\natexlab{}.
\newblock \showarticletitle{Assortment of Encounters and Evolution of
  Cooperativeness}.
\newblock \bibinfo{journal}{\emph{Proceedings of the National Academy of
  Sciences of the United States of America}}  \bibinfo{volume}{79}
  (\bibinfo{year}{1982}), \bibinfo{pages}{1331–1335}.
\newblock
\urldef\tempurl%
\url{https://www.jstor.org/stable/12036}
\showURL{%
\tempurl}


\bibitem[\protect\citeauthoryear{Fehr and Fischbacher}{Fehr and
  Fischbacher}{2004}]%
        {fehr_2004_social}
\bibfield{author}{\bibinfo{person}{Ernst Fehr} {and} \bibinfo{person}{Urs
  Fischbacher}.} \bibinfo{year}{2004}\natexlab{}.
\newblock \showarticletitle{Social norms and human cooperation}.
\newblock \bibinfo{journal}{\emph{Trends in Cognitive Sciences}}
  \bibinfo{volume}{8} (\bibinfo{date}{04} \bibinfo{year}{2004}),
  \bibinfo{pages}{185--190}.
\newblock
\urldef\tempurl%
\url{https://doi.org/10.1016/j.tics.2004.02.007}
\showDOI{\tempurl}


\bibitem[\protect\citeauthoryear{Fehr, Fischbacher, and Gächter}{Fehr
  et~al\mbox{.}}{2002}]%
        {fehr_2002_strong}
\bibfield{author}{\bibinfo{person}{Ernst Fehr}, \bibinfo{person}{Urs
  Fischbacher}, {and} \bibinfo{person}{Simon Gächter}.}
  \bibinfo{year}{2002}\natexlab{}.
\newblock \showarticletitle{Strong reciprocity, human cooperation, and the
  enforcement of social norms}.
\newblock \bibinfo{journal}{\emph{Human Nature}}  \bibinfo{volume}{13}
  (\bibinfo{date}{03} \bibinfo{year}{2002}), \bibinfo{pages}{1--25}.
\newblock
\urldef\tempurl%
\url{https://doi.org/10.1007/s12110-002-1012-7}
\showDOI{\tempurl}


\bibitem[\protect\citeauthoryear{Fehr and Gächter}{Fehr and Gächter}{2000}]%
        {fehr_2000_cooperation}
\bibfield{author}{\bibinfo{person}{Ernst Fehr} {and} \bibinfo{person}{Simon
  Gächter}.} \bibinfo{year}{2000}\natexlab{}.
\newblock \showarticletitle{Cooperation and Punishment in Public Goods
  Experiments}.
\newblock \bibinfo{journal}{\emph{American Economic Review}}
  \bibinfo{volume}{90} (\bibinfo{date}{09} \bibinfo{year}{2000}),
  \bibinfo{pages}{980--994}.
\newblock
\urldef\tempurl%
\url{https://doi.org/10.1257/aer.90.4.980}
\showDOI{\tempurl}


\bibitem[\protect\citeauthoryear{Foundation}{Foundation}{2022}]%
        {ethereumfoundation_2022_solo}
\bibfield{author}{\bibinfo{person}{Ethereum Foundation}.}
  \bibinfo{year}{2022}\natexlab{}.
\newblock \bibinfo{title}{Solo stake your ETH}.
\newblock
\newblock
\urldef\tempurl%
\url{https://ethereum.org/en/staking/solo/}
\showURL{%
\tempurl}


\bibitem[\protect\citeauthoryear{Fudenberg and Tirole}{Fudenberg and
  Tirole}{1991a}]%
        {fudenberg1991game}
\bibfield{author}{\bibinfo{person}{Drew Fudenberg} {and} \bibinfo{person}{Jean
  Tirole}.} \bibinfo{year}{1991}\natexlab{a}.
\newblock \bibinfo{booktitle}{\emph{Game theory}}.
\newblock \bibinfo{publisher}{MIT press}.
\newblock


\bibitem[\protect\citeauthoryear{Fudenberg and Tirole}{Fudenberg and
  Tirole}{1991b}]%
        {fudenberg1991perfect}
\bibfield{author}{\bibinfo{person}{Drew Fudenberg} {and} \bibinfo{person}{Jean
  Tirole}.} \bibinfo{year}{1991}\natexlab{b}.
\newblock \showarticletitle{Perfect Bayesian equilibrium and sequential
  equilibrium}.
\newblock \bibinfo{journal}{\emph{journal of Economic Theory}}
  \bibinfo{volume}{53}, \bibinfo{number}{2} (\bibinfo{year}{1991}),
  \bibinfo{pages}{236--260}.
\newblock


\bibitem[\protect\citeauthoryear{Fudenberg and Tirole}{Fudenberg and
  Tirole}{1991c}]%
        {fudenberg_1991_perfect}
\bibfield{author}{\bibinfo{person}{Drew Fudenberg} {and} \bibinfo{person}{Jean
  Tirole}.} \bibinfo{year}{1991}\natexlab{c}.
\newblock \showarticletitle{Perfect Bayesian equilibrium and sequential
  equilibrium}.
\newblock \bibinfo{journal}{\emph{Journal of Economic Theory}}
  \bibinfo{volume}{53} (\bibinfo{date}{04} \bibinfo{year}{1991}),
  \bibinfo{pages}{236--260}.
\newblock
\urldef\tempurl%
\url{https://doi.org/10.1016/0022-0531(91)90155-w}
\showDOI{\tempurl}


\bibitem[\protect\citeauthoryear{Gagniuc}{Gagniuc}{2017}]%
        {gagniuc_2017_markov}
\bibfield{author}{\bibinfo{person}{Paul~A Gagniuc}.}
  \bibinfo{year}{2017}\natexlab{}.
\newblock \bibinfo{booktitle}{\emph{Markov chains : from theory to
  implementation and experimentation}}.
\newblock \bibinfo{publisher}{John Wiley \& Sons}.
\newblock


\bibitem[\protect\citeauthoryear{Gersbach, Mamageishvili, and
  Schneider}{Gersbach et~al\mbox{.}}{2022a}]%
        {gersbach_2022_risky}
\bibfield{author}{\bibinfo{person}{Hans Gersbach}, \bibinfo{person}{Akaki
  Mamageishvili}, {and} \bibinfo{person}{Manvir Schneider}.}
  \bibinfo{year}{2022}\natexlab{a}.
\newblock \bibinfo{title}{Risky Vote Delegation}.
\newblock
\newblock
\urldef\tempurl%
\url{https://ssrn.com/abstract=4069832}
\showURL{%
\tempurl}


\bibitem[\protect\citeauthoryear{Gersbach, Mamageishvili, and
  Schneider}{Gersbach et~al\mbox{.}}{2022b}]%
        {gersbach_2022_staking}
\bibfield{author}{\bibinfo{person}{Hans Gersbach}, \bibinfo{person}{Akaki
  Mamageishvili}, {and} \bibinfo{person}{Manvir Schneider}.}
  \bibinfo{year}{2022}\natexlab{b}.
\newblock \showarticletitle{Staking Pools on Blockchains}.
\newblock \bibinfo{journal}{\emph{arXiv:2203.05838 [cs]}} (\bibinfo{date}{10}
  \bibinfo{year}{2022}).
\newblock
\urldef\tempurl%
\url{https://arxiv.org/abs/2203.05838}
\showURL{%
\tempurl}


\bibitem[\protect\citeauthoryear{Gilad, Hemo, Micali, Vlachos, and
  Zeldovich}{Gilad et~al\mbox{.}}{2017}]%
        {gilad_2017_algorand}
\bibfield{author}{\bibinfo{person}{Yossi Gilad}, \bibinfo{person}{Rotem Hemo},
  \bibinfo{person}{Silvio Micali}, \bibinfo{person}{Georgios Vlachos}, {and}
  \bibinfo{person}{Nickolai Zeldovich}.} \bibinfo{year}{2017}\natexlab{}.
\newblock \showarticletitle{Algorand: Scaling Byzantine Agreements for
  Cryptocurrencies}.
\newblock \bibinfo{journal}{\emph{Proceedings of the 26th Symposium on
  Operating Systems Principles}} (\bibinfo{date}{10} \bibinfo{year}{2017}).
\newblock
\urldef\tempurl%
\url{https://doi.org/10.1145/3132747.3132757}
\showDOI{\tempurl}


\bibitem[\protect\citeauthoryear{Guo and Ren}{Guo and Ren}{2022}]%
        {guo_2022_bitcoins}
\bibfield{author}{\bibinfo{person}{Dongning Guo} {and} \bibinfo{person}{Ling
  Ren}.} \bibinfo{year}{2022}\natexlab{}.
\newblock \showarticletitle{Bitcoin's Latency--Security Analysis Made Simple}.
\newblock \bibinfo{journal}{\emph{arXiv:2203.06357 [cs]}} (\bibinfo{date}{08}
  \bibinfo{year}{2022}).
\newblock
\urldef\tempurl%
\url{https://arxiv.org/abs/2203.06357}
\showURL{%
\tempurl}


\bibitem[\protect\citeauthoryear{Halaburda, He, and Li}{Halaburda
  et~al\mbox{.}}{2021}]%
        {halaburda_2021_an}
\bibfield{author}{\bibinfo{person}{Hanna Halaburda}, \bibinfo{person}{Zhiguo
  He}, {and} \bibinfo{person}{Jiasun Li}.} \bibinfo{year}{2021}\natexlab{}.
\newblock \bibinfo{title}{An Economic Model of Consensus on Distributed
  Ledgers}.
\newblock
\newblock
\urldef\tempurl%
\url{https://www.nber.org/papers/w29515}
\showURL{%
\tempurl}


\bibitem[\protect\citeauthoryear{Halpern}{Halpern}{2007}]%
        {halpern_2007_computer}
\bibfield{author}{\bibinfo{person}{Joseph~Y. Halpern}.}
  \bibinfo{year}{2007}\natexlab{}.
\newblock \showarticletitle{Computer Science and Game Theory: A Brief Survey}.
\newblock \bibinfo{journal}{\emph{arXiv:cs/0703148}} (\bibinfo{date}{03}
  \bibinfo{year}{2007}).
\newblock
\urldef\tempurl%
\url{https://arxiv.org/abs/cs/0703148}
\showURL{%
\tempurl}


\bibitem[\protect\citeauthoryear{Harris and Waggoner}{Harris and
  Waggoner}{2019}]%
        {harris_2019_decentralized}
\bibfield{author}{\bibinfo{person}{Justin~D. Harris} {and} \bibinfo{person}{Bo
  Waggoner}.} \bibinfo{year}{2019}\natexlab{}.
\newblock \bibinfo{title}{Decentralized and Collaborative AI on Blockchain}.
\newblock , \bibinfo{numpages}{368–375}~pages.
\newblock
\urldef\tempurl%
\url{https://doi.org/10.1109/Blockchain.2019.00057}
\showDOI{\tempurl}


\bibitem[\protect\citeauthoryear{He and Li}{He and Li}{2022}]%
        {he_2022_contract}
\bibfield{author}{\bibinfo{person}{Zhiguo He} {and} \bibinfo{person}{Jiasun
  Li}.} \bibinfo{year}{2022}\natexlab{}.
\newblock \bibinfo{title}{Contract Enforcement and Decentralized Consensus: The
  Case of Slashing}.
\newblock
\newblock
\urldef\tempurl%
\url{https://ssrn.com/abstract=4036000}
\showURL{%
\tempurl}


\bibitem[\protect\citeauthoryear{Henrich, Boyd, Bowles, Camerer, Fehr, Gintis,
  and McElreath}{Henrich et~al\mbox{.}}{2001}]%
        {henrich_2001_in}
\bibfield{author}{\bibinfo{person}{Joseph Henrich}, \bibinfo{person}{Robert
  Boyd}, \bibinfo{person}{Samuel Bowles}, \bibinfo{person}{Colin Camerer},
  \bibinfo{person}{Ernst Fehr}, \bibinfo{person}{Herbert Gintis}, {and}
  \bibinfo{person}{Richard McElreath}.} \bibinfo{year}{2001}\natexlab{}.
\newblock \showarticletitle{In Search of Homo Economicus: Behavioral
  Experiments in 15 Small-Scale Societies}.
\newblock \bibinfo{journal}{\emph{American Economic Review}}
  \bibinfo{volume}{91} (\bibinfo{date}{05} \bibinfo{year}{2001}),
  \bibinfo{pages}{73--78}.
\newblock
\urldef\tempurl%
\url{https://doi.org/10.1257/aer.91.2.73}
\showDOI{\tempurl}


\bibitem[\protect\citeauthoryear{Henrich and Muthukrishna}{Henrich and
  Muthukrishna}{2020}]%
        {henrich_origins_2020}
\bibfield{author}{\bibinfo{person}{Joseph Henrich} {and}
  \bibinfo{person}{Michael Muthukrishna}.} \bibinfo{year}{2020}\natexlab{}.
\newblock \showarticletitle{The Origins and Psychology of Human Cooperation}.
\newblock   \bibinfo{volume}{72} (\bibinfo{year}{2020}).
\newblock
\urldef\tempurl%
\url{https://doi.org/10.1146/annurev-psych-081920-042106}
\showDOI{\tempurl}


\bibitem[\protect\citeauthoryear{Hoffman, Morgan, Raymond, Abaluck, Dellavigna,
  Finan, Gailmard, Gerber, Green, Green, Leon, Myatt, Rabin, and
  Houweling}{Hoffman et~al\mbox{.}}{2013}]%
        {hoffman_2013_one}
\bibfield{author}{\bibinfo{person}{Mitchell Hoffman}, \bibinfo{person}{John
  Morgan}, \bibinfo{person}{Collin Raymond}, \bibinfo{person}{Jason Abaluck},
  \bibinfo{person}{Stefano Dellavigna}, \bibinfo{person}{Fred Finan},
  \bibinfo{person}{Sean Gailmard}, \bibinfo{person}{Alan Gerber},
  \bibinfo{person}{Don Green}, \bibinfo{person}{Jennifer Green},
  \bibinfo{person}{Gianmarco Leon}, \bibinfo{person}{David Myatt},
  \bibinfo{person}{Matthew Rabin}, {and} \bibinfo{person}{Rob Houweling}.}
  \bibinfo{year}{2013}\natexlab{}.
\newblock \bibinfo{title}{One in a Million: A Field Experiment on Belief
  Formation and Pivotal Voting *}.
\newblock
\newblock
\urldef\tempurl%
\url{https://www.haas.berkeley.edu/wp-content/uploads/Morgan.pdf}
\showURL{%
\tempurl}


\bibitem[\protect\citeauthoryear{Horvitz}{Horvitz}{2016}]%
        {horvitz2016one}
\bibfield{author}{\bibinfo{person}{Eric Horvitz}.}
  \bibinfo{year}{2016}\natexlab{}.
\newblock \bibinfo{title}{One hundred year study on artificial intelligence}.
\newblock
\newblock


\bibitem[\protect\citeauthoryear{Hou, Zhou, Ji, Daian, Tramer, Fanti, and
  Juels}{Hou et~al\mbox{.}}{2020}]%
        {hou_2020_squirrl}
\bibfield{author}{\bibinfo{person}{Charlie Hou}, \bibinfo{person}{Mingxun
  Zhou}, \bibinfo{person}{Yan Ji}, \bibinfo{person}{Phil Daian},
  \bibinfo{person}{Florian Tramer}, \bibinfo{person}{Giulia Fanti}, {and}
  \bibinfo{person}{Ari Juels}.} \bibinfo{year}{2020}\natexlab{}.
\newblock \showarticletitle{SquirRL: Automating Attack Analysis on Blockchain
  Incentive Mechanisms with Deep Reinforcement Learning}.
\newblock \bibinfo{journal}{\emph{arXiv:1912.01798 [cs]}} (\bibinfo{date}{08}
  \bibinfo{year}{2020}).
\newblock
\urldef\tempurl%
\url{https://arxiv.org/abs/1912.01798}
\showURL{%
\tempurl}


\bibitem[\protect\citeauthoryear{Ivanov}{Ivanov}{2022}]%
        {ivanov2022reinforcement}
\bibfield{author}{\bibinfo{person}{Sergey Ivanov}.}
  \bibinfo{year}{2022}\natexlab{}.
\newblock \showarticletitle{Reinforcement Learning Textbook}.
\newblock \bibinfo{journal}{\emph{arXiv preprint arXiv:2201.09746}}
  (\bibinfo{year}{2022}).
\newblock


\bibitem[\protect\citeauthoryear{Jackson, Rodriguez-Barraquer, and Tan}{Jackson
  et~al\mbox{.}}{2012}]%
        {jackson2012epsilon}
\bibfield{author}{\bibinfo{person}{Matthew~O Jackson}, \bibinfo{person}{Tomas
  Rodriguez-Barraquer}, {and} \bibinfo{person}{Xu Tan}.}
  \bibinfo{year}{2012}\natexlab{}.
\newblock \showarticletitle{Epsilon-equilibria of perturbed games}.
\newblock \bibinfo{journal}{\emph{Games and Economic Behavior}}
  \bibinfo{volume}{75}, \bibinfo{number}{1} (\bibinfo{year}{2012}),
  \bibinfo{pages}{198--216}.
\newblock


\bibitem[\protect\citeauthoryear{Jehiel}{Jehiel}{2001}]%
        {jehiel_2001_limited}
\bibfield{author}{\bibinfo{person}{Philippe Jehiel}.}
  \bibinfo{year}{2001}\natexlab{}.
\newblock \showarticletitle{Limited Foresight May Force Cooperation}.
\newblock \bibinfo{journal}{\emph{The Review of Economic Studies}}
  \bibinfo{volume}{68} (\bibinfo{year}{2001}), \bibinfo{pages}{369--391}.
\newblock
\urldef\tempurl%
\url{https://www.jstor.org/stable/2695933}
\showURL{%
\tempurl}


\bibitem[\protect\citeauthoryear{Khanchandani and Wattenhofer}{Khanchandani and
  Wattenhofer}{2021}]%
        {khanchandani_2021_byzantine}
\bibfield{author}{\bibinfo{person}{Pankaj Khanchandani} {and}
  \bibinfo{person}{Roger Wattenhofer}.} \bibinfo{year}{2021}\natexlab{}.
\newblock \bibinfo{title}{Byzantine Agreement with Unknown Participants and
  Failures}.
\newblock
\newblock
\urldef\tempurl%
\url{https://arxiv.org/pdf/2102.10442.pdf}
\showURL{%
\tempurl}


\bibitem[\protect\citeauthoryear{Koduri and Lo}{Koduri and Lo}{2021}]%
        {koduri_2021_the}
\bibfield{author}{\bibinfo{person}{Nihal Koduri} {and}
  \bibinfo{person}{Andrew~W. Lo}.} \bibinfo{year}{2021}\natexlab{}.
\newblock \showarticletitle{The origin of cooperation}.
\newblock \bibinfo{journal}{\emph{Proceedings of the National Academy of
  Sciences}}  \bibinfo{volume}{118} (\bibinfo{date}{06} \bibinfo{year}{2021}),
  \bibinfo{pages}{e2015572118}.
\newblock
\urldef\tempurl%
\url{https://doi.org/10.1073/pnas.2015572118}
\showDOI{\tempurl}


\bibitem[\protect\citeauthoryear{Kokoris-Kogias, Jovanovic, Gasser, Gailly,
  Syta, and Ford}{Kokoris-Kogias et~al\mbox{.}}{2018}]%
        {kokoriskogias_2018_omniledger}
\bibfield{author}{\bibinfo{person}{Eleftherios Kokoris-Kogias},
  \bibinfo{person}{Philipp Jovanovic}, \bibinfo{person}{Linus Gasser},
  \bibinfo{person}{Nicolas Gailly}, \bibinfo{person}{Ewa Syta}, {and}
  \bibinfo{person}{Bryan Ford}.} \bibinfo{year}{2018}\natexlab{}.
\newblock \bibinfo{title}{OmniLedger: A Secure, Scale-Out, Decentralized Ledger
  via Sharding}.
\newblock , \bibinfo{numpages}{583–598}~pages.
\newblock
\urldef\tempurl%
\url{https://doi.org/10.1109/SP.2018.000-5}
\showDOI{\tempurl}


\bibitem[\protect\citeauthoryear{Kotla, Clement, Wong, Alvisi, and
  Dahlin}{Kotla et~al\mbox{.}}{2007}]%
        {kotla_2007_zyzzyva}
\bibfield{author}{\bibinfo{person}{Ramakrishna Kotla}, \bibinfo{person}{Allen
  Clement}, \bibinfo{person}{Edmund Wong}, \bibinfo{person}{Lorenzo Alvisi},
  {and} \bibinfo{person}{Mike Dahlin}.} \bibinfo{year}{2007}\natexlab{}.
\newblock \bibinfo{title}{Zyzzyva: Speculative Byzantine Fault Tolerance}.
\newblock
\newblock
\urldef\tempurl%
\url{https://www.cs.utexas.edu/users/dahlin/papers/Zyzzyva-CACM.pdf}
\showURL{%
\tempurl}


\bibitem[\protect\citeauthoryear{Kroer and Sandholm}{Kroer and
  Sandholm}{2014}]%
        {kroer2014extensive}
\bibfield{author}{\bibinfo{person}{Christian Kroer} {and}
  \bibinfo{person}{Tuomas Sandholm}.} \bibinfo{year}{2014}\natexlab{}.
\newblock \showarticletitle{Extensive-form game abstraction with bounds}. In
  \bibinfo{booktitle}{\emph{Proceedings of the fifteenth ACM conference on
  Economics and computation}}. \bibinfo{pages}{621--638}.
\newblock


\bibitem[\protect\citeauthoryear{Lamport, Shostak, and Pease}{Lamport
  et~al\mbox{.}}{1982}]%
        {lamport_1982_the}
\bibfield{author}{\bibinfo{person}{Leslie Lamport}, \bibinfo{person}{Robert
  Shostak}, {and} \bibinfo{person}{Marshall Pease}.}
  \bibinfo{year}{1982}\natexlab{}.
\newblock \showarticletitle{The Byzantine Generals Problem}.
\newblock \bibinfo{journal}{\emph{ACM Transactions on Programming Languages and
  Systems}}  \bibinfo{volume}{4} (\bibinfo{date}{07} \bibinfo{year}{1982}),
  \bibinfo{pages}{382--401}.
\newblock
\urldef\tempurl%
\url{https://doi.org/10.1145/357172.357176}
\showDOI{\tempurl}


\bibitem[\protect\citeauthoryear{Levin and Zhang}{Levin and Zhang}{2020}]%
        {levin_2020_bridging}
\bibfield{author}{\bibinfo{person}{Dan Levin} {and} \bibinfo{person}{Luyao
  Zhang}.} \bibinfo{year}{2020}\natexlab{}.
\newblock \showarticletitle{Bridging Level-K to Nash Equilibrium}.
\newblock \bibinfo{journal}{\emph{The Review of Economics and Statistics}}
  (\bibinfo{date}{10} \bibinfo{year}{2020}), \bibinfo{pages}{1--44}.
\newblock
\urldef\tempurl%
\url{https://doi.org/10.1162/rest_a_00990}
\showDOI{\tempurl}


\bibitem[\protect\citeauthoryear{Levin and Lo}{Levin and Lo}{2021}]%
        {levin_2021_introduction}
\bibfield{author}{\bibinfo{person}{Simon~A. Levin} {and}
  \bibinfo{person}{Andrew~W. Lo}.} \bibinfo{year}{2021}\natexlab{}.
\newblock \showarticletitle{Introduction to PNAS special issue on evolutionary
  models of financial markets}.
\newblock \bibinfo{journal}{\emph{Proceedings of the National Academy of
  Sciences}}  \bibinfo{volume}{118} (\bibinfo{date}{06} \bibinfo{year}{2021}).
\newblock
\urldef\tempurl%
\url{https://doi.org/10.1073/pnas.2104800118}
\showDOI{\tempurl}


\bibitem[\protect\citeauthoryear{Lewis-Pye and Roughgarden}{Lewis-Pye and
  Roughgarden}{2022}]%
        {lewispye_2022_byzantine}
\bibfield{author}{\bibinfo{person}{Andrew Lewis-Pye} {and} \bibinfo{person}{Tim
  Roughgarden}.} \bibinfo{year}{2022}\natexlab{}.
\newblock \bibinfo{title}{Byzantine Generals in the Permissionless Setting}.
\newblock
\newblock
\urldef\tempurl%
\url{https://arxiv.org/pdf/2101.07095.pdf}
\showURL{%
\tempurl}


\bibitem[\protect\citeauthoryear{Lone and Mir}{Lone and Mir}{2019}]%
        {lone_2019_consensus}
\bibfield{author}{\bibinfo{person}{Auqib~Hamid Lone} {and}
  \bibinfo{person}{Roohie~Naaz Mir}.} \bibinfo{year}{2019}\natexlab{}.
\newblock \showarticletitle{Consensus protocols as a model of trust in
  blockchains}.
\newblock \bibinfo{journal}{\emph{International Journal of Blockchains and
  Cryptocurrencies}}  \bibinfo{volume}{1} (\bibinfo{year}{2019}),
  \bibinfo{pages}{7}.
\newblock
\urldef\tempurl%
\url{https://doi.org/10.1504/ijbc.2019.101845}
\showDOI{\tempurl}


\bibitem[\protect\citeauthoryear{Mallouh, Abuzaghleh, and Qawaqneh}{Mallouh
  et~al\mbox{.}}{2021}]%
        {mallouh_2021_a}
\bibfield{author}{\bibinfo{person}{Arafat~Abu Mallouh}, \bibinfo{person}{Omar
  Abuzaghleh}, {and} \bibinfo{person}{Zakariya Qawaqneh}.}
  \bibinfo{year}{2021}\natexlab{}.
\newblock \bibinfo{title}{A Hierarchy of Deep Reinforcement Learning Agents for
  Decision Making in Blockchain Nodes}.
\newblock , \bibinfo{numpages}{197–202}~pages.
\newblock
\urldef\tempurl%
\url{https://doi.org/10.1109/EUROCON52738.2021.9535600}
\showDOI{\tempurl}


\bibitem[\protect\citeauthoryear{Marwala and Xing}{Marwala and Xing}{2018}]%
        {marwala_2018_blockchain}
\bibfield{author}{\bibinfo{person}{Tshilidzi Marwala} {and} \bibinfo{person}{Bo
  Xing}.} \bibinfo{year}{2018}\natexlab{}.
\newblock \showarticletitle{Blockchain and Artificial Intelligence}.
\newblock \bibinfo{journal}{\emph{arXiv:1802.04451 [cs]}} (\bibinfo{date}{10}
  \bibinfo{year}{2018}).
\newblock
\urldef\tempurl%
\url{https://arxiv.org/abs/1802.04451}
\showURL{%
\tempurl}


\bibitem[\protect\citeauthoryear{Maskin and Tirole}{Maskin and Tirole}{2001a}]%
        {maskin_2001_markov}
\bibfield{author}{\bibinfo{person}{Eric Maskin} {and} \bibinfo{person}{Jean
  Tirole}.} \bibinfo{year}{2001}\natexlab{a}.
\newblock \showarticletitle{Markov Perfect Equilibrium}.
\newblock \bibinfo{journal}{\emph{Journal of Economic Theory}}
  \bibinfo{volume}{100} (\bibinfo{date}{10} \bibinfo{year}{2001}),
  \bibinfo{pages}{191--219}.
\newblock
\urldef\tempurl%
\url{https://doi.org/10.1006/jeth.2000.2785}
\showDOI{\tempurl}


\bibitem[\protect\citeauthoryear{Maskin and Tirole}{Maskin and Tirole}{2001b}]%
        {maskin2001markov}
\bibfield{author}{\bibinfo{person}{Eric Maskin} {and} \bibinfo{person}{Jean
  Tirole}.} \bibinfo{year}{2001}\natexlab{b}.
\newblock \showarticletitle{Markov perfect equilibrium: I. Observable actions}.
\newblock \bibinfo{journal}{\emph{Journal of Economic Theory}}
  \bibinfo{volume}{100}, \bibinfo{number}{2} (\bibinfo{year}{2001}),
  \bibinfo{pages}{191--219}.
\newblock


\bibitem[\protect\citeauthoryear{McKenzie}{McKenzie}{2009}]%
        {mckenzie_2009_evolutionary}
\bibfield{author}{\bibinfo{person}{Alexander~J McKenzie}.}
  \bibinfo{year}{2009}\natexlab{}.
\newblock \bibinfo{title}{Evolutionary Game Theory (Stanford Encyclopedia of
  Philosophy)}.
\newblock
\newblock
\urldef\tempurl%
\url{https://plato.stanford.edu/entries/game-evolutionary/}
\showURL{%
\tempurl}


\bibitem[\protect\citeauthoryear{Momose and Ren}{Momose and Ren}{2022}]%
        {cryptoeprint:2022/404}
\bibfield{author}{\bibinfo{person}{Atsuki Momose} {and} \bibinfo{person}{Ling
  Ren}.} \bibinfo{year}{2022}\natexlab{}.
\newblock \bibinfo{title}{Constant Latency in Sleepy Consensus}.
\newblock \bibinfo{howpublished}{Cryptology ePrint Archive, Paper 2022/404}.
\newblock
\urldef\tempurl%
\url{https://eprint.iacr.org/2022/404}
\showURL{%
\tempurl}
\newblock
\shownote{\url{https://eprint.iacr.org/2022/404}}.


\bibitem[\protect\citeauthoryear{Moscibroda, Schmid, and
  Wattenhofer}{Moscibroda et~al\mbox{.}}{2006}]%
        {moscibroda_2006_when}
\bibfield{author}{\bibinfo{person}{Thomas Moscibroda}, \bibinfo{person}{Stefan
  Schmid}, {and} \bibinfo{person}{Roger Wattenhofer}.}
  \bibinfo{year}{2006}\natexlab{}.
\newblock \showarticletitle{When selfish meets evil}.
\newblock \bibinfo{journal}{\emph{Proceedings of the twenty-fifth annual ACM
  symposium on Principles of distributed computing - PODC '06}}
  (\bibinfo{year}{2006}).
\newblock
\urldef\tempurl%
\url{https://doi.org/10.1145/1146381.1146391}
\showDOI{\tempurl}


\bibitem[\protect\citeauthoryear{Nakamoto}{Nakamoto}{2008}]%
        {nakamoto_2008_bitcoin}
\bibfield{author}{\bibinfo{person}{Satoshi Nakamoto}.}
  \bibinfo{year}{2008}\natexlab{}.
\newblock \bibinfo{title}{Bitcoin: a Peer-to-Peer Electronic Cash System}.
\newblock
\newblock
\urldef\tempurl%
\url{https://bitcoin.org/bitcoin.pdf}
\showURL{%
\tempurl}


\bibitem[\protect\citeauthoryear{Neu, Sridhar, Yang, Tse, and Alizadeh}{Neu
  et~al\mbox{.}}{2022}]%
        {neu_2022_longest}
\bibfield{author}{\bibinfo{person}{Joachim Neu}, \bibinfo{person}{Srivatsan
  Sridhar}, \bibinfo{person}{Lei Yang}, \bibinfo{person}{David Tse}, {and}
  \bibinfo{person}{Mohammad Alizadeh}.} \bibinfo{year}{2022}\natexlab{}.
\newblock \showarticletitle{Longest Chain Consensus Under Bandwidth
  Constraint}.
\newblock \bibinfo{journal}{\emph{arXiv:2111.12332 [cs]}} (\bibinfo{date}{05}
  \bibinfo{year}{2022}).
\newblock
\urldef\tempurl%
\url{https://arxiv.org/abs/2111.12332}
\showURL{%
\tempurl}


\bibitem[\protect\citeauthoryear{Nisan}{Nisan}{2007}]%
        {noamnisan_2007_algorithmic}
\bibfield{author}{\bibinfo{person}{Noam Nisan}.}
  \bibinfo{year}{2007}\natexlab{}.
\newblock \bibinfo{booktitle}{\emph{Algorithmic game theory}}.
\newblock \bibinfo{publisher}{Cambridge University Press}.
\newblock


\bibitem[\protect\citeauthoryear{Osborne and Rubinstein}{Osborne and
  Rubinstein}{1994}]%
        {osborne_1994_a}
\bibfield{author}{\bibinfo{person}{Martin~J Osborne} {and}
  \bibinfo{person}{Ariel Rubinstein}.} \bibinfo{year}{1994}\natexlab{}.
\newblock \bibinfo{booktitle}{\emph{A course in game theory}}.
\newblock \bibinfo{publisher}{Mit Press}.
\newblock


\bibitem[\protect\citeauthoryear{Parra~Moyano and Ross}{Parra~Moyano and
  Ross}{2017}]%
        {parramoyano_2017_kyc}
\bibfield{author}{\bibinfo{person}{José Parra~Moyano} {and}
  \bibinfo{person}{Omri Ross}.} \bibinfo{year}{2017}\natexlab{}.
\newblock \showarticletitle{KYC Optimization Using Distributed Ledger
  Technology}.
\newblock \bibinfo{journal}{\emph{Business \& Information Systems Engineering}}
   \bibinfo{volume}{59} (\bibinfo{date}{11} \bibinfo{year}{2017}),
  \bibinfo{pages}{411--423}.
\newblock
\urldef\tempurl%
\url{https://doi.org/10.1007/s12599-017-0504-2}
\showDOI{\tempurl}


\bibitem[\protect\citeauthoryear{Pass and Shi}{Pass and Shi}{2017}]%
        {Pass2017HybridCE}
\bibfield{author}{\bibinfo{person}{Rafael Pass} {and} \bibinfo{person}{Elaine
  Shi}.} \bibinfo{year}{2017}\natexlab{}.
\newblock \showarticletitle{Hybrid Consensus: Efficient Consensus in the
  Permissionless Model}. In \bibinfo{booktitle}{\emph{DISC}}.
\newblock


\bibitem[\protect\citeauthoryear{Pease, Shostak, and Lamport}{Pease
  et~al\mbox{.}}{1980}]%
        {pease_1980_reaching}
\bibfield{author}{\bibinfo{person}{M. Pease}, \bibinfo{person}{R. Shostak},
  {and} \bibinfo{person}{L. Lamport}.} \bibinfo{year}{1980}\natexlab{}.
\newblock \showarticletitle{Reaching Agreement in the Presence of Faults}.
\newblock \bibinfo{journal}{\emph{J. ACM}}  \bibinfo{volume}{27}
  (\bibinfo{date}{04} \bibinfo{year}{1980}), \bibinfo{pages}{228--234}.
\newblock
\urldef\tempurl%
\url{https://doi.org/10.1145/322186.322188}
\showDOI{\tempurl}


\bibitem[\protect\citeauthoryear{Pu, Alvisi, and Eyal}{Pu
  et~al\mbox{.}}{2022}]%
        {cryptoeprint:2022/796}
\bibfield{author}{\bibinfo{person}{Youer Pu}, \bibinfo{person}{Lorenzo Alvisi},
  {and} \bibinfo{person}{Ittay Eyal}.} \bibinfo{year}{2022}\natexlab{}.
\newblock \bibinfo{title}{Safe Permissionless Consensus}.
\newblock \bibinfo{howpublished}{Cryptology ePrint Archive, Paper 2022/796}.
\newblock
\urldef\tempurl%
\url{https://eprint.iacr.org/2022/796}
\showURL{%
\tempurl}


\bibitem[\protect\citeauthoryear{Ross, Greene, and House}{Ross
  et~al\mbox{.}}{1977}]%
        {ross_1977_the}
\bibfield{author}{\bibinfo{person}{Lee Ross}, \bibinfo{person}{David Greene},
  {and} \bibinfo{person}{Pamela House}.} \bibinfo{year}{1977}\natexlab{}.
\newblock \showarticletitle{The “false consensus effect”: An egocentric
  bias in social perception and attribution processes}.
\newblock \bibinfo{journal}{\emph{Journal of Experimental Social Psychology}}
  \bibinfo{volume}{13} (\bibinfo{date}{05} \bibinfo{year}{1977}),
  \bibinfo{pages}{279--301}.
\newblock
\urldef\tempurl%
\url{https://doi.org/10.1016/0022-1031(77)90049-x}
\showDOI{\tempurl}


\bibitem[\protect\citeauthoryear{Rubinstein}{Rubinstein}{2002}]%
        {rubinstein_2002_modeling}
\bibfield{author}{\bibinfo{person}{Ariel Rubinstein}.}
  \bibinfo{year}{2002}\natexlab{}.
\newblock \bibinfo{booktitle}{\emph{Modeling bounded rationality}}.
\newblock \bibinfo{publisher}{Mit Press}.
\newblock


\bibitem[\protect\citeauthoryear{Rubinstein and Salant}{Rubinstein and
  Salant}{2016}]%
        {rubinstein2016}
\bibfield{author}{\bibinfo{person}{Ariel Rubinstein} {and}
  \bibinfo{person}{Yuval Salant}.} \bibinfo{year}{2016}\natexlab{}.
\newblock \showarticletitle{"Isn't everyone like me?": On the presence of
  self-similarity in strategic interactions}.
\newblock \bibinfo{journal}{\emph{Judgment and Decision Making}}
  \bibinfo{volume}{11}, \bibinfo{number}{2} (\bibinfo{year}{2016}),
  \bibinfo{pages}{168--173}.
\newblock
\urldef\tempurl%
\url{https://EconPapers.repec.org/RePEc:jdm:journl:v:11:y:2016:i:2:p:168-173}
\showURL{%
\tempurl}


\bibitem[\protect\citeauthoryear{Salah, Rehman, Nizamuddin, and
  Al-Fuqaha}{Salah et~al\mbox{.}}{2019}]%
        {salah_2019_blockchain}
\bibfield{author}{\bibinfo{person}{Khaled Salah}, \bibinfo{person}{M.~Habib~Ur
  Rehman}, \bibinfo{person}{Nishara Nizamuddin}, {and} \bibinfo{person}{Ala
  Al-Fuqaha}.} \bibinfo{year}{2019}\natexlab{}.
\newblock \showarticletitle{Blockchain for AI: Review and Open Research
  Challenges}.
\newblock \bibinfo{journal}{\emph{IEEE Access}}  \bibinfo{volume}{7}
  (\bibinfo{year}{2019}), \bibinfo{pages}{10127--10149}.
\newblock
\urldef\tempurl%
\url{https://doi.org/10.1109/access.2018.2890507}
\showDOI{\tempurl}


\bibitem[\protect\citeauthoryear{Saleh}{Saleh}{2018}]%
        {saleh_2018_blockchain}
\bibfield{author}{\bibinfo{person}{Fahad Saleh}.}
  \bibinfo{year}{2018}\natexlab{}.
\newblock \showarticletitle{Blockchain Without Waste: Proof-of-Stake}.
\newblock \bibinfo{journal}{\emph{SSRN Electronic Journal}}
  (\bibinfo{year}{2018}).
\newblock
\urldef\tempurl%
\url{https://doi.org/10.2139/ssrn.3183935}
\showDOI{\tempurl}


\bibitem[\protect\citeauthoryear{Saleh}{Saleh}{2021}]%
        {saleh2021blockchain}
\bibfield{author}{\bibinfo{person}{Fahad Saleh}.}
  \bibinfo{year}{2021}\natexlab{}.
\newblock \showarticletitle{Blockchain without waste: Proof-of-stake}.
\newblock \bibinfo{journal}{\emph{The Review of financial studies}}
  \bibinfo{volume}{34}, \bibinfo{number}{3} (\bibinfo{year}{2021}),
  \bibinfo{pages}{1156--1190}.
\newblock


\bibitem[\protect\citeauthoryear{Sankar, Sindhu, and Sethumadhavan}{Sankar
  et~al\mbox{.}}{2017}]%
        {sankar2017survey}
\bibfield{author}{\bibinfo{person}{Lakshmi~Siva Sankar}, \bibinfo{person}{M
  Sindhu}, {and} \bibinfo{person}{M Sethumadhavan}.}
  \bibinfo{year}{2017}\natexlab{}.
\newblock \showarticletitle{Survey of consensus protocols on blockchain
  applications}. In \bibinfo{booktitle}{\emph{2017 4th international conference
  on advanced computing and communication systems (ICACCS)}}. IEEE,
  \bibinfo{pages}{1--5}.
\newblock


\bibitem[\protect\citeauthoryear{Schneider and Weber}{Schneider and
  Weber}{2013}]%
        {schneider_2013_longterm}
\bibfield{author}{\bibinfo{person}{Frrddric Schneider} {and}
  \bibinfo{person}{Roberto~A. Weber}.} \bibinfo{year}{2013}\natexlab{}.
\newblock \showarticletitle{Long-Term Commitment and Cooperation}.
\newblock \bibinfo{journal}{\emph{SSRN Electronic Journal}}
  (\bibinfo{year}{2013}).
\newblock
\urldef\tempurl%
\url{https://doi.org/10.2139/ssrn.2334376}
\showDOI{\tempurl}


\bibitem[\protect\citeauthoryear{Sen}{Sen}{2013}]%
        {sen_2013_a}
\bibfield{author}{\bibinfo{person}{Sandip Sen}.}
  \bibinfo{year}{2013}\natexlab{}.
\newblock \showarticletitle{A comprehensive approach to trust management}.
\newblock  (\bibinfo{year}{2013}), \bibinfo{pages}{797--800}.
\newblock
\urldef\tempurl%
\url{http://www.ifaamas.org/Proceedings/aamas2013/docs/p797.pdf}
\showURL{%
\tempurl}


\bibitem[\protect\citeauthoryear{Shimer and Smith}{Shimer and Smith}{2000}]%
        {shimer_2000_assortative}
\bibfield{author}{\bibinfo{person}{Robert Shimer} {and} \bibinfo{person}{Lones
  Smith}.} \bibinfo{year}{2000}\natexlab{}.
\newblock \showarticletitle{Assortative Matching and Search}.
\newblock \bibinfo{journal}{\emph{Econometrica}}  \bibinfo{volume}{68}
  (\bibinfo{date}{03} \bibinfo{year}{2000}), \bibinfo{pages}{343--369}.
\newblock
\urldef\tempurl%
\url{https://doi.org/10.1111/1468-0262.00112}
\showDOI{\tempurl}


\bibitem[\protect\citeauthoryear{Shoham and Leyton-Brown}{Shoham and
  Leyton-Brown}{2009}]%
        {yoavshoham_2009_multiagent}
\bibfield{author}{\bibinfo{person}{Yoav Shoham} {and} \bibinfo{person}{Kevin
  Leyton-Brown}.} \bibinfo{year}{2009}\natexlab{}.
\newblock \bibinfo{booktitle}{\emph{Multiagent systems: algorithmic,
  game-theoretic, and logical foundations}}.
\newblock \bibinfo{publisher}{Cambridge University Press}.
\newblock


\bibitem[\protect\citeauthoryear{Simon}{Simon}{1955}]%
        {simon_1955_a}
\bibfield{author}{\bibinfo{person}{Herbert~A. Simon}.}
  \bibinfo{year}{1955}\natexlab{}.
\newblock \showarticletitle{A Behavioral Model of Rational Choice}.
\newblock \bibinfo{journal}{\emph{The Quarterly Journal of Economics}}
  \bibinfo{volume}{69} (\bibinfo{year}{1955}), \bibinfo{pages}{99--118}.
\newblock
\urldef\tempurl%
\url{https://doi.org/10.2307/1884852}
\showDOI{\tempurl}


\bibitem[\protect\citeauthoryear{Singh, Rathore, and Park}{Singh
  et~al\mbox{.}}{2019}]%
        {singh_2019_blockiotintelligence}
\bibfield{author}{\bibinfo{person}{Sushil~Kumar Singh},
  \bibinfo{person}{Shailendra Rathore}, {and} \bibinfo{person}{Jong~Hyuk
  Park}.} \bibinfo{year}{2019}\natexlab{}.
\newblock \showarticletitle{BlockIoTIntelligence: A Blockchain-enabled
  Intelligent IoT Architecture with Artificial Intelligence}.
\newblock \bibinfo{journal}{\emph{Future Generation Computer Systems}}
  (\bibinfo{date}{09} \bibinfo{year}{2019}).
\newblock
\urldef\tempurl%
\url{https://doi.org/10.1016/j.future.2019.09.002}
\showDOI{\tempurl}


\bibitem[\protect\citeauthoryear{Smith and Price}{Smith and Price}{1973}]%
        {smith1973logic}
\bibfield{author}{\bibinfo{person}{JMPGR Smith} {and} \bibinfo{person}{George~R
  Price}.} \bibinfo{year}{1973}\natexlab{}.
\newblock \showarticletitle{The logic of animal conflict}.
\newblock \bibinfo{journal}{\emph{Nature}} \bibinfo{volume}{246},
  \bibinfo{number}{5427} (\bibinfo{year}{1973}), \bibinfo{pages}{15--18}.
\newblock


\bibitem[\protect\citeauthoryear{Smith}{Smith}{1979}]%
        {smith1979game}
\bibfield{author}{\bibinfo{person}{John~Maynard Smith}.}
  \bibinfo{year}{1979}\natexlab{}.
\newblock \showarticletitle{Game theory and the evolution of behaviour}.
\newblock \bibinfo{journal}{\emph{Proceedings of the Royal Society of London.
  Series B. Biological Sciences}} \bibinfo{volume}{205}, \bibinfo{number}{1161}
  (\bibinfo{year}{1979}), \bibinfo{pages}{475--488}.
\newblock


\bibitem[\protect\citeauthoryear{SMITH and PRICE}{SMITH and PRICE}{1973}]%
        {smith_1973_the}
\bibfield{author}{\bibinfo{person}{J.~MAYNARD SMITH} {and}
  \bibinfo{person}{G.~R. PRICE}.} \bibinfo{year}{1973}\natexlab{}.
\newblock \showarticletitle{The Logic of Animal Conflict}.
\newblock \bibinfo{journal}{\emph{Nature}}  \bibinfo{volume}{246}
  (\bibinfo{date}{11} \bibinfo{year}{1973}), \bibinfo{pages}{15--18}.
\newblock
\urldef\tempurl%
\url{https://doi.org/10.1038/246015a0}
\showDOI{\tempurl}


\bibitem[\protect\citeauthoryear{Staatz}{Staatz}{1983}]%
        {staatz_1983_the}
\bibfield{author}{\bibinfo{person}{John~M. Staatz}.}
  \bibinfo{year}{1983}\natexlab{}.
\newblock \showarticletitle{The Cooperative as a Coalition: A Game‐Theoretic
  Approach}.
\newblock \bibinfo{journal}{\emph{American Journal of Agricultural Economics}}
  \bibinfo{volume}{65} (\bibinfo{date}{12} \bibinfo{year}{1983}),
  \bibinfo{pages}{1084--1089}.
\newblock
\urldef\tempurl%
\url{https://doi.org/10.2307/1240425}
\showDOI{\tempurl}


\bibitem[\protect\citeauthoryear{Sutton and Barto}{Sutton and Barto}{2018}]%
        {sutton2018reinforcement}
\bibfield{author}{\bibinfo{person}{Richard~S Sutton} {and}
  \bibinfo{person}{Andrew~G Barto}.} \bibinfo{year}{2018}\natexlab{}.
\newblock \bibinfo{booktitle}{\emph{Reinforcement learning: An introduction}}.
\newblock \bibinfo{publisher}{MIT press}.
\newblock


\bibitem[\protect\citeauthoryear{Tardos and Vazirani}{Tardos and
  Vazirani}{2007}]%
        {tardos2007basic}
\bibfield{author}{\bibinfo{person}{Eva Tardos} {and} \bibinfo{person}{Vijay~V
  Vazirani}.} \bibinfo{year}{2007}\natexlab{}.
\newblock \showarticletitle{Basic solution concepts and computational issues}.
\newblock \bibinfo{journal}{\emph{Algorithmic game theory}}
  (\bibinfo{year}{2007}), \bibinfo{pages}{3--28}.
\newblock


\bibitem[\protect\citeauthoryear{Tong, Dong, and Zheng}{Tong
  et~al\mbox{.}}{2019}]%
        {tong_2019_trustpbft}
\bibfield{author}{\bibinfo{person}{Wei Tong}, \bibinfo{person}{Xuewen Dong},
  {and} \bibinfo{person}{Jiawei Zheng}.} \bibinfo{year}{2019}\natexlab{}.
\newblock \bibinfo{title}{Trust-PBFT: A PeerTrust-Based Practical Byzantine
  Consensus Algorithm}.
\newblock , \bibinfo{numpages}{344–349}~pages.
\newblock
\urldef\tempurl%
\url{https://doi.org/10.1109/NaNA.2019.00066}
\showDOI{\tempurl}


\bibitem[\protect\citeauthoryear{TRON}{TRON}{2018}]%
        {tron_2018_advanced}
\bibfield{author}{\bibinfo{person}{TRON}.} \bibinfo{year}{2018}\natexlab{}.
\newblock \bibinfo{title}{Advanced Decentralized Blockchain Platform Whitepaper
  Version: 2.0 San Francisco}.
\newblock
\newblock
\urldef\tempurl%
\url{https://tron.network/static/doc/white_paper_v_2_0.pdf}
\showURL{%
\tempurl}


\bibitem[\protect\citeauthoryear{Tsabary, Yechieli, Manuskin, and Eyal}{Tsabary
  et~al\mbox{.}}{2021}]%
        {tsabary_2021_madhtlc}
\bibfield{author}{\bibinfo{person}{Itay Tsabary}, \bibinfo{person}{Matan
  Yechieli}, \bibinfo{person}{Alex Manuskin}, {and} \bibinfo{person}{Ittay
  Eyal}.} \bibinfo{year}{2021}\natexlab{}.
\newblock \showarticletitle{MAD-HTLC: Because HTLC is Crazy-Cheap to Attack}.
\newblock \bibinfo{journal}{\emph{arXiv:2006.12031 [cs]}} (\bibinfo{date}{03}
  \bibinfo{year}{2021}).
\newblock
\urldef\tempurl%
\url{https://arxiv.org/abs/2006.12031}
\showURL{%
\tempurl}


\bibitem[\protect\citeauthoryear{Veronese, Correia, Bessani, Lung, and
  Verissimo}{Veronese et~al\mbox{.}}{2013}]%
        {veronese_2013_efficient}
\bibfield{author}{\bibinfo{person}{Giuliana~Santos Veronese},
  \bibinfo{person}{Miguel Correia}, \bibinfo{person}{Alysson~Neves Bessani},
  \bibinfo{person}{Lau~Cheuk Lung}, {and} \bibinfo{person}{Paulo Verissimo}.}
  \bibinfo{year}{2013}\natexlab{}.
\newblock \showarticletitle{Efficient Byzantine Fault-Tolerance}.
\newblock \bibinfo{journal}{\emph{IEEE Trans. Comput.}}  \bibinfo{volume}{62}
  (\bibinfo{date}{01} \bibinfo{year}{2013}), \bibinfo{pages}{16--30}.
\newblock
\urldef\tempurl%
\url{https://doi.org/10.1109/tc.2011.221}
\showDOI{\tempurl}


\bibitem[\protect\citeauthoryear{Wackerow}{Wackerow}{2022}]%
        {wackerow_2022_proofofstake}
\bibfield{author}{\bibinfo{person}{Paul Wackerow}.}
  \bibinfo{year}{2022}\natexlab{}.
\newblock \bibinfo{title}{Proof-of-stake (PoS)}.
\newblock
\newblock
\urldef\tempurl%
\url{https://ethereum.org/en/developers/docs/consensus-mechanisms/pos/}
\showURL{%
\tempurl}


\bibitem[\protect\citeauthoryear{Wadhwa, Stoeter, Zhang, and Nayak}{Wadhwa
  et~al\mbox{.}}{2022}]%
        {cryptoeprint:2022/546}
\bibfield{author}{\bibinfo{person}{Sarisht Wadhwa}, \bibinfo{person}{Jannis
  Stoeter}, \bibinfo{person}{Fan Zhang}, {and} \bibinfo{person}{Kartik Nayak}.}
  \bibinfo{year}{2022}\natexlab{}.
\newblock \bibinfo{title}{He-HTLC: Revisiting Incentives in HTLC}.
\newblock \bibinfo{howpublished}{Cryptology ePrint Archive, Paper 2022/546}.
\newblock
\urldef\tempurl%
\url{https://eprint.iacr.org/2022/546}
\showURL{%
\tempurl}
\newblock
\shownote{\url{https://eprint.iacr.org/2022/546}}.


\bibitem[\protect\citeauthoryear{Wang, Ji, Liu, Li, Li, Zhang, and Shi}{Wang
  et~al\mbox{.}}{2021}]%
        {wang_2021_an}
\bibfield{author}{\bibinfo{person}{Feilong Wang}, \bibinfo{person}{Yipeng Ji},
  \bibinfo{person}{Mingsheng Liu}, \bibinfo{person}{Yangyang Li},
  \bibinfo{person}{Xiong Li}, \bibinfo{person}{Xu Zhang}, {and}
  \bibinfo{person}{Xiaojun Shi}.} \bibinfo{year}{2021}\natexlab{}.
\newblock \showarticletitle{An Optimization Strategy for PBFT Consensus
  Mechanism Based On Consortium Blockchain}.
\newblock \bibinfo{journal}{\emph{Proceedings of the 3rd ACM International
  Symposium on Blockchain and Secure Critical Infrastructure}}
  (\bibinfo{date}{05} \bibinfo{year}{2021}).
\newblock
\urldef\tempurl%
\url{https://doi.org/10.1145/3457337.3457843}
\showDOI{\tempurl}


\bibitem[\protect\citeauthoryear{Wang, Yang, Li, Chen, and Hu}{Wang
  et~al\mbox{.}}{2019}]%
        {wang2019data}
\bibfield{author}{\bibinfo{person}{Puming Wang}, \bibinfo{person}{Laurence~T
  Yang}, \bibinfo{person}{Jintao Li}, \bibinfo{person}{Jinjun Chen}, {and}
  \bibinfo{person}{Shangqing Hu}.} \bibinfo{year}{2019}\natexlab{}.
\newblock \showarticletitle{Data fusion in cyber-physical-social systems:
  State-of-the-art and perspectives}.
\newblock \bibinfo{journal}{\emph{Information Fusion}}  \bibinfo{volume}{51}
  (\bibinfo{year}{2019}), \bibinfo{pages}{42--57}.
\newblock


\bibitem[\protect\citeauthoryear{Wang and Singh}{Wang and Singh}{2007}]%
        {wang_2007_formal}
\bibfield{author}{\bibinfo{person}{Yonghong Wang} {and}
  \bibinfo{person}{Munindar~P Singh}.} \bibinfo{year}{2007}\natexlab{}.
\newblock \showarticletitle{Formal trust model for multiagent systems}.
\newblock  (\bibinfo{year}{2007}), \bibinfo{pages}{1551–1556}.
\newblock
\urldef\tempurl%
\url{https://www.aaai.org/Papers/IJCAI/2007/IJCAI07-250.pdf}
\showURL{%
\tempurl}


\bibitem[\protect\citeauthoryear{Wang and Wattenhofer}{Wang and
  Wattenhofer}{2020}]%
        {wang_2020_asynchronous}
\bibfield{author}{\bibinfo{person}{Ye Wang} {and} \bibinfo{person}{Roger
  Wattenhofer}.} \bibinfo{year}{2020}\natexlab{}.
\newblock \showarticletitle{Asynchronous Byzantine Agreement in Incomplete
  Networks}.
\newblock \bibinfo{journal}{\emph{Proceedings of the 2nd ACM Conference on
  Advances in Financial Technologies}} (\bibinfo{date}{10}
  \bibinfo{year}{2020}).
\newblock
\urldef\tempurl%
\url{https://doi.org/10.1145/3419614.3423250}
\showDOI{\tempurl}


\bibitem[\protect\citeauthoryear{Xiao, Zhang, Lou, and Hou}{Xiao
  et~al\mbox{.}}{2020}]%
        {xiao2020survey}
\bibfield{author}{\bibinfo{person}{Yang Xiao}, \bibinfo{person}{Ning Zhang},
  \bibinfo{person}{Wenjing Lou}, {and} \bibinfo{person}{Y~Thomas Hou}.}
  \bibinfo{year}{2020}\natexlab{}.
\newblock \showarticletitle{A survey of distributed consensus protocols for
  blockchain networks}.
\newblock \bibinfo{journal}{\emph{IEEE Communications Surveys \& Tutorials}}
  \bibinfo{volume}{22}, \bibinfo{number}{2} (\bibinfo{year}{2020}),
  \bibinfo{pages}{1432--1465}.
\newblock


\bibitem[\protect\citeauthoryear{Xing and Marwala}{Xing and Marwala}{2018}]%
        {xing_2018_the}
\bibfield{author}{\bibinfo{person}{Bo Xing} {and} \bibinfo{person}{Tshilidzi
  Marwala}.} \bibinfo{year}{2018}\natexlab{}.
\newblock \showarticletitle{The Synergy of Blockchain and Artificial
  Intelligence}.
\newblock \bibinfo{journal}{\emph{SSRN Electronic Journal}}
  (\bibinfo{year}{2018}).
\newblock
\urldef\tempurl%
\url{https://doi.org/10.2139/ssrn.3225357}
\showDOI{\tempurl}


\bibitem[\protect\citeauthoryear{Xue, Xu, Wu, Lu, and Xu}{Xue
  et~al\mbox{.}}{2020}]%
        {xue_2020_incentive}
\bibfield{author}{\bibinfo{person}{Gang Xue}, \bibinfo{person}{Jia Xu},
  \bibinfo{person}{Hanwen Wu}, \bibinfo{person}{Weifeng Lu}, {and}
  \bibinfo{person}{Lijie Xu}.} \bibinfo{year}{2020}\natexlab{}.
\newblock \showarticletitle{Incentive Mechanism for Rational Miners in Bitcoin
  Mining Pool}.
\newblock \bibinfo{journal}{\emph{Information Systems Frontiers}}
  (\bibinfo{date}{06} \bibinfo{year}{2020}).
\newblock
\urldef\tempurl%
\url{https://doi.org/10.1007/s10796-020-10019-2}
\showDOI{\tempurl}


\bibitem[\protect\citeauthoryear{Yaish, Stern, and Zohar}{Yaish
  et~al\mbox{.}}{2022}]%
        {cryptoeprint:2022/1020}
\bibfield{author}{\bibinfo{person}{Aviv Yaish}, \bibinfo{person}{Gilad Stern},
  {and} \bibinfo{person}{Aviv Zohar}.} \bibinfo{year}{2022}\natexlab{}.
\newblock \bibinfo{title}{Uncle Maker: (Time)Stamping Out The Competition in
  Ethereum}.
\newblock \bibinfo{howpublished}{Cryptology ePrint Archive, Paper 2022/1020}.
\newblock
\urldef\tempurl%
\url{https://eprint.iacr.org/2022/1020}
\showURL{%
\tempurl}
\newblock
\shownote{\url{https://eprint.iacr.org/2022/1020}}.


\bibitem[\protect\citeauthoryear{Yang and Yue}{Yang and Yue}{2019}]%
        {yang_2019_cooperation}
\bibfield{author}{\bibinfo{person}{Chun-Lei Yang} {and}
  \bibinfo{person}{Ching-Syang~Jack Yue}.} \bibinfo{year}{2019}\natexlab{}.
\newblock \showarticletitle{Cooperation in an Assortative Matching Prisoners
  Dilemma Experiment with Pro-Social Dummies}.
\newblock \bibinfo{journal}{\emph{Scientific Reports}}  \bibinfo{volume}{9}
  (\bibinfo{date}{09} \bibinfo{year}{2019}).
\newblock
\urldef\tempurl%
\url{https://doi.org/10.1038/s41598-019-50083-6}
\showDOI{\tempurl}


\bibitem[\protect\citeauthoryear{Yin, Malkhi, Reiter, Gueta, and Abraham}{Yin
  et~al\mbox{.}}{2019}]%
        {yin_2019_hotstuff}
\bibfield{author}{\bibinfo{person}{Maofan Yin}, \bibinfo{person}{Dahlia
  Malkhi}, \bibinfo{person}{Michael~K. Reiter}, \bibinfo{person}{Guy~Golan
  Gueta}, {and} \bibinfo{person}{Ittai Abraham}.}
  \bibinfo{year}{2019}\natexlab{}.
\newblock \showarticletitle{HotStuff}.
\newblock \bibinfo{journal}{\emph{Proceedings of the 2019 ACM Symposium on
  Principles of Distributed Computing}} (\bibinfo{date}{07}
  \bibinfo{year}{2019}).
\newblock
\urldef\tempurl%
\url{https://doi.org/10.1145/3293611.3331591}
\showDOI{\tempurl}


\bibitem[\protect\citeauthoryear{Zhang, Ma, and Liu}{Zhang
  et~al\mbox{.}}{2022}]%
        {zhang2022sok}
\bibfield{author}{\bibinfo{person}{Luyao Zhang}, \bibinfo{person}{Xinshi Ma},
  {and} \bibinfo{person}{Yulin Liu}.} \bibinfo{year}{2022}\natexlab{}.
\newblock \showarticletitle{SoK: Blockchain Decentralization}.
\newblock \bibinfo{journal}{\emph{arXiv preprint arXiv:2205.04256}}
  (\bibinfo{year}{2022}).
\newblock


\bibitem[\protect\citeauthoryear{Zhang, Xue, and Liu}{Zhang
  et~al\mbox{.}}{2019}]%
        {zhang_2019_security}
\bibfield{author}{\bibinfo{person}{Rui Zhang}, \bibinfo{person}{Rui Xue}, {and}
  \bibinfo{person}{Ling Liu}.} \bibinfo{year}{2019}\natexlab{}.
\newblock \showarticletitle{Security and Privacy on Blockchain}.
\newblock \bibinfo{journal}{\emph{Comput. Surveys}}  \bibinfo{volume}{52}
  (\bibinfo{date}{07} \bibinfo{year}{2019}), \bibinfo{pages}{1--34}.
\newblock
\urldef\tempurl%
\url{https://doi.org/10.1145/3316481}
\showDOI{\tempurl}


\bibitem[\protect\citeauthoryear{Zhang}{Zhang}{2023}]%
        {zhang2023design}
\bibfield{author}{\bibinfo{person}{Sunshine Zhang}.}
  \bibinfo{year}{2023}\natexlab{}.
\newblock \showarticletitle{The Design Principle of Blockchain: An Initiative
  for the SoK of SoKs}.
\newblock \bibinfo{journal}{\emph{arXiv preprint arXiv:2301.00479}}
  (\bibinfo{year}{2023}).
\newblock


\end{thebibliography}

\onecolumn
\section{Tables and Figures}
\begin{table}[!htbp]
\caption{Game theory studies on blockchain consensus: dynamic game of imperfect information~\cite{fudenberg1991game}}
\begin{tabular}{|m{1.8cm}|c|c|c|c|m{1.8cm}|m{1.8cm}|}
\hline
   Literature & \makecell{Consensus\\ protocol} & \makecell{Game Theory \\ Solution Concept} & Agent Types & Evaluation \\ \hline\hline
   
    \citet{moscibroda_2006_when} & \makecell{PBFT variant\\ (non-forkable)} & PBE  & \makecell{rational or\\byzantine} & \makecell{consensus\\ social welfare} \\ \hline
    
    \citet{article} & \makecell{PoW (election)\\ PBFT variant\\(transaction)} & none & \makecell{honest or\\byzantine} & \makecell{agreement\\safty and liveness}\\ \hline
    
    \citet{saleh_2018_blockchain} & \makecell{PoS \\ (forkable)} & \makecell{PBE\\~\citet{fudenberg1991perfect}} & rational & \makecell{consensus\\ social welfare}\\ \hline

    \citet{cong_2019_decentralized} & \makecell{PBFT variant\\ (non-forkable)} & PBE  & \makecell{rational or\\byzantine} & \makecell{consensus\\ social welfare} \\ \hline
    
    \citet{biais_2019_the}& \makecell{PoW\\ (forkable)} & \makecell{MPE\\~\citet{maskin2001markov}} & rational & \makecell{consensus\\ social welfare} \\ \hline
    
     \citet{10.5555/3398761.3398772} & \makecell{PBFT variant\\ (non-forkable)} & \makecell{PBE} & \makecell{rational or\\byzantine} &\makecell{consensus\\termination\\validity} \\ \hline
    
    \citet{halaburda_2021_an} & \makecell{PBFT variant\\ (non-forkable)} & PBE  & \makecell{rational or\\byzantine} & \makecell{consensus\\ social welfare} \\ \hline 
    
    \citet{gersbach_2022_risky} & \makecell{PBFT variant\\ (non-forkable)} & PBE  & \makecell{rational or\\byzantine} & \makecell{consensus\\ social welfare} \\ \hline
    
    \citet{gersbach_2022_staking} & \makecell{PBFT variant\\ (non-forkable)} & PBE  & \makecell{rational or\\byzantine} & \makecell{consensus\\ social welfare} \\ \hline
    
    Ours & \makecell{pBFT variant\\ (non-forkable)}& \makecell{ESS\\~\citet{smith1979game}\\~\citet{{smith1973logic}}} & \makecell{bounded \\ rational} & \makecell{safety\\liveness\\validity\\social welfare}\\ \hline
\end{tabular}
\label{tab:1}
\end{table}

\begin{table}[!htbp]
\caption{Stable equilibria with reparametrization}
\begin{tabular}{|c|c|c|}
\hline
    Equilibrium Type & Conditions & Agents Payoff \\ \hline\hline
   
    Honest Stable Equilibrium & \makecell{$1 - \gamma \geqslant x_{1} > \max(\frac{1}{2}+\frac{\frac{1}{2}}{2\alpha-2\beta+1}, \gamma)$\\or $\quad x_{1} > \max(1-\gamma, \gamma)$} & \makecell{Equation ~\ref{eq:7}\\Equation ~\ref{eq:13}}\\ \hline
    
    Byzantine Stable Equilibrium & \makecell{$\gamma \leqslant x_{1} < min(\frac{1}{2}+\frac{\frac{1}{2}}{2\alpha-2\beta+1},1 - \gamma)$\\or $\quad x_{1} < \min(1-\gamma, \gamma)$} & \makecell{Equation ~\ref{eq:17}}\\ \hline
    
    Pooling Stable Equilibrium & \makecell{$1-\gamma \geqslant x_{1} = \frac{1}{2}+\frac{\frac{1}{2}}{2\alpha-2\beta+1} \geqslant \gamma$ \\or $\quad m = 1$} & \makecell{Equation ~\ref{eq:7}}\\ \hline
\end{tabular}
\label{tab:5}
\end{table}

\begin{table}[!htbp]
\caption{Evaluations with safety, liveness, and validity (The Honest (1) refers to the equilibrium condition in Lemma 3(1) and Honest (2) refers to the equilibrium condition in Lemma 3(2); The Byzantine (1) and (2) each refers to the two equilibrium conditions in Lemma 3(3).)}
\begin{tabular}{|c|c|c|c|c|c|c|c|c|}
\hline

\makecell{Stable\\Equilibium} & \makecell{Initial\\Honest\\Pivotal} & \makecell{Initial\\Byzantine\\Pivotal} & \makecell{Immediate\\Safety} & \makecell{Eventual\\Safety} & \makecell{Immediate\\Liveness} & \makecell{Eventual\\Liveness} & \makecell{Eventual\\Validity} & \makecell{Social\\Welfare\\for agents\\with honest\\strategy}\\
\hline \hline

\makecell{Honest (1)} & yes & yes & no & yes & yes & yes & yes & $R-c_{check}-c_{send}$ \\\hline

\makecell{Honest (2)} & yes & no & yes & yes & yes & yes & yes & $R-c_{check}-c_{send}$ \\\hline

\makecell{Byzantine (1)} & yes & yes & no & no & yes & no & no & $0$ \\\hline

\makecell{Byzantine (2)} & yes & yes & no & no & no & no & no & $0$ \\\hline

\makecell{Pooling} & yes & yes & no & no & yes & yes & no & \makecell{$R-c_{check}-c_{send}-\frac{(R-c_{send})(R-c_{send}+\kappa)}{2R-2c_{send}+\kappa}$\\(less than optimal)} \\\hline

\hline
\end{tabular}
\label{tab:6}
\end{table}

\begin{table}[!htbp]
\caption{Policy parameters that affect equilibrium outcomes in steady states}
\begin{tabular}{|c|m{7cm}|m{4.5cm}|}
\hline

\makecell{notation} & \makecell{definition in our model}  & \makecell{corresponding parameters\\in Ethereum 2.0} \\
\hline \hline

\makecell{$R$ or $\alpha$} & the reward to the validators who send a message when the block is accepted, or the reward-punishment ratio & \multirow{2}{*}{\makecell[l]{Reward Per Block (about 2ETH)\\\cite{bitinfocharts_2019_ethereum}}}\\\cline{1-2}

\makecell{$c_{check}$} & the cost to the validators who check the validity of the proposed block & ~\\\cline{1-2}

\makecell{$c_{send}$ or $\beta$} & the cost of the validators who send a message, or the cost-punishment ratio & ~\\\hline

\makecell{$\kappa$} & the cost occurs to all validators with the honest strategy when an invalid block is accepted & slashing\cite{he_2022_contract} \\ \hline

\makecell{$\nu$ or $\gamma$} & the majority threshold of the votes, or the pivotality rate & Depositing 32 ETH to activate a validator\cite{ethereumfoundation_2022_solo}\\\hline

\makecell{$x_{1}$} & the initial number of agents with honest strategies & KYC problem \cite{parramoyano_2017_kyc}\\\hline

\hline
\end{tabular}
\label{tab:7}
\end{table}

\newpage
\appendix
\section{Proofs for proposition}

\begin{proof}[Proof for Lemma 3(1)]

    When $N(1-x_{1}) \geqslant \nu$, agents playing $S_{B}$ has pivotality in the voting process. Therefore, to reach the honest stable equilibrium, the initial number of agents playing $S_{H}$ should at least reach $\nu$, meaning $Nx_{1} \geqslant \nu$.
    
    \textit{Validator behaviors and proposal outcome} In round $1$, if the proposer $P_{1}$ plays $S_{B}$, she proposes an invalid proposal $h_{1}$. After all the agents check the validity of $h_{t}$, agents with $S_{B}$ vote for $h_{1}$ while agents with $S_{H}$ do nothing. $h_{1}$ will be accepted since $N(1-x_{1}) \geqslant \nu$. 
    
    Therefore, the payoffs in round $1$ for the validators with $P_{1}$ playing $S_{B}$ are:
    \begin{align}\label{eq:5}
    \begin{aligned}
        V_{BB}(x_{1})&=R-c_{check}-c_{send}\\
        V_{HB}(x_{1})&=-c_{check}-\kappa
    \end{aligned}
    \end{align}

    In round $1$, if the proposer $P_{1}$ plays $S_{H}$, she proposes a valid proposal $h_{1}$. After all the agents check the validity of $h_{1}$, agents with $S_{H}$ vote for $h_{1}$ while agents with $S_{B}$ do nothing. $h_{1}$ will be accepted since $Nx_{1} \geqslant \nu$. 
    
	Therefore, the payoffs in round $1$ for the validators with $P_{1}$ playing $S_{H}$ are:
	\begin{align}\label{eq:6}
	\begin{aligned}
	    V_{HH}(x_{1})&=R-c_{check}-c_{send}\\
	    V_{BH}(x_{1})&=-c_{check}
	\end{aligned}
	\end{align}
	
	\textit{Validator Payoff matrix} The payoff matrix for validators in round 1 when initially $n(1-x_{1}) \geqslant \nu$ and $nx_{1} \geqslant \nu$ is shown in Table~\ref{tab:2}.
	
	\begin{table}[H]
    \caption{Payoff matrix when $n(1-x_{1}) \geqslant \nu$ and $nx_{1} \geqslant \nu$}
    \begin{tabular}{|m{4cm}||p{4cm}|m{4cm}|}
    \hline
        ~ & Proposer -- Honest & Proposer -- Byzantine \\ \hline\hline
        Validator -- Honest & $V_{HH}(x_{1})=R-c_{check}-c_{send}$ & $V_{HB}(x_{1})=-c_{check}-\kappa$ \\ \hline
        Validator -- Byzantine & $V_{BH}(x_{1})=-c_{check}$ & $V_{BB}(x_{1})=R-c_{check}-c_{send}$ \\ \hline
    \end{tabular}
    \label{tab:2}
    \end{table}
	
    From equations ~\ref{eq:1} from lemma 1, ~\ref{eq:2} from lemma 2, ~\ref{eq:5}, and ~\ref{eq:6}, we can get:
    \begin{align}\label{eq:7}
    \begin{aligned}
        V_{H}(x_{1})=&[m+(1-m)x_{1}](R-c_{check}-c_{send})\\
        &+(1-m)(1-x_{1})(-c_{check}-\kappa)\\
        V_{B}(x_{1})=&x_{1}(1-m)(-c_{check})\\
        &+[m+(1-m)(1-x_{1})](R-c_{check}-c_{send})
    \end{aligned}
    \end{align}

    To achieve the honest stable equilibrium, $x_{t}$ should converge to 1. Then for round $t$ and $t-1$ ($t \geqslant 2$) before convergence,
    \begin{align}\label{eq:8}
        x_{t} = \frac{x_{t-1}V_{H}(x_{t-1})}{x_{t-1}V_{H}(x_{t-1})+(1-x_{t-1})V_{B}(x_{t-1})} > x_{t-1}
    \end{align}
    
    Formula ~\ref{eq:8} can be simplified to (since $x_{t} \neq x_{t-1}$ and $x_{t-1} \in (0,1)$):
    \begin{align}\label{eq:9}
        V_{H}(x_{t-1}) > V_{B}(x_{t-1})
    \end{align}
    
    And by combining equation ~\ref{eq:7} with ~\ref{eq:9} and deduction with $t = 2$:
    \begin{align}\label{eq:10}
        V_{H}(x_{1})-V_{B}(x_{1})=-(1-m)[(1-2x_{1})(R-c_{send})+(1-x_{1})\kappa]
    \end{align}
    
    Equation ~\ref{eq:10} implies that, if $m \neq 1$ and $x_{1}>\frac{R-c_{send}+\kappa}{2R-2c_{send}+\kappa}$, $x_{t}$ would converge to 1, and the system will reach the honest stable equilibrium. The backward proof can be implemented by taking $x_{1} > \max(\frac{R-c_{send}+\kappa}{2R-2c_{send}+\kappa}, \frac{\nu}{N})$ and $m \neq 1$ into formula ~\ref{eq:8}. Therefore, when $N(1-x_{1})\geqslant \nu$, the honest stable equilibrium can be reached if and only if $x_{1} > \max(\frac{R-c_{send}+\kappa}{2R-2c_{send}+\kappa}, \frac{\nu}{N})$ and $m \neq 1$.

\end{proof}

\begin{proof}[Proof for Lemma 3(2)]

    When $N(1-x_{1})<\nu$, agents playing $S_{B}$ do not have pivotality in the voting process. Therefore, to reach the honest stable equilibrium, the initial number of agents playing $S_{H}$ should at least reach $\nu$, meaning $Nx_{1} \geqslant \nu$.
    
    \textit{Validator behaviors and proposal outcome}  In round $1$, if the proposer $P_{1}$ plays $S_{B}$, she proposes an invalid proposal $h_{1}$. Agents with $S_{H}$ would check the validity of $h_{t}$ but not vote for it, while agents with $S_{B}$ do nothing because they do not have pivotality in the voting process. $h_{1}$ will not be accepted since nobody votes for it.
    
    Therefore, the payoffs in round $1$ for the validators with $P_{1}$ playing $S_{B}$ are:
    \begin{align}\label{eq:11}
    \begin{aligned}
        V_{BB}(x_{1})&=0\\
        V_{HB}(x_{1})&=-c_{check}
    \end{aligned}
    \end{align}
    
    In round $1$, if the proposer $P_{1}$ plays $S_{H}$, she proposes a valid proposal $h_{1}$. Agents with $S_{H}$ would check the validity of $h_{t}$ and vote for it, while agents with $S_{B}$ do nothing because they do not have pivotality in the voting process. $h_{1}$ will be accepted since $Nx_{1} \geqslant \nu$.
    
    Therefore, the payoffs in round $1$ for the validators with $P_{1}$ playing $S_{H}$ are:
    \begin{align}\label{eq:12}
    \begin{aligned}
        V_{HH}(x_{1})&=R-c_{check}-c_{send}\\
        V_{BH}(x_{1})&=0
    \end{aligned}
    \end{align}
    
    \textit{Validator Payoff matrix} The payoff matrix for validators in round 1 when initially $n(1-x_{1}) < \nu$ and $nx_{1} \geqslant \nu$ is shown in Table~\ref{tab:3}.
    
    \begin{table}[H]
    \caption{Payoff matrix when $n(1-x_{1}) < \nu$ and $nx_{1} \geqslant \nu$}
    \begin{tabular}{|m{4cm}||p{4cm}|m{4cm}|}
    \hline
        ~ & Proposer -- Honest & Proposer -- Byzantine \\ \hline\hline
        Validator -- Honest & $V_{HH}(x_{1})=R-c_{check}-c_{send}$ & $V_{HB}(x_{1})=-c_{check}$ \\ \hline
        Validator -- Byzantine & $V_{BH}(x_{1})=0$ & $V_{BB}(x_{1})=0$ \\ \hline
    \end{tabular}
    \label{tab:3}
    \end{table}
	
    From equations ~\ref{eq:1} from lemma 1, ~\ref{eq:2} from lemma 2, ~\ref{eq:11}, and ~\ref{eq:12}, we can get:
    \begin{align}\label{eq:13}
    \begin{aligned}
        V_{H}(x_{1})=&[m+(1-m)x_{1}](R-c_{check}-c_{send})\\
        &+(1-m)(1-x_{1})(-c_{check})\\
        V_{B}(x_{1})=&0
    \end{aligned}
    \end{align}
    
    Similar as we have proved in proof of lemma 3(1), formula ~\ref{eq:8} is satisfied with equation ~\ref{eq:13} with $t = 2$. And the backward proof can be implemented by taking $x_{1} \geqslant \frac{\nu}{N}$ with formula ~\ref{eq:8} and equation ~\ref{eq:13}. Therefore, when $N(1-x_{1}) < \nu$, the honest stable equilibrium can be reached if and only if $x_1 \geqslant \frac{\nu}{N}$.

\end{proof}

\begin{proof}[Proof for Lemma 3(3)]
    The proof for Byzantine stable equilibrium achievement \\when $x_{1} < \max(\frac{R-c_{send}+\kappa}{2R-2c_{send}+\kappa}, \frac{\nu}{N})$ and $m \neq 1$ is similar to the proofs of lemma 3(1) by changing formula ~\ref{eq:8} into:
    \begin{align}\label{eq:14}
    \begin{aligned}
        x_{t} = \frac{x_{t}V_{H}(x_{t-1})}{x_{t}V_{H}(x_{t-1})+(1-x_{t})V_{B}(x_{t-1})} < x_{t-1}
    \end{aligned}
    \end{align} 
    
    And the proof for Byzantine stable equilibrium achievement when $x < min(1-\frac{\nu}{N}, \frac{\nu}{N})$ is as below:
    
    In contract with the assumption that $Nx_{1} \geqslant \nu$, when $Nx_{1} < \nu$, the agents with $S_{H}$ do not have pivotality in the voting process.
    
    \textit{Validator behaviors and proposal outcome} In round $1$, if the proposer $P_{1}$ plays $S_{B}$, she proposes an invalid proposal $h_{1}$. Agents with $S_{B}$ would check the validity of $h_{1}$ and vote for it, while agents with $S_{H}$ do nothing because they don't have pivotality in the voting process. $h_{1}$ will be accepted since $N(1-x_{1}) \geqslant \nu$. 
    
    Therefore, the payoffs in round $1$ for the validators with $P_{1}$ playing $S_{B}$ are:
    \begin{align}\label{eq:15}
    \begin{aligned}
        V_{BB}(x_{1})&=R-c_{check}-c_{send}\\
        V_{HB}(x_{1})&=-\kappa
    \end{aligned}
    \end{align}

    In round $1$, if the proposer $P_{1}$ plays $S_{H}$, she proposes a valid proposal $h_{1}$. Agents with $S_{B}$ would check the validity of $h_{1}$ but not vote for it, while agents with $S_{H}$ do nothing because they don't have pivotality in the voting process. $h_{1}$ will not be accepted since nobody votes for it.
    
	Therefore, the payoffs in round $1$ for the validators with $P_{1}$ playing $S_{H}$ are:
	\begin{align}\label{eq:16}
	\begin{aligned}
	    V_{HH}(x_{1})&=0\\
	    V_{BH}(x_{1})&=-c_{check}
	\end{aligned}
	\end{align}
	
	\textit{Validator Payoff matrix} The payoff matrix for validators in round 1 when initially $n(1-x_{1}) \geqslant \nu$ and $nx_{1} < \nu$ is shown in Table~\ref{tab:4}.
	
	\begin{table}[H]
    \caption{Payoff matrix when $n(1-x_{1}) \geqslant \nu$ and $nx_{1} < \nu$}
    \begin{tabular}{|m{4cm}||p{4cm}|m{4cm}|}
    \hline
        ~ & Proposer -- Honest & Proposer -- Byzantine \\ \hline\hline
        Validator -- Honest & $V_{HH}(x_{1})=0$ & $V_{HB}(x_{1})=-\kappa$ \\ \hline
        Validator -- Byzantine & $V_{BH}(x_{1})=-c_{check}$ & $V_{BB}(x_{1})=R-c_{check}-c_{send}$ \\ \hline
    \end{tabular}
    \label{tab:4}
    \end{table}

    From equations ~\ref{eq:1} from lemma 1, ~\ref{eq:2} from lemma 2, ~\ref{eq:15}, and ~\ref{eq:16}, we can get:
    \begin{align}\label{eq:17}
    \begin{aligned}
        V_{H}(x_{1})=&(1-m)(1-x_{1})(-\kappa)\\
        V_{B}(x_{1})=&x_{1}(1-m)(-c_{check})\\
        &+[m+(1-m)(1-x_{1})](R-c_{check}-c_{send})
    \end{aligned}
    \end{align}
    
    Equation ~\ref{eq:17} shows that, the expected payoff of a validator with $S_{H}$ is smaller than a validator with $S_{B}$, and since $N(1-x_{1}) \geqslant \nu$ and $Nx_{1} < \nu$, $1-x_{1}>x_{1}$. Therefore, 
    \begin{align}\label{eq:18}
        x_{t} = \frac{x_{t}V_{H}(x_{t-1})}{x_{t}V_{H}(x_{t-1})+(1-x_{t})V_{B}(x_{t-1})} > x_{t-1}
    \end{align}
    is achieved with any $t$ before convergence. The system will converge to Byzantine equilibrium. The backward proof can be implemented by taking the corresponding values of $x_{1}$ into equations ~\ref{eq:14} and ~\ref{eq:17}. Therefore, if $N(1-x_{1}) \geqslant \nu$, only if $x_{1} < \min(\frac{R-c_{send}+\kappa}{2R-2c_{send}+\kappa}, \frac{\nu}{N})$ and $m \neq 1$, or $x < min(1-\frac{\nu}{N}, \frac{\nu}{N})$ the Byzantine stable equilibrium can be reached.

\end{proof}

\begin{proof}[Proof for Lemma 3(4)]
    In the proof for lemma 3(1), equation ~\ref{eq:10} implies that, if $m = 1$ or $x = \frac{R-c_{send}+\kappa}{2R-2c_{send}+\kappa}$, $V_{H}(x_{1}) = V_{B}(x_{1})$, and therefore formula ~\ref{eq:8} would become $x_{t} = \frac{x_{t}V_{H}(x_{t-1})}{x_{t}V_{H}(x_{t-1})+(1-x_{t})V_{B}(x_{t-1})} = x_{t-1}$. Since $x_{t} = x_{t-1} \in (0,1)$, the pooling stable equilibrium is reached. Proof for lemma 3(2) shows that the pooling stable equilibrium is never reached when initially $N(1-x_{1}) < \nu$ or $Nx_{1} \leqslant \nu$. The backward proof can be implemented by taking $m = 1$ or $x = \frac{R-c_{send}+\kappa}{2R-2c_{send}+\kappa}$ into formula ~\ref{eq:8} and it would become an equation.
\end{proof}

\begin{proof}[Proof for Lemma 3(5)]
    Since neither the number of miners with the honest strategy nor the number of miners with the Byzantine strategy is larger than the threshold, no block will be built, and none of the bounded-rational miners will check the validity nor send a message. Therefore, the payoff for all two strategies is zero and the block-building process is stopped thus the protocol meets a failure.
\end{proof}

\newpage
\section{Glossary Table}

\begin{table}[H]
\caption{Glossary Table}
\begin{tabular}{|m{2cm}|m{6cm}|m{2cm}|m{2cm}|}
\hline
    Glossary & Definition & Reference & Type \\ \hline\hline
   
    \makecell[l]{Practice\\ Byzantine\\ Fault\\ Tolerance \\(pBFT)}  & A consensus algorithm to deal with the Byzantine General Problem. & \citet{castro_2002_practical} & \makecell[l]{Consensus\\protocol}  \\ \hline
    
    Proof-of-Work (PoW) & A form of cryptographic proof in which one party (the prover) proves to others (the verifiers) that a certain amount of a specific computational effort has been expended. & \citet{nakamoto_2008_bitcoin} & \makecell[l]{Consensus\\protocol} \\ \hline
    
    Proof-of-Stake (PoS) & A class of consensus mechanisms for blockchains that work by selecting validators in proportion to their quantity of holdings in the associated cryptocurrency. & \citet{saleh_2018_blockchain} & \makecell[l]{Consensus\\protocol} \\ \hline
    
    Player/ Miner/ Agent / Node & The objects mining in the blockchain. (These three words have the same representations in our paper) & \cite{castro_2002_practical}  & \makecell[l]{Distributed\\computing/\\ Game theory} \\ \hline

    Agreement & \makecell[l]{(1)The nonfaulty processors compute \\exactly the same vector.\\ (2)The element of this vector \\corresponding to a given nonfaulty \\processor is the private value of that \\processor.} & \citet{pease_1980_reaching} & \makecell[l]{Evaluation\\criteria} \\ \hline
    
    Validity & A decided value by any rational player is valid; it satisfies the predefined predicate. & \citet{amoussouguenou_2020_rational} & \makecell[l]{Evaluation\\criteria} \\ \hline 
    
    Termination & Every rational player decides on a value (a block). & \citet{amoussouguenou_2020_rational} & \makecell[l]{Evaluation\\criteria} \\ \hline
    
    Consensus & At least 51\% of the nodes on the network agree on the next global state of the network. & ~ & \makecell[l]{Evaluation\\criteria} \\ \hline
    
    Safety & All non-faulty replicas agree on the sequence numbers of requests that commit locally. & \citet{castro_2002_practical} & \makecell[l]{Evaluation\\criteria} \\ \hline

    Liveness & Replicas must move to a new view if they are unable to execute a request & \citet{castro_2002_practical} & \makecell[l]{Evaluation\\criteria} \\ \hline
    
    Strategy & A function that assigns an action to each nonterminal history & \cite{osborne_1994_a} & Game theory \\ \hline
    
    Perfect Bayesian Equilibrium (PBE) & An equilibrium concept relevant for dynamic games with incomplete information (sequential Bayesian games) & \cite{fudenberg_1991_perfect} & \makecell[l]{Game theory\\ equilibrium\\ solution \\concept} \\ \hline
    
    Markov Perfect Equilibrium (MPE) & A set of mixed strategies for each of the players that satisfies some criteria & \cite{maskin_2001_markov} & \makecell[l]{Game theory\\ equilibrium\\ solution \\concept} \\ \hline
    
\end{tabular}
\label{tab:8}
\end{table}

\section{Notation Table}
\begin{table}[H]
\caption{Notation table}
\begin{tabular}{|c|m{12cm}|}
\hline
\makecell{notation} & definition \\
\hline \hline

\makecell{$\mathcal{A}$} & a set $\{A_{i}\}_{i=1}^{N}$ with $N$ elements, the committee of $N$ miners established at time $t = 0$\\

\makecell{$A_{i}$} & an agent with order $i$\\

\makecell{$N$} & the number of agents in $\mathcal{A}$, the maximum value of $i$\\

\makecell{$t$} & the enumeration of game rounds\\

\makecell{$P_{t}$} & the selected proposer for round $t$\\

\makecell{$h_{i}$} & the proposal made by $P_{t}$ at round $t$\\

\makecell{$R$} & the reward to the validators who send a message when the block is accepted\\

\makecell{$c_{check}$} & the cost to the validators who check the validity of the proposed block\\

\makecell{$c_{send}$} & the cost of the validators who send a message\\

\makecell{$\kappa$} & the cost occurs to all validators with the honest strategy when an invalid block is accepted\\

\makecell{$S_{H}$} & the honest strategy\\

\makecell{$S_{B}$} & the Byzantine strategy\\

\makecell{$s_{i}$} & the strategy chosen by agent $A_{i}$ at round $t$, $s_{i} \in \{S_{H}, S_{B}\}$\\

\makecell{$\nu$} & the majority threshold of the votes\\

\makecell{$x_{t}$} & the proportion of agents with $S_{H}$ in round $t$\\

\makecell{$x_{1}$} & the initial proportion of agents with honest strategies\\

\makecell{$m$} & the portion of the rounds that a validator believes to meet a proposer with the same strategy in the game \\

\makecell{$\pi_{ij}(x_{t})$} & the subjective meeting probability of one validator with strategy $S_{i}$ meet one proposer with strategy $S_{j}$ in function of $x_{t}$, where $i, j \in \{H,B\}$\\

\makecell{$V_{ij}(x_{t})$} & the expected payoff of one validator with strategy $S_{i}$ meet one proposer with strategy $S_{j}$ in function of $x_{t}$, where $i, j \in \{H,B\}$\\

\makecell{$V_{i}(x_{t})$} & the expected validator payoff with the subjective belief of one validator with strategy $S_{i}$ in the function of $x_{t}$\\

\makecell{$P_{H_{t}}$} & the probability that an agent chooses $S_{H}$ in round $t$\\

\makecell{$P_{B_{t}}$} & the probability that an agent chooses $S_{B}$ in round $t$\\

\makecell{$\alpha$} & the proportion of $R$ to $\kappa$\\

\makecell{$\beta$} & the proportion of $c_{check}$ to $\kappa$\\

\makecell{$\gamma$} & the proportion of $\nu$ to $N$\\

\makecell{$n$} & the number of evolutionary rounds in a game\\

\hline

\end{tabular}
\label{tab:9}
\end{table}

\end{document}